\documentclass[twocolumn, aps, 10pt, english, superscriptaddress, prl, 
amsmath, 
amssymb, floatfix, longbibliography
]{revtex4-1}
\usepackage[export]{adjustbox}
\usepackage{subcaption}
\usepackage{standalone}
\usepackage{svg}
\usepackage[table]{xcolor}

\usepackage{multirow}
\usepackage{graphicx}
\usepackage{makecell}
\usepackage[T1]{fontenc}
\usepackage[utf8]{inputenc}
\usepackage{color}
\usepackage[english]{babel}
\usepackage{graphicx}
\usepackage{cancel}
\usepackage{esint}
\usepackage{float}
\usepackage{tikz}
\usepackage{empheq}

\makeatletter

\usepackage{dcolumn}
\usepackage{bm}
\usepackage{soul}
\usepackage{amsfonts}
\usepackage{slashed}
\usepackage{enumerate}
\usepackage{mathtools}
\usepackage{physics}
\usepackage{amsthm}
\usepackage{inputenc}
\usepackage[normalem]{ulem}
\usepackage{svg}
\usepackage{comment}
\usepackage{bbm}

\newcounter{claim} \newcommand{\claim}{\refstepcounter{claim}\textit{Claim~\theclaim.}}

\let\old@makecaption=\@makecaption
\usepackage{subcaption}
\let\@makecaption=\old@makecaption

\usepackage{epsfig}
\usepackage{dsfont}
\usepackage{xcolor}
\newcommand{\Ient}{\mathfrak{I}_N^{(\text{ent})}}
\newcommand{\Iprod}{\mathfrak{I}_N^{(\text{prod})}}
\newcommand{\Icd}{\mathfrak{I}_N^{(\text{cd})}}
\newcommand{\Iult}{\mathfrak{I}^{(\text{ult})}}
\newcommand{\rev}[1]{#1}

\usepackage[unicode=true,pdfusetitle,bookmarks=true,bookmarksnumbered=false,bookmarksopen=false,breaklinks=false,pdfborder={0 0 0},pdfborderstyle={},backref=false,colorlinks=true]{hyperref}


\begin{document}

\title{Optimal quantum metrology protocols with erasure qubits}

\author{Michal Arieli}
\email{arieli.michal@gmail.com} 
\affiliation{Racah Institute of Physics, The Hebrew University of Jerusalem, Jerusalem 91904, Givat Ram, Israel}
\author{Alex Retzker}
\affiliation{Racah Institute of Physics, The Hebrew University of Jerusalem, Jerusalem 91904, Givat Ram, Israel}
\author{Tuvia Gefen}
\email{tuvia.gefen@mail.huji.ac.il}
\affiliation{Racah Institute of Physics, The Hebrew University of Jerusalem, Jerusalem 91904, Givat Ram, Israel}

\date{\today}

\begin{abstract}

We investigate the 
precision limits and optimal protocols for sensing 
single qubit signals in the presence of erasure noise.
We study a hierarchy of precision limits achievable with metrological strategies of differing complexity, and identify the optimal protocol for each.
The detectability of erasure noise is shown to lead to enhanced precision limits and simplified sensing protocols. 
For energy gap estimation, we demonstrate that
a simple product-state \rev{strategy
with} continuous erasure detection yields significant improvements, outperforming optimal passive entangled protocols \rev{over a broad range of finite system sizes}.
We 
show that
for 
other single-qubit signals, quantum error correction provides a substantial advantage by correcting the dominant erasure processes, and can restore Heisenberg-limited precision in certain erasure configurations. 
 As a byproduct of our analysis, we 
 find erasure-conversion schemes for qubits subject to thermal noise that attain the corresponding ultimate precision limits.

\end{abstract}

\maketitle

{\bf{Introduction}}--- Several quantum technologies rely on precise sensing of single-qubit signals in the presence of noise. Examples include energy gap estimation in atomic clocks 
\cite{optical_atomic_clocks, robinson2024direct, kaubruegger2021quantum, kaubruegger2025progress, kielinski2025bayesian, direkci2026heisenberg},
Zeeman shift estimation in magnetometers
\cite{schmitt2017submillihertz, boss2017quantum,magneto1,magneto2},  nano-NMR \cite{aslam2017nanoscale,glenn2018high, cohen2020achieving, nano}, and Rabi frequency estimation \cite{gefen2016parameter,plenio2016sensing,harutyunyan2025qubit}. In the absence of noise,
optimal protocols typically exploit entanglement to achieve quantum enhanced precision, where
the fundamental limit in frequency estimation is known as the Heisenberg limit (HL) \cite{giovannetti2011advances}. 
Noise \rev{can} significantly degrade precision and \rev{preclude} HL \cite{huelga1997improvement,demkowicz2012elusive}. 
In recent years, there has been a tremendous progress
in
\rev{determining optimal
sensing protocols and precision bounds across noisy quantum metrological}
settings
\cite{demkowicz2014using, herrera2015quantum,sekatski2017quantum, demkowicz2017adaptive, ZhouNC18AchievingHeisenberg,yang2019memory, zhou2020optimal, zhou2021asymptotic, liu2023optimal, kurdzialek2023using,liu2024efficient, sekatski2022optimal,gardner2025lindblad, mondal2026optimal, kielinski2024ghz, mann2025quantum, dulian2025qmetro++}.
Several methods of noise mitigation and correction for quantum metrology were devised, yet for most relevant cases, e.g., amplitude damping noise, dephasing, and coupling to a thermal bath, Heisenberg scaling cannot be \rev{recovered} \cite{demkowicz2012elusive,sekatski2017quantum, demkowicz2017adaptive, ZhouNC18AchievingHeisenberg}.



In many physical systems, the main noise mechanism can be engineered or converted to an erasure noise, which is a detectable error \cite{wu2022erasure,kubica2023erasure,scholl2023erasure, levine2024demonstrating, chow2024circuit,teoh2023dual,koottandavida2024erasure,kang2023quantum,ma2023high}. This feature has already proven highly valuable in quantum computing, where erasure conversion substantially increases error correction thresholds and improves noisy performance \cite{wu2022erasure, kubica2023erasure,gu2025fault,jacoby2025magic}. 
For quantum metrology, recent works have also demonstrated improved performance in the presence of erasure noise or via erasure conversion \cite{ouyang2021robust,ouyang2023describing,ma2019improving,kielinski2024ghz,niroula2024quantum, chen2024quantum,ma2025enhancing,lin2025covariant}. 
However, a systematic \rev{understanding} of 
the precision \rev{limits} and optimal \rev{sensing} protocols \rev{under erasure} noise is 
missing. 
In particular,
\rev{how does error detectability improve the attainable precision, and which sensing resources are required to exploit it.}


\rev{In this work, we show that the detectability of erasure noise changes both the attainable precision and the resources required to achieve it.}
For \rev{transverse} signals, such as Rabi-frequency estimation,
this detectability \rev{can} lead to  substantially improved precision limits.
In these cases, quantum error correction (QEC) can correct the dominant erasure noise and provide a significant metrological advantage, reaching HL in \rev{suitable} erasure configurations.    
For energy gap estimation, 
\rev{the improvement is bounded, but error detectability enables simpler sensing protocols.
We show 
that a continuous erasure detection protocol applied to product states yields a significant improvement, outperforming optimal passive entangled strategies for systems of up to $\sim 100$ qubits.
In the asymptotic limit, spin-squeezed states \cite{kitagawa1993squeezed} with an orientation adapted to the erasure rates attain the ultimate precision bound without intermediate control.} 
\rev{Finally, we use erasure conversion to construct  asymptotically optimal protocols for qubits subject to thermal noise that require only spin squeezing of the effective erasure qubits, without any noiseless ancillas.}





\begin{figure}[ht]
    \centering
    \includegraphics[width=\linewidth]{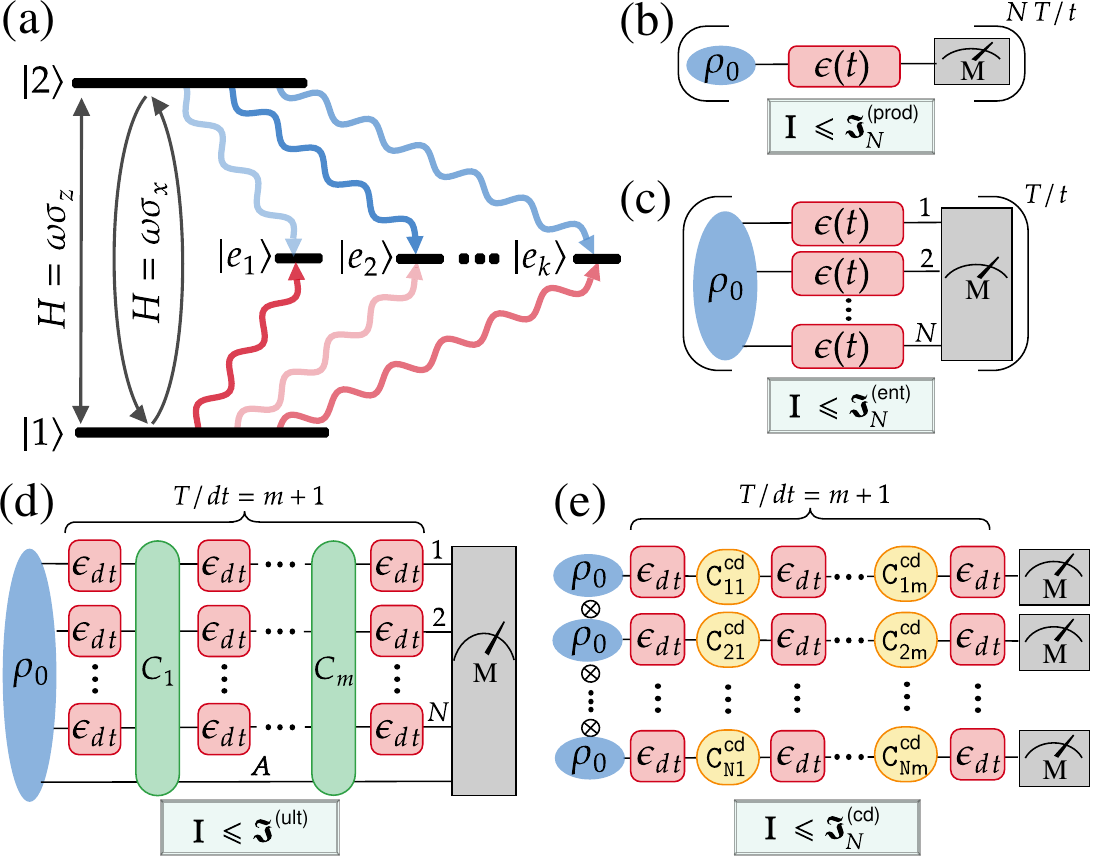}
    \caption{ (a) Schematic of the erasure qubit.
    Each level of the qubit probe $\ket{i}$ ($i=1,2$) can decay to an erasure state $\ket{e_j}$ with rate $\gamma_{i,j}$. 
    (b) 
    Product protocols (class (i)): initializing and measuring each qubit individually.
    (c) Entangled protocols (class (ii)): includes also ancilla-free entangled states and measurements. 
    (d) 
    Sequential protocols (class (iii)):
    the most general entangled and adaptive strategies, in which we allow any fast quantum  control and arbitrary entanglement of the probes with ancillas. (e) 
    Continuous erasure detection protocols (class (iv)):
    the qubits are initialized into a product state and subjected to restricted controls, allowing for erasure detection and a fast reset. 
     The achievable precision in classes (i)–(iv) is bounded by $\Iprod$, $\Ient$, $\Iult$ (asymptotically in large $N T$), and  $\Icd$, respectively.
     }    \label{fig:1}
\end{figure}

{\bf{Erasure qubit}}---
We consider a qubit probe,  and denote its computational basis as $\left\{ |1\rangle,|2\rangle\right\}.$
The Hamiltonian acting on this probe is given by $H=\omega G,$
where $\omega$ is the parameter we want to estimate and $G$ is a generic single qubit operator, and thus takes the form of 
$G = \Vec{n}\cdot \Vec{\sigma}$, where $\Vec{n}$ is a unit vector and $\Vec{\sigma}$ is the vector of Pauli matrices.
The qubit is subject to an erasure noise; that is,
 a noise that maps the qubit subspace into auxiliary states, denoted by $\{ |e_{i}\rangle\} _{i=1}^{k}$.
More concretely, we consider a time evolution given by the following Lindblad master equation:
\begin{align}
\overset{\cdot}{\rho}=-i\left[\omega G,\rho\right]+\sum_{i,j}\left(L_{i,j}\rho L_{i,j}^{\dagger}-\frac{1}{2}\left\{ L_{i,j}^{\dagger}L_{i,j},\rho\right\} \right),
\label{eq:time_evol_erasure_qubit}
\end{align}
where $L_{i,j}=\sqrt{\gamma_{i,j}}|e_{j}\rangle\langle i|$
are the erasure jump operators 
that
correspond to a decay from the qubit state $|i\rangle$ ($i=1,2$) to the erasure states $|e_j\rangle,$ and $\gamma_{i,j}$ are the decay rates. This setting is illustrated in Fig. \ref{fig:1} (a).
We denote by $\mathcal{E}\left(t\right)$ the quantum channel given by evolving the qubit under this Lindblad equation.  


{\bf{Protocol classes and their precision limits}}---
Given this erasure qubit setting, we consider estimating the parameter $\omega$ and organize quantum metrology strategies into classes of varying complexity. For each class of strategies, we define the associated precision limit as the optimal precision achievable within that class.

In general, these precision limits rely on the quantum (classical) Cram\'er-Rao bounds, which state that $\Delta\omega\geq I^{-1/2}$ $(\Delta\omega\geq I_{C}^{-1/2}),$
where $I$ ($I_C$) corresponds to the quantum (classical) Fisher information.
The classical Fisher information (CFI) can be calculated directly from the probability distribution of the outcomes, $\left\{ p_{i}(\omega)\right\} _{i}$: $I_C=\sum_{i}\left(\partial_{\omega}p_{i}\right)^{2}/p_{i}.$
Similarly the quantum Fisher information (QFI) can be calculated directly from the density matrix of the probe, $\rho \left( \omega \right)$ \cite{helstrom_quantum_1969, braunstein1994statistical}:
$I\bm{(}\rho\bm{)} = \underset{ \rev{j,k;p_{j}+p_{k}>0} }{\sum}\dfrac{2}{p_{j}+p_{k}}\abs{\bra{\psi_{j}}\partial_{\omega}\rho\ket{\psi_{k}}}^{2}$,
where $\{p_k \}_k$ are the  eigenvalues of the density matrix $\rho$ and $\ket{\psi_k}$ are the corresponding eigenstates.


Assuming that we have $N$ erasure qubits that evolve according to Eq. \ref{eq:time_evol_erasure_qubit}, and a total time $T,$ we consider the different classes of metrological strategies that are illustrated in Fig. \ref{fig:1}.

 The simplest strategy class is (i) where the  $N$ qubit probes are initialized in a product state and allowed to evolve freely for a duration $t$, after which a measurement is performed. This procedure is repeated $T/t$ times. The maximal QFI achievable within this class is denoted by 
$\Iprod$. This bound 
is attained when both the product initial state and $t$ are optimized. 
It is well established that in the noiseless case $\Iprod \propto NT^2,$ while in the noisy case  it typically scales as $\Iprod \propto NT$ \cite{demkowicz2012elusive,demkowicz2014using, zhou2021asymptotic}.

The second strategy class (ii) consists of {\it{entangled}} (ancilla-free) strategies. 
These extend class (i) by allowing for a general entangled input state and measurement. The optimal QFI achievable within this class is denoted by $\Ient$. In the noiseless case,  $\Ient \propto N^2T^2$ while in the presence of noise \rev{the typical scaling is} $\Ient \propto NT$ \cite{demkowicz2012elusive,demkowicz2014using, zhou2021asymptotic}. 

We further define $\mathcal{I}_N^{\left(\text{prod}\right)}$ and $\mathcal{I}_N^{\left(\text{ent}\right)}$ as the maximal QFI within these classes when only the input state is optimized. For class (i) this can be calculated efficiently using the single-qubit extended-channel QFI (ECQFI) technique, whereas for class (ii) the calculation is in general more complex. Specifically,  
it can be calculated with an iterative see-saw (ISS) method \cite{macieszczak2013quantum, dulian2025qmetro++, lukacs2026iterative}, while the ECQFI method is not necessarily tight. Calculation of this $\mathcal{I}_N^{\left(\text{ent}\right)}$ might be also intractable in the large $N$ limit.

The most general strategy class (iii) consists of sequential protocols, in which we allow any possible sequential control, any input state and any number of noiseless ancillas. The optimal QFI within this class in the asymptotic limit of large $T$ (or $N$) is denoted by $\Iult$. It represents the ultimate precision bound, since it includes optimization over all possible protocols.
$\Iult$ can be 
calculated using the sequential ECQFI bound that was developed in Refs. \cite{sekatski2017quantum,ZhouNC18AchievingHeisenberg, demkowicz2017adaptive}.
It was further shown in Refs. \cite{sekatski2017quantum,demkowicz2017adaptive, ZhouNC18AchievingHeisenberg}
that $\Iult \propto N^2T^{2}$ iff $G\notin\text{span}\left\{ \mathbbm{1},L_{i,j},L_{i,j}^{\dagger},L_{i,j}^{\dagger}L_{k,m}\right\}_{i,j,k,m}, $
where $G$ and $\left\{L_{i,j} \right\}$ are the Hamiltonian and jump operators in Eq. \ref{eq:time_evol_erasure_qubit} respectively.
Otherwise, $\Iult \propto NT.$ 


These classes of metrological strategies satisfy $\left(i\right)\subseteq\left(ii\right)\subseteq\left(iii\right),$ and thus the corresponding QFI bounds satisfy $\Iprod\leq \Ient \leq \Iult.$ This chain of inequalities is true for every $N$ and $T$.

We study in this Letter another class of protocols (iv), which is particularly relevant to erasure noise: continuous erasure detection (CD). In this strategy class, the input state can be any product state and the erasure levels are being continuously monitored \cite{clark2019quantum, yang2023efficient,radaelli2026parameter}. A continuous feedback and control is applied based on the erasure detection outcomes, e.g., the qubit is reset following detection events.
This strategy is clearly still limited by $\Iult.$ However, we show that it can be considerably better than strategy class (i) (optimal product state), and furthermore in some cases it also outperforms strategy class (ii) (optimal entangled states protocols) up to a large $N$.

For erasure qubits, this hierarchy of precision limits strongly depends on $G$ and on the erasure decay rates $\left\{ \gamma_{i,j} \right\}.$
In what follows we explore these limits and the optimal protocols. We split into cases of \rev{longitudinal signals} ($G=\sigma_{z}$), 
and \rev{transverse signals} ($G \perp \sigma_{z}$).




{\bf{Precision limits for $\mathbf{G=\sigma_{z}}$}}---
For $G=\sigma_{z}$,
\rev{we characterize the hierarchy of precision limits and show how erasure detectability enables simpler sensing strategies in both the asymptotic and finite $N$ regimes.} 
For convenience, we denote $\Gamma_1:=\sum_{j=1}^{k}\gamma_{1,j},$ $\Gamma_2:=\sum_{j=1}^{k}\gamma_{2,j},$
and remark that in this case all  bounds depend solely on these sums.

We begin with the ultimate limit, given by \cite{sekatski2017quantum,demkowicz2017adaptive,ZhouNC18AchievingHeisenberg}:\\ 
$\Iult=4\, \underset{h}{\min}\!\left[\, \left(\frac{NT}{\dd t} \right)\!\norm{\alpha}\!+\! \left(\frac{NT}{\dd t} \right)^2\,\norm{\beta}\!\left(  \norm{\beta} + 2\sqrt{\norm {\alpha}} \!\right)\,\right]$,
where $\alpha = (\dot{\mathbf{K}}_{dt}- i h \mathbf{K}_{dt})^\dagger (\dot{\mathbf{K}}_{dt}- i h \mathbf{K}_{dt}),$ $\beta = i\mathbf{K}^\dagger_{dt} (\dot{\mathbf{K}}_{dt}- i h \mathbf{K}_{dt})
$, 
and $\mathbf{K}_{dt}$ 
is a vector of the Kraus operators of the infinitesimal evolution channel $\mathcal{E}\left( dt\right)$.
Applying this to our case (Eq. \ref{eq:time_evol_erasure_qubit} with $G=\sigma_z$) leads to \cite{supp}
\begin{equation}\label{eq:secqfi_bound_sz}
    \Iult = \dfrac{16 NT}{\left(\sqrt{\Gamma_1} + \sqrt{\Gamma_2}\right)^2}.
\end{equation}
\rev{This}
is the ultimate precision limit, 
\rev{hence the linear scaling with $NT$ rules out 
Heisenberg scaling
for this case.}

\begin{figure}
\centering
\begin{tikzpicture}

\node[anchor=south west, inner sep=0] (A)
{\includegraphics[width=\linewidth]{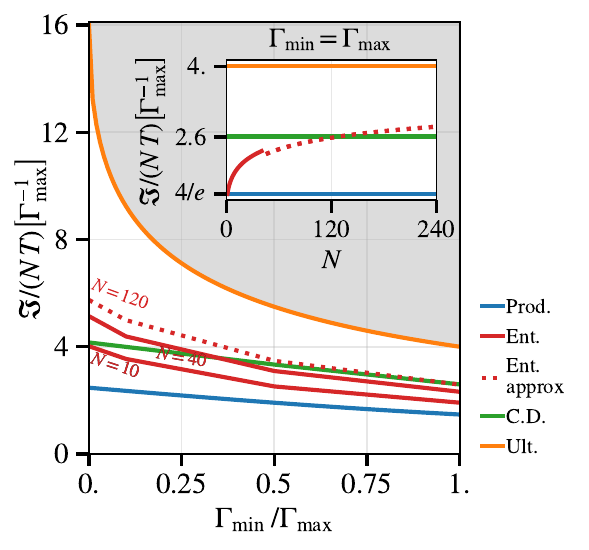}};

\begin{scope}[x={(A.south east)}, y={(A.north west)}]

\node[anchor=south west] at (0.79,0.46) {\renewcommand{\arraystretch}{1.3}
\begin{tabular}{|c|}
\hline \rowcolor{gray!20}
$\Iprod$ \\ Eq.~\ref{eq:ecqfi_sz_product} \\
\hline \rowcolor{gray!20}
$\Ient$ \\ Claim~\ref{cl:claim_1} \\
\hline \rowcolor{gray!20}
$\Icd$ \\ Eq.~\ref{eq:cd_sz} \\
\hline \rowcolor{gray!20}
$\Iult$ \\ Eq.~\ref{eq:secqfi_bound_sz} \\
\hline
\end{tabular}
};

\node at (0.45,1.01) {\large (a)};
\node at (0.87,1.01) {\large (b)};

\end{scope}

\end{tikzpicture}
 \caption{ 
Hierarchy of precision limits for $G=\sigma_{z}.$
(a) Main plot: optimal QFI rate with different strategies as a function of $\Gamma_{\min}/\Gamma_{\max}$. 
This figure presents the gap between the ultimate limit (orange line), optimal product strategies (blue line) and optimal entangled strategies for different $N$ (red solid and dotted lines correspond to exact and approximate values respectively \cite{supp}).
Continuous erasure detection (green line) outperforms entangled strategies up to a certain $N$, where this crossover depends on the decay rates.
Inset: optimal QFI with the different strategies as a function of $N$ for the symmetric case $\Gamma_{\max}=\Gamma_{\min}$. Continuous erasure detection outperforms entangled strategies up to $N \sim 120$. (b) Summary of the precision limits achieved with the different metrological strategies for $G =\sigma_z.$ }
\label{fig:sigma_z_different_strategies}
\end{figure}
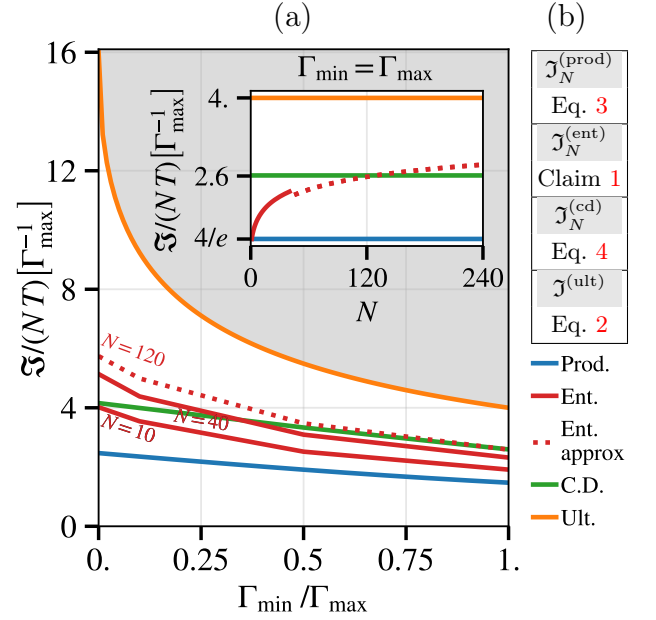

Let us compare this bound to the QFI of the simplest strategy: optimal product states ($\Iprod$). 
We calculate $\mathcal{I}_N^{\left(\text{prod}\right)}$ using the single-qubit ECQFI bound \cite{demkowicz2012elusive,KolodynskiNJP13EfficientTools, fujiwara2008fibre}, which optimizes the QFI for a quantum channel $\mathcal{E} \left( t\right)$ over all 
input states. 
Applying this bound to our problem yields \cite{supp}:  
\begin{align}\label{eq:ecqfi_sz_product}
\begin{split}
\Iprod=\underset{t}{\text{max}}\frac{16tNT}{\left(e^{\frac{\Gamma_1t}{2}}+e^{\frac{\Gamma_2 t}{2}}\right)^{2}}
\approx\begin{cases}
\frac{2.47}{\Gamma_{\max}}NT & \Gamma_{\min}=0\\
\frac{4}{e\Gamma_{\max}}NT & \Gamma_{\min}=\Gamma_{\max},
\end{cases}
\end{split}
\end{align}
where $\Gamma_{\min, \max}$ are the extrema of $\{\Gamma_1, \Gamma_2\}.$ 
This bound is saturated by the single-qubit input state
$\ket{\psi_{\text{opt}}} = \sqrt{p_{\text{opt}}}\ket{1}  + \sqrt{1-p_{\text{opt}}}\ket{2},$
where
$p_{\text{opt}}=\big[1+\tanh\frac{t_{\text{opt}}}{4}(\Gamma_{1}-\Gamma_{2})\big]/2,$
\rev{and $t_{\text{opt}}$ is the numerically optimized interrogation time.}
Comparing to the ultimate bound we observe $2.7\leq\Iult/\Iprod \leq 6.47,$ where the maximal metrological gain is obtained for $\Gamma_{\min}=0$, and the minimal is for $\Gamma_{\min}=\Gamma_{\max}$. This comparison is presented in Fig. \ref{fig:sigma_z_different_strategies}.

It is now desirable to understand which protocols attain $\Iult$ and how much improvement over $\Iprod$ can be obtained using entanglement and CD strategies.
Let us first explore the performance of optimal entangled states (strategy class (ii)). 
We 
show that this strategy is optimal in the asymptotic limit:\\
\claim \label{cl:claim_1} 
$\underset{N \rightarrow \infty}{\text{lim}}\Ient=\Iult$ for any \rev{erasure rates satisfying $\Gamma_1+ \Gamma_2>0,$} 
and this saturation is \rev{attained} using optimal spin-squeezed states and optimizing over \rev{the} evolution time $t$. 

The case of $\Gamma_{1}=\Gamma_{2}$ was proven in Ref. \cite{demkowicz2015quantum}; we generalize this proof in the supplemental material to any $\Gamma_1,\Gamma_2$ \cite{supp}.
This is done by calculating an error propagation \rev{lower bound on}  the QFI and \rev{showing that
after optimizing both the squeezing and its orientation}
it converges to $\Iult$.
\rev{
The mean-spin direction of the optimal spin-squeezed state in this case 
is shifted away from the equator by an angle $\phi$ determined by $\Gamma_1, \Gamma_2$:
$\langle J_x \rangle \approx \frac{N}{2} \cos \left( \phi \right), \langle J_z \rangle \approx \frac{N}{2} \sin \left( \phi \right),$ with $\phi=\arcsin\left(\frac{\sqrt{\Gamma_{1}}-\sqrt{\Gamma_{2}}}{\sqrt{\Gamma_{1}}+\sqrt{\Gamma_{2}}}\right).$
Hence for $\Gamma_{1}\neq\Gamma_{2}$, the
optimal states are not the standard equator-centered spin-squeezed
states commonly used for dephasing \cite{kitagawa1993squeezed,zhou2021asymptotic}.}

Claim \ref{cl:claim_1}, however, corresponds only to the asymptotic $N \rightarrow \infty$ limit. To examine the performance in the finite $N$ regime, we plot $\Ient$
as a function of $N$ in Fig. \ref{fig:sigma_z_different_strategies}.
It can be observed that, while this strategy is asymptotically optimal, its convergence to $\Iult$ is slow.
We find that the optimal entangled states correspond to
spin squeezed states for $N \geq 40$. 
These numerical calculations are done using iterative see-saw (ISS)  methods \cite{macieszczak2013quantum, dulian2025qmetro++, lukacs2026iterative}. 

Let us now analyze the performance of CD protocols (strategy class (iv)), where we continuously measure the erasure states, detecting whether a decay occurred.
A control is implemented on the qubits based on the erasure detection outcomes. If an erasure is detected, the optimal control is to reset the qubit to an optimal input state. If
the probe evolves without an erasure for a duration $\tau$, an optimal measurement is performed and a new cycle begins.
Optimizing this strategy gives \cite{supp}
\begin{align}\begin{split}\label{eq:cd_sz}
\Icd&=\underset{\tau}{\text{max}}\frac{16\tau^2 NT}{\left(\sqrt{\int_{0}^{\tau}e^{\Gamma_{1}t}\,\text{d}t}+\sqrt{\int_{0}^{\tau}e^{\Gamma_{2}t}\,\text{d}t}\right)^{2}}   
\\ & \approx\begin{cases}
\frac{4.16}{\Gamma_{\max}}NT & \Gamma_{\min}=0\\
\frac{2.6}{\Gamma_{\max}}NT & \Gamma_{\min}=\Gamma_{\max}.
\end{cases}
\end{split}
\end{align}
\rev{$\Icd$ has the same form as   $\Iprod$ in Eq. \ref{eq:ecqfi_sz_product}, but the decay terms $e^{\Gamma_{j}\tau}$ are replaced by $\int_{0}^{\tau}e^{\Gamma_{j}t}\,\text{d}t/\tau.$
This corresponds to an effective reduced decay rate: $ \frac{1}{\tau}\text{log}\left(\int_{0}^{\tau}e^{\Gamma_{j}t}\,\text{d}t/\tau\right) \leq \Gamma_{j},$
which captures the advantage in using CD.
The gain arises because an erased probe is reset immediately, avoiding the otherwise unproductive remainder of a failed interrogation.}
\rev{Consequently,} $\Icd$ 
exceeds $\Iprod$
by a factor of $1.68-1.76,$ 
\rev{depending on the asymmetry between $\Gamma_\text{min,max}$.}
\rev{This bound is shown in
Fig. \ref{fig:sigma_z_different_strategies}
for different $\Gamma_{\min},$ $\Gamma_{\max}$.}
To better appreciate the performance of this strategy, we compare its QFI to that of the optimal entangled strategy for finite $N$. We observe that in the symmetric case ($\Gamma_{\min}= \Gamma_{\max}$),  $\Icd$ outperforms  $\Ient$  up to  $N\approx120$ qubits (Fig. \ref{fig:sigma_z_different_strategies} inset). 
Therefore, although the CD strategy is asymptotically inferior to the optimal entangled strategy as $N \rightarrow \infty,$ it  considerably outperforms it in the small $N$ regime.
This comparison highlights the effectiveness of CD strategies.
For asymmetric decay, $\Gamma_{\min}< \Gamma_{\max},$
the crossover between $\Icd$ and $\Ient$ occurs at smaller $N$
(see Fig. \ref{fig:sigma_z_different_strategies}). 
We anticipate that combining entanglement with CD can lead to further improvement,
\rev{but this analysis is left} to future work.

\rev{{\textbf{ Application to optimal sensing under thermal noise--}}}
\rev{Claim \ref{cl:claim_1} also provides a constructive asymptotically optimal protocol
for thermal noise and amplitude damping.}
These noises
can be converted to erasure noise using the following code space $\{|0_L\rangle=|11\rangle,|1_L\rangle=|22\rangle\},$ which constitutes a logical erasure qubit;
\rev{a single excitation or decay takes the pair into the odd-parity subspace, making the event detectable by parity monitoring \cite{kubica2023erasure, supp}.}
Applying Claim \ref{cl:claim_1} to this logical erasure qubit, we get that an optimal spin squeezing \rev{of these effective erasure qubits} 
attains $\Iult$
\footnote{ \rev{$\Iult$
for thermal noise and amplitude damping 
coincides with the erasure qubit bound of Eq. \ref{eq:secqfi_bound_sz}, where $\Gamma_{1},\Gamma_{2}$ correspond to the decay and excitation rates \cite{sekatski2017quantum}.}}. \rev{The full derivation is shown in the supplemental}.
\rev{This protocol requires no noiseless} ancillas, and \rev{provides} an  alternative to 
\rev{approaches based} on conversion to dephasing \cite{zhou2021asymptotic}.
\rev{We also remark that optimal spin squeezing of the physical qubits alone does not saturate $\Iult$ for these noises \cite{demkowicz2014using,kielinski2024ghz}, whereas logical spin squeezing of the effective erasure qubits attains this limit.}

{\bf{Precision limits for $\mathbf{G \perp \sigma_{z}}$}}---
For \rev{transverse signals, such as} $G=\sigma_{x}$ or $G=\sigma_{y},$
\rev{erasure noise produces a different hierarchy of sensing strategies, in which QEC can yield a significant metrological gain.}



\begin{figure}
\centering
\begin{tikzpicture}

\node[anchor=south west, inner sep=0] (A)
{\includegraphics[width=.66\linewidth]{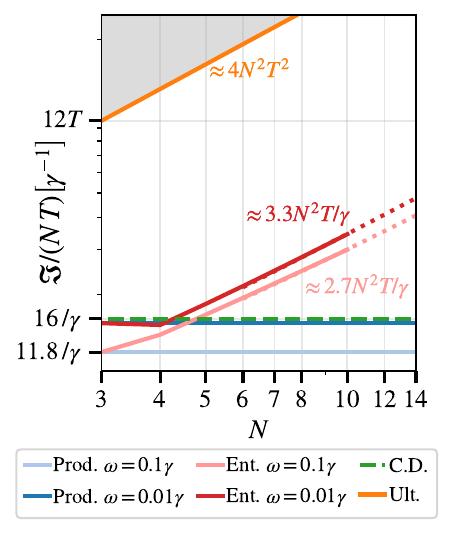}};
\begin{scope}[x={(A.south east)}, y={(A.north west)}]

    \node[anchor=south west, inner sep=0] at (-0.5,0.6)
      {\includegraphics[width=0.33\linewidth]{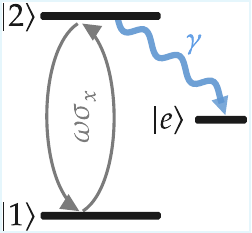}};
\node at (-.25,1.03) {\large (a)};
\node at (0.6,1.03) {\large (b)};
\node[anchor=south west] at (-.31,.51) 
{\large (c)};
\node[anchor=south west] at (-.5,0.01) {\renewcommand{\arraystretch}{2}
\begin{tabular}{|>{\columncolor{gray!20}}c| c|}
\hline 
$\Iprod$ & \makecell{$\leq \frac{16NT}{\gamma}$\\\scriptsize ($\omega \rightarrow 0$)}  \\
\hline 
$\Ient$ & \makecell{$\leq \frac{3.3 N^2 T}{\gamma}$\\\scriptsize ($\omega \rightarrow 0$)} \\
\hline 
$\Icd$ & $ \frac{16NT}{\gamma}$  \\
\hline 
$\Iult$& Eq.~\ref{eq:sigma_x_ultimate_seq_bounds}\\
\hline
\end{tabular}
};
\end{scope}

\end{tikzpicture}
 \caption{ Hierarchy of precision limits for $G=\sigma_{x}$ under single erasure noise, as sketched in (a). (b) Optimal QFI rate as a function of $N$. 
The optimal product strategies for $\omega=0.1 \gamma$ and $0.01\gamma$ (bright and dark blue lines, respectively) approach $16/\gamma$ in the limit $\omega\rightarrow 0.$ In contrast, the optimal continuous detection strategy (dashed green line) attains this value for any $\omega$. While these protocols are limited to SQL scaling, the optimal entangled strategies ($\omega=0.1 \gamma,\: 0.01 \gamma$, given by the pink, red lines respectively) achieve a Heisenberg scaling in $N$, but not in $T$. The ultimate bound (orange line) is the noiseless HL. (c) Summary of
the precision limits.}
\label{fig:sigma_x_different_strategies}
\end{figure}
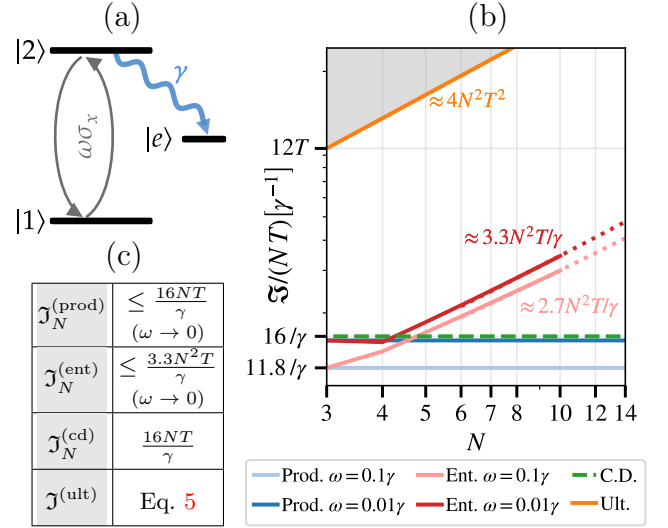

\rev{Let us define the overlap rate $\Gamma_\text{ov}=\sum_{j=1}^{k}\text{min }\left\{ \gamma_{1,j},\gamma_{2,j}\right\}.$}
$\Iult$ \rev{then satisfies
\begin{align}\label{eq:sigma_x_ultimate_seq_bounds}
\begin{split}
&\dfrac{2NT}{\Gamma_\text{ov} }\leq\Iult\leq\dfrac{4NT}{\Gamma_\text{ov} } \;\;\text{ (for $\Gamma_\text{ov}>0$)},\\
& \Iult=4N^2T^2 \hspace{1.2cm} \text{ (for $\Gamma_\text{ov}=0$)}.
\end{split}
\end{align}}
\rev{These bounds 
identify $\Gamma_{\text{ov}}$ as the effective noise scale limiting precision. }
Thus, in the asymmetric erasure regime,
where \rev{$\Gamma_\text{ov} \ll \Gamma_1+\Gamma_2,$ }
this bound \rev{becomes} significantly larger than the
corresponding thermal noise bound \footnote{For thermal noise with $G=\sigma_{x}$ and decay and excitation rates of $\Gamma_2,\Gamma_1$, the bound is $\Iult=\frac{4NT}{\Gamma_1+\Gamma_2}.$ \cite{sekatski2017quantum} }
as well as the \rev{longitudinal} erasure  bound (Eq. \ref{eq:secqfi_bound_sz}).  
Furthermore, in the extreme asymmetric case \rev{of $\Gamma_\text{ov}=0$}  the decay can be fully corrected using QEC, allowing the recovery of the noiseless HL.

\rev{For $\Gamma_\text{ov}=0,$ the} HL is attainable using 
\rev{a simple}
QEC code 
$\left\{ |0_{L}\rangle=|+\rangle|+\rangle,|1_{L}\rangle=|-\rangle|-\right\}, $
where $|\pm\rangle=\sqrt{1/2}\left(|1\rangle \pm |2\rangle\right)$. 
Preparing the $N$ qubits in the logical GHZ state $\sqrt{1/2} (|0_L\rangle^{N/2}+ |1_L\rangle^{N/2})$ and applying continuous QEC restores the noiseless HL \cite{gefen2016parameter,plenio2016sensing,chen2024quantum,kwon2026restoring}.
In general, using this QEC code yields an effective logical Hamiltonian $H_L=2\omega \left(|0_{L}\rangle\langle0_{L}|-|1_{L}\rangle\langle1_{L}|\right)$ together with a logical dephasing rate
\rev{ $\Gamma_\text{ov}.$ }
Consequently, 
\rev{for} $\rev{\Gamma_\text{ov}} >0$
, 
an optimal logical spin-squeezed state of this QEC code asymptotically achieves
a QFI of $\frac{2NT}{ \rev{\Gamma_\text{ov} } }$ \cite{zhou2021asymptotic,supp}, 
 which matches the lower bound of Eq. \ref{eq:sigma_x_ultimate_seq_bounds}.

In most erasure configurations, however, $\Iult >\frac{2NT}{\rev{\Gamma_\text{ov}} },$
indicating that alternative QEC codes are required to saturate $\Iult$. 
\rev{For}
a single erasure state 
with $\gamma_{1},\gamma_{2} >0$ 
and without loss of generality $\gamma_1\leq \gamma_2,$
\rev{the ultimate precision limit is}
$\Iult=\frac{4NT}{\gamma_{1}},$
which corresponds to the upper bound of Eq. \ref{eq:sigma_x_ultimate_seq_bounds}.
This $\Iult$ is saturable asymptotically with the following ``zero signal, zero noise'' code:
$\left\{ |0_{L}\rangle=|\epsilon_{+}\rangle|0\rangle_{a},|1_{L}\rangle=|\epsilon_{-}\rangle|1\rangle_{a}\right\} ,$
where
$|0/1\rangle_a$ denotes noiseless ancillary states,
$|\epsilon_{\pm}\rangle=\epsilon|1\rangle\pm\sqrt{1-\epsilon^{2}}|2\rangle,$
and  $\epsilon \rightarrow0,$ which is the limit of vanishing signal and vanishing noise. 
A derivation of the optimality of these QEC codes
\rev{and} further numerical calculations of $\Iult$
\rev{for} general erasure configurations 
are provided in the supplemental \cite{supp}.

To illustrate the significant gain of QEC protocols, let us compare $\Iult$ to the optimal product state QFI, $\Iprod$.  
For simplicity, we restrict ourselves to the simplest case of a single erasure decay from $|2\rangle,$ i.e., $\gamma_1=0, \gamma\coloneqq \gamma_{2} >0,$
as illustrated in Fig. \ref{fig:sigma_x_different_strategies} (a). 
While QEC protocols achieve HL, 
we find that 
$\Iprod\leq 16NT/\gamma,$ where equality
is achieved only in the limit of $\omega \rightarrow0.$ 
In this limit,
$\Iprod$ is obtained by initializing every qubit to $|1\rangle,$ which becomes a noiseless state as $\omega \rightarrow0$.
Given that $\omega \ll \gamma,$
we can adiabatically eliminate the fast decaying state $|2\rangle,$ such that the dynamics is described by an effective erasure decay from $|1\rangle$ with decay rate of $\gamma_{\text{eff}}=4\omega^2/\gamma.$ This effective erasure dynamics then accounts for a QFI of $16NT/\gamma$ \cite{supp, sekatski2022optimal, gardner2025lindblad}.

We further study the performance of CD protocols, and find that they also fall substantially short of $\Iult$
in the asymmetric erasure regime.
Focusing again on the case of a single erasure decay,
we find that the optimal strategy is to reset the state only after erasure detection, so that 
all the information is encoded in the decay times. 
The QFI of this protocol can be calculated analytically and is equal to $\Icd=16NT/\gamma$.
It thus reaches the same limit as $\Iprod,$ but does so for any $\omega$, not only in the regime of $\omega \ll \gamma.$

Lastly, we discuss the entangled strategy and $\Ient.$ 
In the limit of large $N$ this strategy attains Heisenberg scaling with $N$,
but not with $T,$ since Heisenberg scaling with $T$ requires sequential QEC.
From numerical simulations, we obtain that for large enough $N$ the optimal entangled state is the GHZ state $\sqrt{1/2}\left(|+\rangle^{N}+|-\rangle^{N}\right),$
and that $\Ient\rightarrow 3.3 N^{2}T/\gamma$ in the limit of large $N$ and $\omega \rightarrow0.$
Hence, in the limit of large $N,T$ it considerably outperforms all product strategies, yet it is much smaller than $\Iult$, i.e., 
$\Iprod\leq\Icd \ll \Ient\ll\Iult.$
This hierarchy of precision limits is shown and summarized in Fig. \ref{fig:sigma_x_different_strategies}.

{\bf{Discussion and Conclusions}}---
\rev{We have characterized precision limits and sensing protocols under erasure noise.} 
We have shown that for 
\rev{transverse signals}, QEC can introduce a significant metrological gain by correcting the dominant erasure processes.
For 
\rev{longitudinal signals, 
the ultimate precision bound is attained asymptotically by suitable
spin squeezed states.} 
\rev{At finite system sizes}, 
product CD protocols outperform optimal passive entangled strategies \rev{over a substantial range of qubits number}, 
highlighting the power of CD strategies.
\rev{We further use erasure conversion to construct asymptotically optimal protocols under thermal noise without noiseless ancillas.}
Interesting future research directions include combining CD with entanglement 
\rev{or} coherent control,
\rev{and extending the construction based on erasure conversion and logical spin squeezing to other noise models}.




{\textit{Acknowledgements---}}
The authors are thankful to Ran Finkelstein, Nelson Darkwah Oppong, Su Direkci, James W. Gardner, and Yanbei Chen for useful discussions. T.G. acknowledges funding from the quantum science and technology early-career grant of the Israeli council for higher education and ISF Grant No. 3302/25. A. R. acknowledges the support  of the Israeli Science Foundation, the Israeli Innovation Authority and the Schwartzmann university chair.

\clearpage
\onecolumngrid
\setcounter{section}{0}
\setcounter{subsection}{0}
\setcounter{equation}{0}
\setcounter{figure}{0}
\setcounter{table}{0}
\renewcommand{\theequation}{S\arabic{equation}}
\renewcommand{\thefigure}{S\arabic{figure}}
\renewcommand{\thetable}{S\arabic{table}}
\renewcommand{\theHequation}{supp.\arabic{equation}}
\renewcommand{\theHfigure}{supp.\arabic{figure}}
\renewcommand{\theHtable}{supp.\arabic{table}}
\renewcommand{\theHsection}{supp.\arabic{section}}
\begin{center}
\textbf{\large Supplemental Material}
\end{center}
\tableofcontents

\section{Ultimate asymptotic precision limits of erasure qubits}\label{app:sec_sequential_ecqfi_bounds}

Let us review the sequential ECQFI bounds that were derived in Refs. \cite{sekatski2017quantum, demkowicz2017adaptive,ZhouNC18AchievingHeisenberg}. These bounds are used to calculate $\Iult$ for the various scenarios.
We assume a Markovian time evolution given by a Lindblad master equation
\begin{align*}
\frac{d\rho}{dt}=-i\left[H,\rho\right]+\underset{k=1}{\overset{r}{\sum}}L_{k}\rho L_{k}^{\dagger}-\frac{1}{2}\left\{ L_{k}^{\dagger}L_{k},\rho\right\},    
\end{align*}
where $H$ is the Hamiltonian and $L_k$ are the Lindblad jump operators.
The Hamiltonian is $H=\omega {G},$ such that $\omega$ is the parameter to be estimated.
To find the most general sequential extended-channel QFI (ECQFI) we consider infinitesimal-time evolution Kraus operators given by
\begin{align*}
\mathbf{K}=\left(\begin{array}{c}
\mathbbm{1}-i Hdt-\frac{1}{2}\underset{k}{\sum}L_{k}^{\dagger}L_{k}dt\\
L_{1}\sqrt{dt}\\
\vdots\\
L_{r}\sqrt{dt}
\end{array}\right).    
\end{align*}
The sequential ECQFI is then given by
\begin{align*}
\mathcal{I}=4\underset{h}{\min}\left(\left(\frac{NT}{dt}\right)||\alpha||+\left(\frac{NT}{dt}\right)^{2}||\beta||\left(||\beta||+2\sqrt{||\alpha||}\right)\right),    
\end{align*}
where $\alpha=(\overset{\cdot}{\mathbf{K}}-ih\mathbf{K})^{\dagger}(\overset{\cdot}{\mathbf{K}}-ih\mathbf{K})$, $\beta=i\mathbf{K}^{\dagger}(\overset{\cdot}{\mathbf{K}}-ih\mathbf{K})$
and $h$ is a Hermitian $\left(r+1\right)\times\left(r+1\right)$
matrix that acts on the $\mathbf{K}$ vector.
It was shown in Refs. \cite{sekatski2017quantum,demkowicz2017adaptive,ZhouNC18AchievingHeisenberg} that, if $G\in\text{span}\{ \mathbbm{1},L_{i},L_{i}^{\dagger},L_{i}^{\dagger}L_{j}\} _{i,j}$ then there exists $h$ such that $\beta=0$. The asymptotic bound would then be equal to
\begin{align*}
\mathcal{I}=4\left(\frac{NT}{dt} \right)\underset{h}{\min} ||\alpha|| \\
{\text{subject to  }} \; \beta=0.
\end{align*}
This expression can be further simplified in the case of infinitesimal channels.
We expand $\alpha,$ $\beta,$ $h$ and $\mathbf{K}$ in powers of $\sqrt{dt}$
\begin{align*}
&\alpha=\alpha^{\left(0\right)}+\alpha^{\left(1\right)}\sqrt{dt}+\alpha^{\left(2\right)}dt+O\left(dt^{3/2}\right)\\
&\beta=\beta^{\left(0\right)}+\beta^{\left(1\right)}\sqrt{dt}+\beta^{\left(2\right)}dt+\beta^{\left(3\right)}dt^{3/2}+O\left(dt^{2}\right)\\
&h=h^{\left(0\right)}+h^{\left(1\right)}\sqrt{dt}+h^{\left(2\right)}dt+h^{\left(3\right)}dt^{3/2}+O\left(dt^{2}\right)\\
&\mathbf{K}=\mathbf{K}^{\left(0\right)}+\mathbf{K}^{\left(1\right)}\sqrt{dt}+\mathbf{K}^{\left(2\right)}dt. 
\end{align*}
Note that
\begin{align*}
 \mathbf{K}^{\left(0\right)}=\left(\begin{array}{c}
\mathbbm{1}\\
0\\
\vdots\\
0
\end{array}\right), \quad\mathbf{K}^{\left(1\right)}=\left(\begin{array}{c}
0\\
L_{1}\\
\vdots\\
L_{r}
\end{array}\right), \quad\mathbf{K}^{\left(2\right)}=\left(\begin{array}{c}
-i H-\frac{1}{2}\underset{j}{\sum}L_{j}^{\dagger}L_{j}\\
0\\
\vdots\\
0
\end{array}\right).   
\end{align*}
It can be shown that $\beta=0$ iff $\beta^{\left(2\right)}=G-\underset{j,k}{\sum}h_{j,k}^{(0)}L_{k}^{\dagger}L_{j}-\underset{k}{\sum}h_{0,k}^{(1)}\left(L_{k}+L_{k}^{\dagger}\right)-h_{0,0}^{(2)}I$ can be set to zero. This is satisfied iff $G\in\text{span}\{ \mathbbm{1},L_{i},L_{i}^{\dagger},L_{i}^{\dagger}L_{j}\} _{i,j}.$
The leading and relevant order of $\alpha$ is $\alpha^{(2)}=\left(h^{\left(1\right)}\mathbf{K}^{\left(0\right)}+h^{\left(0\right)}\mathbf{K}^{\left(1\right)}\right)^{\dagger}\left(h^{\left(1\right)}\mathbf{K}^{\left(0\right)}+h^{\left(0\right)}\mathbf{K}^{\left(1\right)}\right).$
Therefore given that $G\in\text{span}\{ \mathbbm{1},L_{i},L_{i}^{\dagger},L_{i}^{\dagger}L_{j}\} _{i,j}$ the asymptotic sequential ECQFI is
$4\mathfrak{a}NT,$ with
\begin{align}\label{eq: fundamental_ecqfi}
\begin{split}
 &\mathfrak{a}=\underset{h}{\min}\left(h^{\left(1\right)}\mathbf{K}^{\left(0\right)}+h^{\left(0\right)}\mathbf{K}^{\left(1\right)}\right)^{\dagger}\left(h^{\left(1\right)}\mathbf{K}^{\left(0\right)}+h^{\left(0\right)}\mathbf{K}^{\left(1\right)}\right)\\
 & \text{subject to } G=\underset{j,k}{\sum}h_{j,k}^{(0)}L_{k}^{\dagger}L_{j}+\underset{k}{\sum}h_{0,k}^{(1)}\left(L_{k}+L_{k}^{\dagger}\right)+h_{0,0}^{(2)}\mathbbm{1}.
 \end{split}
\end{align}


We consider the qubit case where the Hamiltonian $H=\omega G$ acts on a two-level system $\text{span }\{ |1\rangle,|2\rangle\} $.
Besides of the two level system, there are $k$ erasure states 
$\{ |e_{i}\rangle\} _{i=1}^{k}$ to which the qubit decays. 
We therefore have $2 k$ jump operators
\begin{align*}
\left\{ L_{1,j}=\sqrt{\gamma_{1,j}} \sigma_{1}^{(j)}, \;L_{2,j}=\sqrt{\gamma_{2,j}} \sigma_{2}^{(j)} \right\} _{j=1}^k,    
\end{align*}
where $\sigma_{1}^{(j)}=|e_{j}\rangle\langle1|,$
$\sigma_{2}^{(j)}=|e_{j}\rangle\langle2|.$

The precision bound depends on the direction of $G$. Hence,
we split the analysis to the following three cases.

\subsection{\texorpdfstring{$\Iult$}{Iult} for erasure qubits with \texorpdfstring{$G \parallel \sigma_z$}{G parallel to sigma\_z}}\label{app:A1}
Without loss of generality, $G=\sigma_z$ (for any $G=\alpha \sigma_{z}$ the QFI is just multiplied by $\alpha^2$).
Let us first consider the case of a single erasure state $|e\rangle.$
Note that $\sigma_{z}=\sigma_{1}^{\dagger}\sigma_{1}-\sigma_{2}^{\dagger}\sigma_{2}$.
The constraint in Eq. \ref{eq: fundamental_ecqfi} is thus
\begin{align*}
\sigma_{z}=h_{1,1}^{(0)}\gamma_{1}\sigma_{1}^{\dagger}\sigma_{1}+h_{2,2}^{(0)}\gamma_{2}\sigma_{2}^{\dagger}\sigma_{2}+h_{0,0}^{(2)}\mathbbm{1},    
\end{align*}
which implies
\begin{align*}
&h_{1,1}^{(0)}\gamma_{1}+h_{0,0}^{(2)}=1\\
&h_{2,2}^{(0)}\gamma_{2}+h_{0,0}^{(2)}=-1.
\end{align*}
We can thus express all $h^{(0)}$ terms using $h_{0,0}^{(2)}$. Denoting $\mathfrak{h}:=h_{0,0}^{(2)}$ we get

\begin{align*}
& h^{(0)}\mathbf{K}^{(1)}= \begin{pmatrix}
        0 &  & & \\
        & \frac{(1-\mathfrak{h})}{\gamma_{1}} & & \\
        & & & -\frac{(1+\mathfrak{h})}{\gamma_{2}}
        \end{pmatrix}\begin{pmatrix}
        0 \\
        \sqrt{\gamma_{1}}\sigma_{1} \\
       \sqrt{\gamma_{2}}\sigma_{2}
\end{pmatrix}=\left(\begin{array}{c}
0\\
\sigma_{1}\frac{\left(1-\mathfrak{h}\right)}{\sqrt{\gamma_{1}}}\\
-\sigma_{2}\frac{\left(1+\mathfrak{h}\right)}{\sqrt{\gamma_{2}}}
\end{array}\right)\\
&\Rightarrow\alpha^{(2)}=\sigma_{1}^{\dagger}\sigma_{1}\frac{(1-\mathfrak{h})^{2}}{\gamma_{1}}+\sigma_{2}^{\dagger}\sigma_{2}\frac{(1+\mathfrak{h})^{2}}{\gamma_{2}}.
\end{align*}
The optimization problem is thus
\begin{align*}
\underset{\mathfrak{h}}{\min} \; \max\left\{ \frac{(1-\mathfrak{h})^{2}}{\gamma_{1}},\frac{(1+\mathfrak{h})^{2}}{\gamma_{2}}\right\}.     
\end{align*}
Observe that the minmax is obtained when $\frac{(1-\mathfrak{h})^{2}}{\gamma_{1}}=\frac{(1+\mathfrak{h})^{2}}{\gamma_{2}}.$ There are two values of $\mathfrak{h}$ for which this equality is attained. The minmax corresponds to the solution at $-1\leq\mathfrak{h} \leq1$
\begin{align*}
&\mathfrak{h}=\frac{\sqrt{\gamma_{2}}-\sqrt{\gamma_{1}}}{\sqrt{\gamma_{2}}+\sqrt{\gamma_{1}}}\;\;
\Rightarrow\;\;
{\min}||\alpha^{(2)}||=\frac{4}{(\sqrt{\gamma_{1}}+\sqrt{\gamma_{2}})^{2}}\Rightarrow \boxed{\Iult=\frac{16}{(\sqrt{\gamma_{1}}+\sqrt{\gamma_{2}})^{2}}NT.}
\end{align*}

We now generalize to multiple erasure states, where the jump operators are
$\{ L_{2j-1}=\sqrt{\gamma_{1,j}}\sigma_{1,j},\,L_{2j}=\sqrt{\gamma_{2,j}}\sigma_{2,j}\} _{j=1}^{k}.$ The constraint becomes $\mathbf{K}^{\left(1\right)\dagger}h^{\left(0\right)}\mathbf{K}^{\left(1\right)}=\sigma_{z},$ i.e.,
\begin{align*}
\sigma_{z}&=\sum_{j=1}^{k}\left(h_{2j-1,2j-1}^{(0)}\gamma_{1,j}\sigma_{1,j}^{\dagger}\sigma_{1,j}+h_{2j,2j}^{(0)}\gamma_{2,j}\sigma_{2,j}^{\dagger}\sigma_{2,j}\right)+h_{0,0}^{(2)}\mathbbm{1}
\end{align*}
and the other terms of $h$ can be set to zero. Denoting $h_{1, j}:=h_{2j-1,2j-1}^{(0)}$  and $h_{2, j}:=h_{2j,2j}^{(0)}$ we get

\begin{align*}
&\sum_{j=1}^{k}h_{1, j}\gamma_{1, j}=1-h_{0,0}^{(2)}\\
&\sum_{j=1}^{k}h_{2, j}\gamma_{2, j}=-1-h_{0,0}^{(2)}
\end{align*}

which together imply
\begin{align*}
\sum_{j=1}^{k}(h_{1, j}\gamma_{1, j}-h_{2, j}\gamma_{2, j}) = 2.
\end{align*}

To compute $\alpha^{(2)}$ we evaluate
\begin{align*}
&h^{(0)}\mathbf{K}^{(1)}=\left(\begin{array}{c}
h_{1, j}\sqrt{\gamma_{1,j}}\sigma_{1,j}\\
h_{2, j}\sqrt{\gamma_{2,j}}\sigma_{2,j}
\end{array}\right)_{j=1}^{k}\Rightarrow \alpha^{(2)}=\Pi_1\left(\sum_{j=1}^{k}h_{1, j}^2\gamma_{1,j}\right)+\Pi_2\left(\sum_{j=1}^{k}h_{2, j}^2\gamma_{2,j}\right). 
\end{align*}

The optimization problem is
\begin{align*}
&\underset{\{  h_{1, j}, h_{2, j}\} _{j=1}^k}{\min} \max\left\{\sum_{j=1}^{k}h_{1, j}^2\gamma_{1,j},\sum_{j=1}^{k}h_{2, j}^2\gamma_{2,j}\right\}\\
&\text{subject to } \sum_{j=1}^{k}(h_{1, j}\gamma_{1, j}-h_{2, j}\gamma_{2, j}) = 2.
\end{align*}

Note that for smooth and convex functions the minimizer of $\max\big(f_1(\textbf{x}), f_2(\textbf{x})\big)$ lies among
\begin{equation*}
\min\,f_1(\textbf{x}), \;\;\min\,f_2(\textbf{x}),\;\; \text{or}\;\;\underset{x: f_1(\textbf{x})=f_2(\textbf{x})}{\min}\,f_2(\textbf{x}).
\end{equation*}
We first consider the case when both terms are equal. Let $\lambda,\mu$ be Lagrange multipliers, and define the Lagrangian
 \begin{align*}
 &\mathfrak{L}=\sum_{j=1}^{k}h_{1, j}^2\gamma_{1, j}+\lambda\left(\sum_{j=1}^{k}(h_{1, j}\gamma_{1, j}+h_{2, j}\gamma_{2, j}) -2\right)+\mu\sum_{j=1}^{k}\left(h_{2, j}^2\gamma_{2, j}-h_{1, j}^2\gamma_{1, j}\right).
\end{align*}
Setting gradients to zero
\begin{align*}
&\underset{h_{1, j}}{\grad{\mathfrak{L}}}=0 
\;\Rightarrow\;
 2 h_{1, j}+\lambda - 2 \mu h_{1, j} = 0\\
&\underset{h_{2, j}}{\grad{\mathfrak{L}}}=0
\;\Rightarrow\;
 \lambda + 2 \mu h_{1, j} = 0. 
\end{align*}
Notably, all $h_{1,j}$ are equal, and likewise all $h_{2,j}$. This holds not only at the intersection of the two terms, but also when minimizing each term individually, as can be seen by repeating the Lagrangian process for each.  Using this, define $\mathfrak{h}:= 1 - h_{1,j}\sum_{j=1}^{k}\gamma_{1, j}$, and using our constraint we express all terms using $\mathfrak{h}$

\begin{align*}
&\underset{\{ h\}}{\min}||\alpha^{(2)}||=\underset{\{  h_{1, j}, h_{2, j}\} _{j=1}^k}{\min} \max\left\{\frac{(\mathfrak{h}-1)^2}{\sum_{j=1}^{k}\gamma_{1,j}},\frac{(\mathfrak{h}+1)^2}{\sum_{j=1}^{k}\gamma_{2,j}}\right\}\Rightarrow \boxed{\Iult=\frac{16}{\left(\sqrt{\Gamma_{1}}+\sqrt{\Gamma_{2}}\right)^{2}}NT.}
\end{align*}
where $\Gamma_i= \sum_{j=1}^{k}\gamma_{i,j}$, as defined also in the main text.

\subsection{\texorpdfstring{$\Iult$}{Iult} for erasure qubits with \texorpdfstring{$G \perp \sigma_z$}{G perpendicular to sigma\_z}}\label{app:A2}
Without loss of generality we can take $G=\sigma_x.$
Let us first consider the case of a single erasure state $|e\rangle.$
Note that $\sigma_{x}=\sigma_{1}^{\dagger}\sigma_{2}+\text{h.c.}$.
Hence the constraint of Eq. \ref{eq: fundamental_ecqfi} is
\begin{align*}
\sigma_{x}=h_{1,2}^{(0)}\sqrt{\gamma_{2}\gamma_{1}}\sigma_1^{\dagger}\sigma_2+h_{2,1}^{(0)}\sqrt{\gamma_{2}\gamma_{1}}\sigma_{2}^{\dagger}\sigma_{1},    
\end{align*}
which implies $h_{1,2}^{(0)}=1/\sqrt{\gamma_{2}\gamma_{1}}.$
Therefore,
\begin{align*}
&h^{(0)}\mathbf{K}^{(1)}=\left(\begin{array}{c}
\frac{\sqrt{\gamma_{1}}}{\sqrt{\gamma_{1}\gamma_{2}}}\sigma_{1}\\
\frac{\sqrt{\gamma_{2}}}{\sqrt{\gamma_{2}\gamma_{1}}}\sigma_{2}
\end{array}\right) \Rightarrow \alpha^{(2)}=(h^{(0)}\mathbf{K}^{(1)})^{\dagger}(h^{(0)}\mathbf{K}^{(1)})=\frac{1}{\gamma_{2}}\sigma_{1}^{\dagger}\sigma_{1}+\frac{1}{\gamma_{1}}\sigma_{2}^{\dagger}\sigma_{2}.
\end{align*}
We therefore get that 
\begin{align}
\begin{split}
&\underset{h}{\min}||\alpha^{(2)}||=\frac{1}{\min\{ \gamma_{1},\gamma_{2}\} } \Rightarrow \boxed{\Iult=\frac{4}{\min\{ \gamma_{1},\gamma_{2}\}}NT.}
\label{supp_eq:asym_bound_sigma_x}
\end{split}
\end{align}

Let us now consider the case of several erasure states, i.e., we have the following jump operators:
$\{ L_{2j-1}=\sqrt{\gamma_{1,j}}\sigma_{1,j},\,L_{2j}=\sqrt{\gamma_{2,j}}\sigma_{2,j}\} _{j=1}^{k}.$
The constraint of Eq. \ref{eq: fundamental_ecqfi} is thus $\mathbf{K}^{\left(1\right)\dagger}h^{\left(0\right)}\mathbf{K}^{\left(1\right)}=\sigma_{x},$
i.e., 
\begin{align*}
\sigma_{x}&=\sum_{j=1}^{k}h_{2j-1,2j}^{(0)}\sqrt{\gamma_{1,j}\gamma_{2,j}}\sigma_{1,j}^{\dagger}\sigma_{2,j}+h_{2j,2j-1}^{(0)}\sqrt{\gamma_{1,j}\gamma_{2,j}}\sigma_{2,j}^{\dagger}\sigma_{1,j}=\sum_{j=1}^{k}
h_{2j-1,2j}^{(0)}\sqrt{\gamma_{1,j}\gamma_{2,j}}\sigma_{x}.
\end{align*}
Denoting $h_{j}:=h_{2j-1,2j}^{(0)}$ we get that the constraint is
\begin{align*}
\sum_{j=1}^{k}h_{j}\sqrt{\gamma_{1,j}\gamma_{2,j}}=1.    
\end{align*}
To derive the objective note that
\begin{align*}
&h^{(0)}\mathbf{K}^{(1)}=\left(\begin{array}{c}
h_{j}\sqrt{\gamma_{1,j}}\sigma_{1,j}\\
h_{j}\sqrt{\gamma_{2,j}}\sigma_{2,j}
\end{array}\right)_{j=1}^{k} \Rightarrow \alpha^{(2)}=\Pi_{1}\left(\sum_{j=1}^{k}h_{j}^{2}\gamma_{1,j}\right)+\Pi_{2}\left(\sum_{j=1}^{k}h_{j}^{2}\gamma_{2,j}\right). 
\end{align*}
The optimization is thus
\begin{align}
\begin{split}
&\underset{\{ h_{j}\} _{j}}{\min}||\alpha^{(2)}||=\underset{\{ h_{j}\} _{j}}{\min}\max\left\{\sum_{j=1}^{k}h_{j}^{2}\gamma_{2,j},\sum_{j=1}^{k}h_{j}^{2}\gamma_{1,j}\right\}\\
&\text{subject to } \sum_{j=1}^{k}h_{j}\sqrt{\gamma_{1,j}\gamma_{2,j}}=1.
\label{eq:ecqfi_sigma_x_multi}
\end{split}
\end{align}
This problem can be solved in general using an SDP. Defining $A=\mathfrak{m} \mathbbm{1}-\text{diag}\left(\sum_{j=1}^{k}h_{j}^{2}\gamma_{2,j},\sum_{j=1}^{k}h_{j}^{2}\gamma_{1,j}\right),$
the SDP reads
\begin{align*}
&\text{min }\mathfrak{m}\\
&\text{subject to }\text{A} \geq0\\
&\sum_{j=1}^{k}h_{j}\sqrt{\gamma_{1,j}\gamma_{2,j}}=1.
\end{align*}
The solution then corresponds to $\underset{\{ h_{j}\} _{j}}{\min}||\alpha^{(2)}||.$
By solving this problem numerically for different cases it can be observed that $\Iult$ in general lies within this range
\begin{align}
\rev{\frac{1}{2\Gamma_\text{ov}}}\leq \underset{\left\{ h_{j}\right\} _{j}}{\min} ||\alpha^{\left(2\right)}|| \leq \rev{\frac{1}{\Gamma_\text{ov}}},
\label{eq:sx_general_ecqfi_bounds}
\end{align}
\rev{where $\Gamma_\text{ov}=\sum_{j=1}^{k}\min\left(\gamma_{1,j},\gamma_{2,j}\right)$}.
Let us prove this observation.

\textit{Proof:} We first prove the inequality:
$\underset{\left\{ h_{j}\right\} _{j}}{\min}||\alpha^{\left(2\right)}||\leq \rev{ \frac{1}{\Gamma_\text{ov}} }.$
Note that 
\begin{align*}
&\underset{\left\{ h_{j}\right\} _{j}}{\min}\left[\max\left(\sum_{j=1}^{k}h_{j}^{2}\gamma_{1,j},\sum_{j=1}^{k}h_{j}^{2}\gamma_{2,j}\right)\right]\leq\underset{\left\{ h_{j}\right\} _{j}}{\min}\left[\sum_{j=1}^{k}h_{j}^{2}\max\left(\gamma_{1,j},\gamma_{2,j}\right)\right]\leq\left[\sum_{j=1}^{k}\min\left(\gamma_{1,j},\gamma_{2,j}\right)\right]^{-1},    
\end{align*}
where the last inequality is obtained by choosing $h_{j}=\frac{1}{\sum_{m}\min\left(\gamma_{1,m},\gamma_{2,m}\right)}\sqrt{\frac{\min\left(\gamma_{1,j},\gamma_{2,j}\right)}{\max\left(\gamma_{1,j},\gamma_{2,j}\right)}},$ which satisfies the constraint in Eq. \ref{eq:ecqfi_sigma_x_multi}.

The inequality $\rev{\frac{1}{2 \Gamma_\text{ov}}}\leq\underset{\left\{ h_{j}\right\} _{j}}{\min}||\alpha^{\left(2\right)}||$
is proven by constructing a QEC protocol that achieves this QFI.
Consider the code space spanned by $\left\{ |0_{L}\rangle=|+\rangle|0\rangle_{a},|1_{L}\rangle=|-\rangle|1\rangle_{a}\right\}, $ 
\rev{where $|\pm\rangle=\frac{1}{\sqrt{2}} (|1\rangle \pm |2\rangle),$ and}
with a recovery operation
\begin{align*}
&|e_{j}\rangle|0\rangle_a\rightarrow|+\rangle|0\rangle_a, \;  |e_{j}\rangle|1\rangle_a\rightarrow\left(-1\right)^{f_{j}}|-\rangle|1\rangle_a\\
& \text{where } f_{j}=\begin{cases}
2 & \gamma_{1,j}\geq\gamma_{2,j}\\
1 & \text{otherwise}.
\end{cases}
\end{align*}
Performing QEC continuously leads to the following logical dephasing dynamics
\begin{align*}
\frac{d\rho}{dt}=-i\omega\left[\sigma_{z}^{L},\rho\right]+\rev{\frac{\Gamma_\text{ov}}{2}}\left(\sigma_{z}^{L}\rho\sigma_{z}^{L}-\rho\right).    
\end{align*}
We refer to Appendix \ref{app:optimal_qec_for_sigma_x}
and to Ref. \cite{zhou2021asymptotic}
for a derivation of this effective dynamics.
It was shown that, under this dynamics, an optimal logical spin squeezed state achieves a QFI of $I=\frac{NT}{\Gamma_\text{ov}/2}$\cite{zhou2021asymptotic}. 
Hence, we obtain a QFI of \rev{$I=\frac{NT}{\Gamma_\text{ov}/2 },$}
and thus $||\alpha^{\left(2\right)}||\geq \rev{ \frac{1}{2 \Gamma_\text{ov}} }.$

We remark that the \rev{QFI} upper bound of $\frac{4T}{\rev{ \Gamma_\text{ov} } }$ is saturable in the case
where the decay from one of the states is dominant,
i.e., where $\forall j\;\gamma_{1,j}\leq\gamma_{2,j}$ and equivalently $\forall j\;\gamma_{1,j}\geq\gamma_{2,j}.$
This can be seen explicitly from the optimization problem of Eq. \ref{eq:ecqfi_sigma_x_multi}: assuming wlog that $\gamma_{1,j}\leq\gamma_{2,j}$
the minimization becomes 
$\underset{{\bf h}}{\min}\;\sum_{j=1}^{k}h_{j}^{2}\gamma_{2,j}$ subject to $\sum_{j=1}^{k}h_{j}\sqrt{\gamma_{2,j}\gamma_{1,j}}=1$.
Using Cauchy-Schwartz inequality
\begin{align*}
\left(\sum_{j=1}^{k}h_{j}\sqrt{\gamma_{2,j}\gamma_{1,j}}\right)^{2}\leq\left(\sum_{j=1}^{k}h_{j}^{2}\gamma_{2,j}\right)\left(\sum_{j=1}^{k}\gamma_{1,j}\right)    
\end{align*}
we obtain that the minimum is indeed $\frac{1}{\sum_{j=1}^{k}\gamma_{1,j}}.$
This also implies that for $\gamma_{1,j}=\gamma_{2,j}=\gamma_j$ the ECQFI is
$\frac{4T}{\sum_{j=1}^{k}\gamma_{j}},$
which is the same as in the case of $G \parallel \sigma_z,$ since given identical decays from $|1\rangle$ and $|2\rangle$ there is no difference between the different Pauli operators.

\begin{figure}[h]
 \begin{subfigure}[t]{0.35\textwidth}
        \centering
\includegraphics[width=\linewidth]{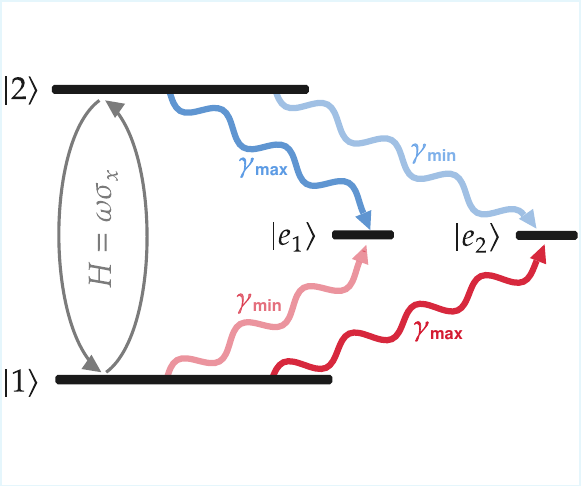}\caption{}
    \end{subfigure}
 \begin{subfigure}[t]{0.45\textwidth}
        \centering
\includegraphics[width=\linewidth]{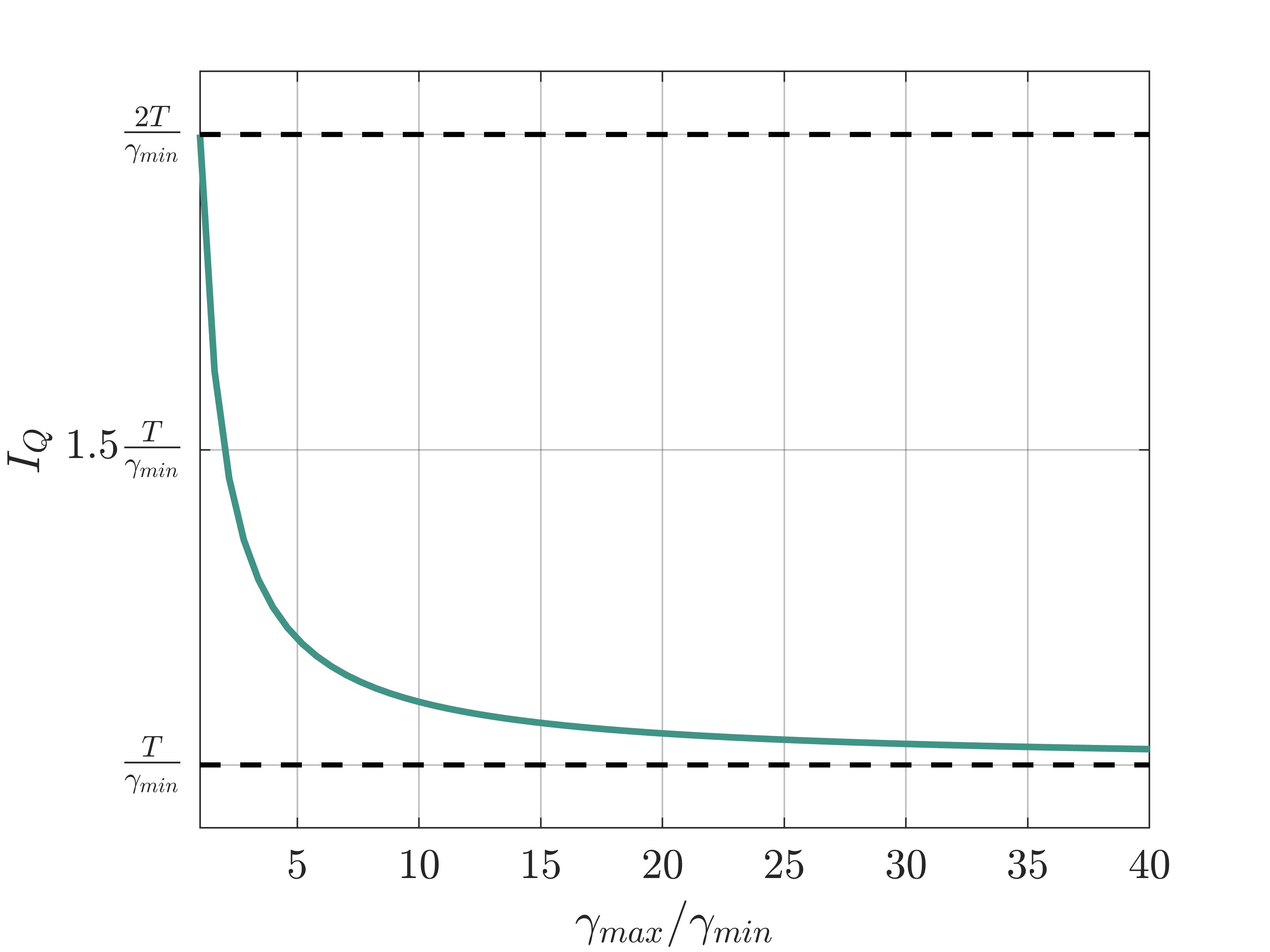}\caption{}
    \end{subfigure}%
\caption{ (a)
Sketch of the erasure configuration: $G=\sigma_{x}$ with two erasure states such that $\gamma_{\min}:=\gamma_{1,1}=\gamma_{2,2}<\gamma_{\max}:=\gamma_{1,2}=\gamma_{2,1}.$
(b) $\Iult$ given this erasure configuration, it can be seen that the precision limits lie between the lower and upper bounds of Eq. \ref{eq:sx_general_ecqfi_bounds} depending on $\gamma_{\max}/\gamma_{\min}$. }
\label{fig:sx_multi_erasure_bounds}
\end{figure}

We can further illustrate the bounds by considering
the special case illustrated in Fig. \ref{fig:sx_multi_erasure_bounds} (a):
two erasure states with decay rates of 
$\gamma_{\min}:=\gamma_{1,1}=\gamma_{2,2}<\gamma_{\max}:=\gamma_{1,2}=\gamma_{2,1}$. The dominant decay to $|e_1\rangle$ is from $|2\rangle$ while the dominant decay to $|e_2\rangle$ is from $|1\rangle.$
The resulting sequential ECQFI bounds are illustrated in Fig. \ref{fig:sx_multi_erasure_bounds} (b). It can be seen that as $\gamma_{\max}\rightarrow\gamma_{\min}$ the bound converges to $4\frac{T}{\sum_{j}\min\left\{ \gamma_{1,j},\gamma_{2,j}\right\} }=\frac{2T}{\gamma_{\min}},$
\rev{whereas} in the limit of $\gamma_{\max} \gg \gamma_{\min}$  
the bound converges to 
$2\frac{T}{\sum_{j}\min\left\{ \gamma_{1,j},\gamma_{2,j}\right\} }=\frac{T}{\gamma_{\min}}.$

\subsection{\texorpdfstring{$\Iult$}{Iult} for erasure qubits with general \texorpdfstring{$G$}{G}}

\begin{figure}[t!]
\centering
    \begin{subfigure}[t]{0.4\textwidth}
        \centering
\includegraphics[width=\linewidth]{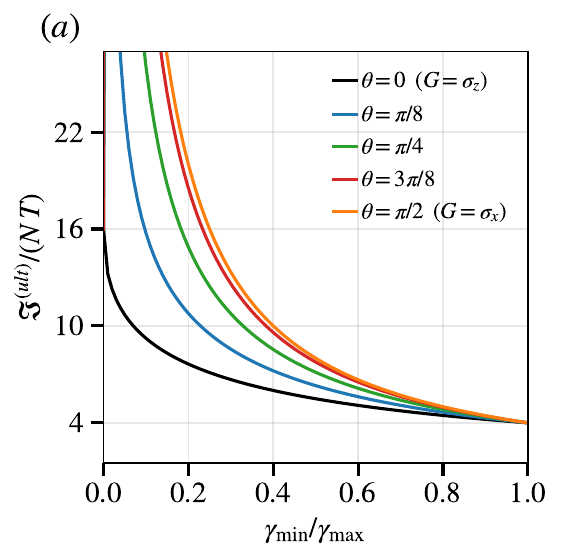}
    \end{subfigure}%
    \begin{subfigure}[t]{0.4\textwidth}
        \centering
\includegraphics[width=\linewidth]{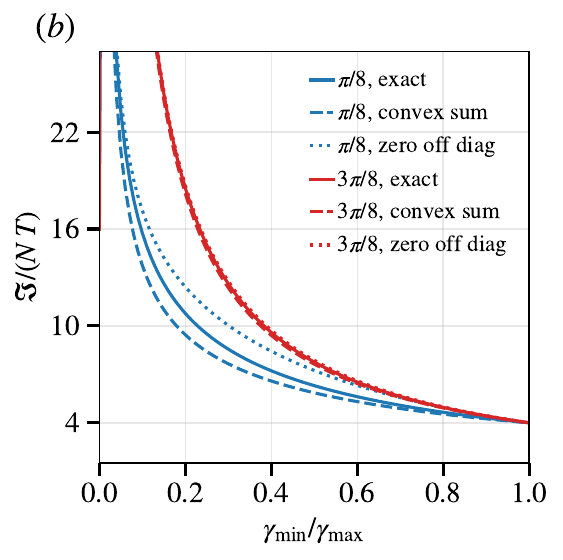}
    \end{subfigure}
\caption{
$\Iult$ for $G=\cos{\theta}\sigma_z+\sin{\theta}\sigma_x$ as a function of $\gamma_{\min}/\gamma_{\max}$. (a) Exact bounds  for several values of $\theta$. (b) For $\theta=\pi/8$ (blue) and $3\pi/8$ (red), comparison of the exact bound (solid) with the convex-sum lower bound of Eq. \ref{eq:convex_sum} (dashed), and with the analytic upper bound of Eq. \ref{eq:zero_off_diag} (dotted).}\label{general_g_upper_bound}
\end{figure}

Without loss of generality we can take $G=\cos{\theta}\sigma_z+\sin{\theta}\sigma_x$ with $\theta \in [0, \frac{\pi}{2}]$.
Our analysis of this case assumes a single erasure state $|e\rangle.$ The constraint in Eq.  \ref{eq: fundamental_ecqfi} is thus
\begin{align*}
\cos{\theta}\sigma_z+\sin{\theta}\sigma_x=h_{1,1}^{(0)}\gamma_{1}\sigma_{1}^{\dagger}\sigma_{1}+h_{2,2}^{(0)}\gamma_{2}\sigma_{2}^{\dagger}\sigma_{2}+h_{1,2}^{(0)}\sqrt{\gamma_{2}\gamma_{1}}\sigma_1^{\dagger}\sigma_0+h_{2,1}^{(0)}\sqrt{\gamma_{2}\gamma_{1}}\sigma_{2}^{\dagger}\sigma_{1}+h_{0,0}^{(2)}\mathbbm{1} 
\end{align*}
which implies
\begin{align*}
&h_{1,1}^{(0)}\gamma_{1}+h_{0,0}^{(2)}=\cos{\theta}\\
&h_{2,2}^{(0)}\gamma_{2}+h_{0,0}^{(2)}=-\cos{\theta}\\
&h_{1, 2}^{(0)}=h_{2, 1}^{(0)}=\sin{\theta}/\sqrt{\gamma_{1}\gamma_{2}}.
\end{align*}
We can thus express all $h^{(0)}$ terms using $h_{0,0}^{(2)}$. Denoting $\mathfrak{h}:=h_{0,0}^{(2)}$ we get
\begin{align*}
\begin{split}
&    h^{(0)}\mathbf{K}^{(1)}= \begin{pmatrix}
        0 &  &  \\
        & \frac{(\cos{\theta}-\mathfrak{h})}{\gamma_{1}} &\frac{\sin{\theta}}{\sqrt{\gamma_{1}\gamma_{1}}}  \\
        & \frac{\sin{\theta}}{\sqrt{\gamma_{1}\gamma_{1}}} & -\frac{(\cos{\theta}+\mathfrak{h})}{\gamma_{2}}
    \end{pmatrix}\begin{pmatrix}
        0 \\
        \sqrt{\gamma_{1}}\sigma_{1} \\
       \sqrt{\gamma_{2}}\sigma_{2}
\end{pmatrix}=\left(\begin{array}{c}
0\\
\frac{(\cos{\theta}-\mathfrak{h})\sigma_1+\sin{\theta}\sigma_2}{\sqrt{\gamma_{1}}}\\
\frac{-(\cos{\theta}+\mathfrak{h})\sigma_2+\sin{\theta}\sigma_1}{\sqrt{\gamma_{2}}}
\end{array}\right)\\
&\Rightarrow\alpha^{(2)}=\Pi_1\left(\frac{(\cos{\theta}-\mathfrak{h})^2}{\gamma_1}+\frac{\sin{\theta}^2}{\gamma_2}\right)+\Pi_2\left(\frac{(\cos{\theta}+\mathfrak{h})^2}{\gamma_2}+\frac{\sin{\theta}^2}{\gamma_1}\right)+\sigma_x\left(\frac{\sin{\theta}(\cos{\theta}-\mathfrak{h})}{\gamma_1}-\frac{\sin{\theta}(\cos{\theta}+\mathfrak{h})}{\gamma_2}\right),
\end{split}
\end{align*}
where $\Pi_j=|j\rangle\langle j|.$
The eigenvalues of this $\alpha^{(2)}$ are
\begin{multline*}
   \lambda_{\pm}= \frac{1}{2 \gamma_1 \gamma_2}\Big[( 1+\mathfrak{h}^2)(\gamma_1 + \gamma_2) + 
  2 \mathfrak{h} \cos{\theta}(\gamma_1 - \gamma_2 ) \pm 
  \sqrt{-4 (-1 + 
      \mathfrak{h}^2)^2 \gamma_1 \gamma_2 + \left[(1 + 
        \mathfrak{h}^2) (\gamma_1 + \gamma_2) + 
     2 \mathfrak{h} \cos{\theta} (\gamma_1 - \gamma_2)\right]^2} \Big]
\end{multline*}

We thus obtain
\begin{align*}
&\underset{h}{\min}||\alpha^{(2)}||=
\underset{\mathfrak{h}}{\min} \; \max\left\{ \lambda_{+}, \lambda_{-}\right\}=\underset{\mathfrak{h}}{\min} \; \lambda_{+} \Rightarrow \boxed{\Iult=4NT\underset{\mathfrak{h}}{\min} \; \lambda_{+}.}    
\end{align*}
We do not have a closed-form solution to this minimization problem and solve it only numerically. The results are plotted in Fig. \ref{general_g_upper_bound}.
To derive an analytic upper bound, we set the off diagonal terms of $\alpha^{(2)}$ to zero. This occurs when $h = -\frac{\gamma_1 - \gamma_2}{\gamma_1 + \gamma_2}\cos{\theta}$ which yields
\begin{equation}\label{eq:zero_off_diag}
\mathcal{I}_{\text{zero}}=4T \left(\frac{4 \gamma_1}{(\gamma_1+ \gamma_2)^2}\cos{\theta}^2+ 
   \frac{\sin{\theta}^2}{\gamma_2}\right). 
\end{equation}
This upper bound is shown in dotted lines in Fig. \ref{general_g_upper_bound}. For comparison, we also include (in dashed lines) the convex sum lower bound: 
\begin{equation}\label{eq:convex_sum}
\mathcal{I}_{\text{convex}} = \cos{\theta}^2 \mathfrak{I}_{ \sigma_z}^{(\text{ult})}+\sin{\theta}^2\mathfrak{I}_{\sigma_{x}}^{(\text{ult})},
\end{equation}
where $\mathfrak{I}_{ \sigma_m}^{(\text{ult})}$ corresponds to the ultimate limit given $G=\sigma_{m}.$
We observe that these lower and upper bounds are more accurate as $\theta$ approaches $\pi/2$ and when $\gamma_{\min}/\gamma_{\max}$ is close to $0$ or $1$.

\section{Optimal entangled strategies for \texorpdfstring{$G \parallel \sigma_{z}$}{G parallel to sigma\_z}: saturation of \texorpdfstring{$\Iult$}{Iult} (proof of Claim 1)}
\label{sec:appendix_squeezed_spin_erasure}
In this section we prove Claim 1 of the main text, i.e., we show that $\Iult$ is saturated in the limit of $N \rightarrow \infty$
for $G =\sigma_{z}$ with the strategy class $(ii)$ (entangled states) by using an optimal ancilla-free spin squeezed state. 
Considering $N$ qubits, let us define the total spin operators: $J_{m}:=\frac{1}{2}\underset{i=1}{\overset{N}{\sum}}\sigma_{m}^{(i)}$, where $m\in\left\{ x,y,z\right\}$.
We take as an initial state the following spin squeezed state \cite{kitagawa1993squeezed}
\begin{align}
|\psi_{\phi,\chi}^{\text{sq}}\rangle=e^{-i\phi J_{y}}e^{i\chi\left(J_{z}J_{y}+J_{y}J_{z}\right)}e^{i\frac{\pi}{2}J_{y}}|1\rangle^{\otimes N}.
\label{eq:squeezed_state_params_def}
\end{align}
The first pulse $e^{i\frac{\pi}{2}J_{y}}$ maps $|1\rangle^{\otimes N}$ into the spin coherent state $|+\rangle^{\otimes N}$. The qubits then evolve under the two-axis twisting Hamiltonian $\chi\left(J_{z}J_{y}+J_{y}J_{z}\right)$ which squeezes $\text{var}\left(J_{y}\right)$ . The final pulse $e^{-i\phi J_{y}}$ is a rotation \rev{about} the y-axis, after which $\langle J_{x}\rangle\approx\frac{N}{2}\cos\left(\phi\right)$ , $\langle J_{z}\rangle\approx\frac{N}{2}\sin\left(\phi\right)$.
At large $N$, the QFI of these states is lower bounded by the error propagation formula \cite{demkowicz2015quantum, pezze2018quantum}
\begin{align}
I\geq\frac{\left(\partial_{\omega}\langle J_{x}\rangle_{t}\right)^{2}}{\text{var}\left(J_{x}\right)_{t}}.
\label{eq:error_propagation_formula}
\end{align}
It therefore suffices to show that the error propagation expression converges to $\Iult.$
To calculate this bound we need to find the time evolution of $\langle J_{x}\rangle$ , $\text{var}\left(J_{x}\right)$ given our erasure channel.
Defining $\eta_{i}=e^{-\Gamma_{i}t}$ (for $i=1,2$) we obtain
\begin{align}
\begin{split}
&\langle J_{x}\rangle_{t}=\sqrt{\eta_{1}\eta_{2}}\left(\cos\left(\varphi\right)\langle J_{x}\rangle_{0}-\sin\left(\varphi\right)\langle J_{y}\rangle_{0}\right), \\
&\partial_{\omega}\langle J_{x}\rangle_{t}=-\sqrt{\eta_{1}\eta_{2}}2t\left(\sin\left(\varphi\right)\langle J_{x}\rangle_{0}+\cos\left(\varphi\right)\langle J_{y}\rangle_{0}\right),
\label{eq:expectation_x_squeezed_t}
\end{split}
\end{align}
where $\varphi=2\omega t$.
$\text{var} \left( J_x \right)$ evolves as
\begin{align}
\begin{split}
&\text{var}\left(J_{x}\right)_{t}=\left(\frac{\eta_{1}\left(1-\eta_{2}\right)+\eta_{2}\left(1-\eta_{1}\right)}{8}\right)N\\
&+\frac{1}{4}\left[\eta_{1}\left(1-\eta_{2}\right)-\eta_{2}\left(1-\eta_{1}\right)\right]\langle J_{z}\rangle_{0}\\
&+ \eta_{1}\eta_{2}\left[\cos^{2}\left(\varphi\right)\text{var}\left(J_{x}\right)_{0}+\sin^{2}\left(\varphi\right)\text{var}\left(J_{y}\right)_{0}-2\sin\left(\varphi\right)\cos\left(\varphi\right)\text{cov}\left(J_{x},J_{y}\right)_{0}\right].
\end{split}
\label{eq:var_x_squeezed_t}
\end{align}
From Eqs. \ref{eq:expectation_x_squeezed_t}, \ref{eq:var_x_squeezed_t} we observe that the maximal signal and minimal noise are obtained at  $\varphi= \pi/2.$
Since $\varphi$ is tunable (we can vary the measurement basis in the $J_x-J_y$ plane), we can assume 
this value of $\varphi$. Inserting Eqs. \ref{eq:expectation_x_squeezed_t}, \ref{eq:var_x_squeezed_t} in the error propagation formula (Eq. \ref{eq:error_propagation_formula}) yields
\begin{align}
\frac{\eta_{1}\eta_{2}\langle J_{x}\rangle_{0}^{2}}{\left(\frac{\eta_{1}\left(1-\eta_{2}\right)+\eta_{2}\left(1-\eta_{1}\right)}{8}\right)N+\frac{1}{4}\left[\eta_{1}\left(1-\eta_{2}\right)-\eta_{2}\left(1-\eta_{1}\right)\right]\langle J_{z}\rangle_{0}+\eta_{1}\eta_{2}\text{var}\left(J_{y}\right)_{0}}4t^{2}.
\label{supp_eq:error_propagation_with_var_jy_term}
\end{align}
In the limit of large $N$, we can neglect the squeezed $\text{var}\left(J_{y}\right)_{0}$ (since $\text{var}\left(J_{y}\right)_{0}/N \rightarrow 0$) and insert 
$\langle J_{x}\rangle\approx\frac{N}{2}\cos\left(\phi\right)$ , $\langle J_{z}\rangle\approx\frac{N}{2}\sin\left(\phi\right),$
to get
\begin{align}
&\frac{\eta_{1}\eta_{2}\cos\left(\phi\right)^{2}/4}{\left(\frac{\eta_{1}\left(1-\eta_{2}\right)+\eta_{2}\left(1-\eta_{1}\right)}{8}\right)+\frac{1}{8}\left[\eta_{1}\left(1-\eta_{2}\right)-\eta_{2}\left(1-\eta_{1}\right)\right]\sin\left(\phi\right)}4Nt^{2}\\
&=\frac{\cos\left(\phi\right)^{2}}{\left(l_{1}+l_{2}\right)+\left(l_{2}-l_{1}\right)\sin\left(\phi\right)}8Nt^{2},
\end{align}
where $l_{1}=\frac{1-\eta_{1}}{\eta_{1}},$ $l_{2}=\frac{1-\eta_{2}}{\eta_{2}}$.
We now need to minimize over both $\phi, t$, this optimization can be performed separately: we first find optimal $\phi$, and then optimal $t$. The maximization over $\phi$  is equivalent to
 \begin{align}
 \underset{\phi}{\min}\frac{l_{1}}{1+\sin\left(\phi\right)}+\frac{l_{2}}{1-\sin\left(\phi\right)}.    
 \end{align}
The optimal $\phi$  satisfies $\sin\left(\phi_{\text{opt}}\right)=\frac{\sqrt{l_{1}}-\sqrt{l_{2}}}{\sqrt{l_{2}}+\sqrt{l_{1}}}=\frac{\sqrt{\eta_{2}\left(1-\eta_{1}\right)}-\sqrt{\eta_{1}\left(1-\eta_{2}\right)}}{\sqrt{\eta_{2}\left(1-\eta_{1}\right)}+\sqrt{\eta_{1}\left(1-\eta_{2}\right)}},$
and the minimum is $\frac{\left(\sqrt{l_{1}}+\sqrt{l_{2}}\right)^{2}}{2}$. The error propagation expression then reduces to
\rev{
\begin{align}
\frac{2}{\left(\sqrt{l_{1}}+\sqrt{l_{2}}\right)^{2}}8Nt^{2}=\frac{2}{\left(\sqrt{\frac{1-\eta_{1}}{\eta_{1}}}+\sqrt{\frac{1-\eta_{2}}{\eta_{2}}}\right)^{2}}8Nt^{2}=\frac{16Nt^{2}}{\left(\sqrt{e^{\Gamma_{1}t}-1}+\sqrt{e^{\Gamma_{2}t}-1}\right)^{2}}.   
\end{align}}
The optimization of the rate over $t$ is thus
\begin{align}
\rev{\underset{t>0}{\sup}\frac{16Nt}{\left(\sqrt{e^{\Gamma_{1}t}-1}+\sqrt{e^{\Gamma_{2}t}-1}\right)^{2}}}=
\boxed{\frac{16}{\left(\sqrt{\Gamma_{1}}+\sqrt{\Gamma_{2}}\right)^{2}}N.}
\label{supp_eq:squeezed_convergence_optimal_time}
\end{align}
We thus obtained $\Iult/T,$
and this establishes the convergence of spin squeezed states to the ultimate asymptotic precision bound and proves Claim 1. $\square$

\rev{Note that the optimal interrogation time in Eq. \ref{supp_eq:squeezed_convergence_optimal_time} is $t \rightarrow 0^{+}.$ This limit must
be taken only after the large-$N$ limit.
For any finite $N$, there is a non-zero residual squeezed
variance, $\operatorname{var}(J_y)_0$ (Eq. \ref{supp_eq:error_propagation_with_var_jy_term}). Because of this term the optimal interrogation time for any finite $N$ is finite. In the limit of $N \rightarrow \infty$ this residual variance can be neglected and the optimal interrogation time converges to $0$. 
}

\rev{Let us also comment on the orientation of the optimal spin-squeezed state. In the asymptotic large $N$ limit the optimal interrogation time goes to $0$. Hence the optimal orientation corresponds to 
\begin{align}
\sin\left(\phi_{\text{opt}}\right)=\frac{\sqrt{\Gamma_{1}}-\sqrt{\Gamma_{2}}}{\sqrt{\Gamma_{1}}+\sqrt{\Gamma_{2}}}.
\end{align}
Thus, unlike the standard equator-centered spin-squeezed states commonly used for dephasing, the
optimal states under asymmetric erasure noise require a
rate-dependent rotation of their mean-spin direction. The equatorial
orientation, $\phi_{\mathrm{opt}}=0,$ is recovered only in the symmetric
case $\Gamma_1=\Gamma_2$.}

\begin{figure}[h!]
    \centering
    \begin{subfigure}[t]{0.45\textwidth}
        \centering
        \includegraphics[width=\linewidth]{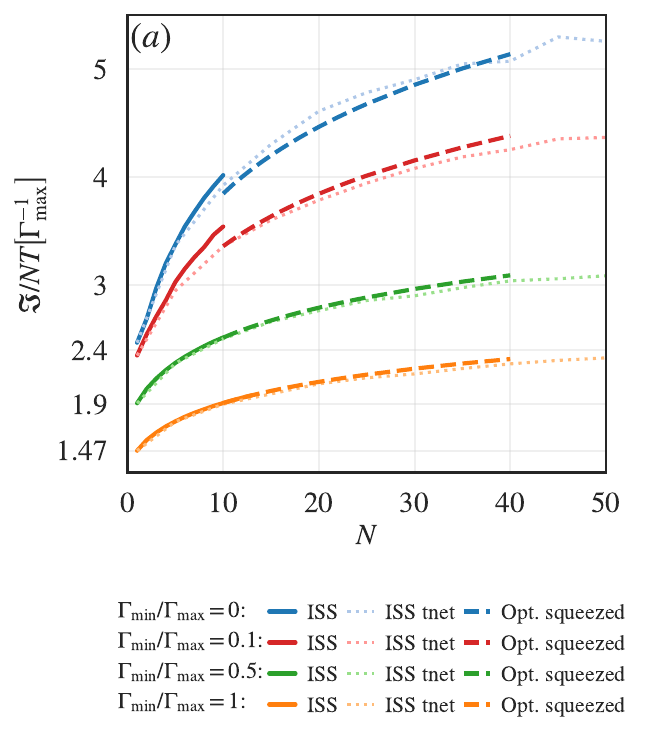}
        \caption*{}
        \label{fig:opt_ent_a}
    \end{subfigure}%
    \begin{subfigure}[t]{0.5
    \textwidth}
        \centering
        \includegraphics[width=\linewidth]{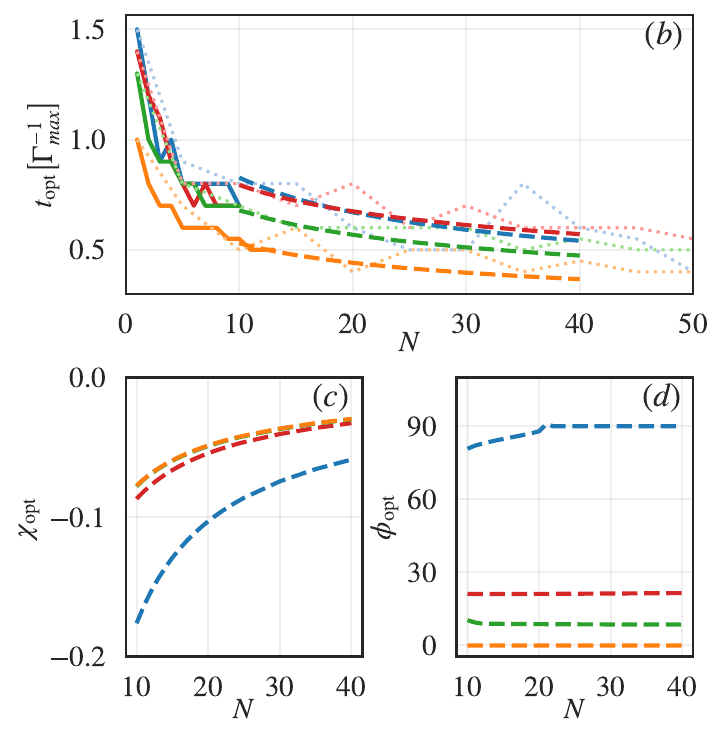}
        \caption*{    }
        \label{fig:opt_ent_bc}
    \end{subfigure}
    \caption{Optimal entangled strategies and spin squeezed state strategies given $G=\sigma_z$ and for different $N, \Gamma_1,\Gamma_2.$
        (a) The QFI with optimal entangled state strategies and optimal spin squeezed state strategies as a function of $N.$ The solid lines correspond to exact optimal entangled states strategies, calculated using ISS. The dotted lines correspond to the optimal entangled QFI obtained with ISS-tnet numerical calculation. The ISS-tnet values provide an approximation of $\Ient.$ The dashed lines correspond to the QFI with the optimal spin-squeezed states.
        (b) Optimal measurement time as a function of $N$.
        (c) Optimal spin squeezing rate $\chi$ as a function of $N$.
        (d) Optimal rotation angle of the spin squeezed state, $\phi,$ as a function of $N.$} \label{fig:opt_ent}
\end{figure}

\begin{figure}[t]
    \centering

    \begin{subfigure}[t]{0.25\textwidth}
        \includegraphics[width=\linewidth]{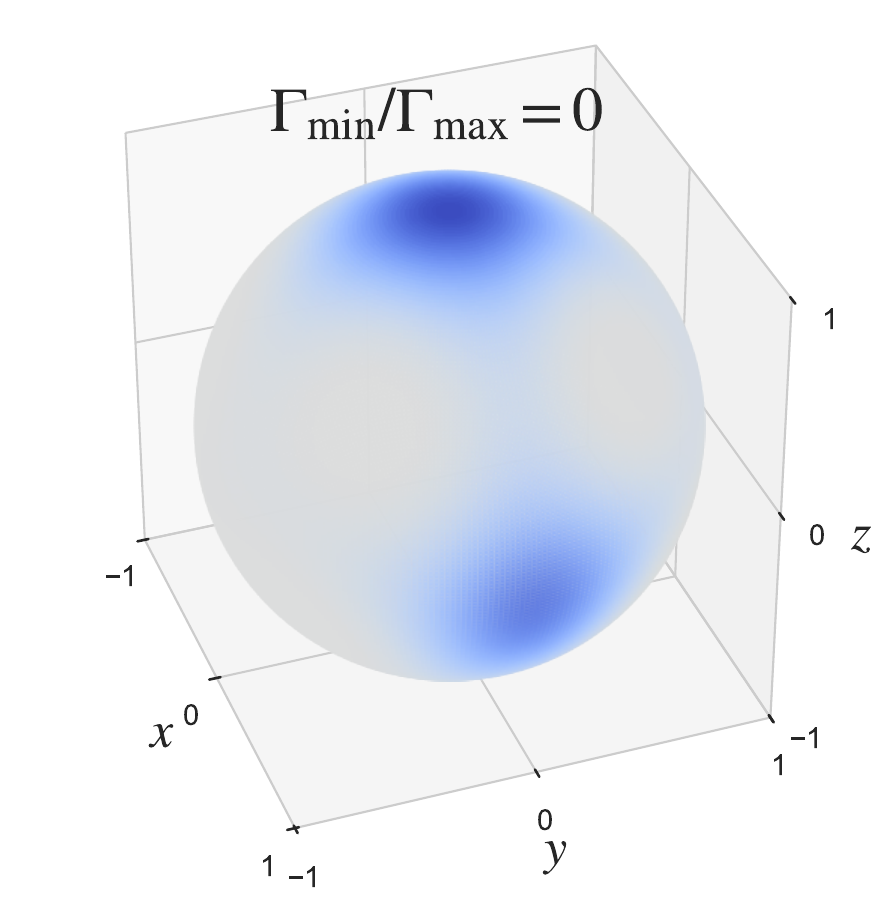}
        \caption{}
    \end{subfigure}\hfill
    \begin{subfigure}[t]{0.25\textwidth}
        \includegraphics[width=\linewidth]{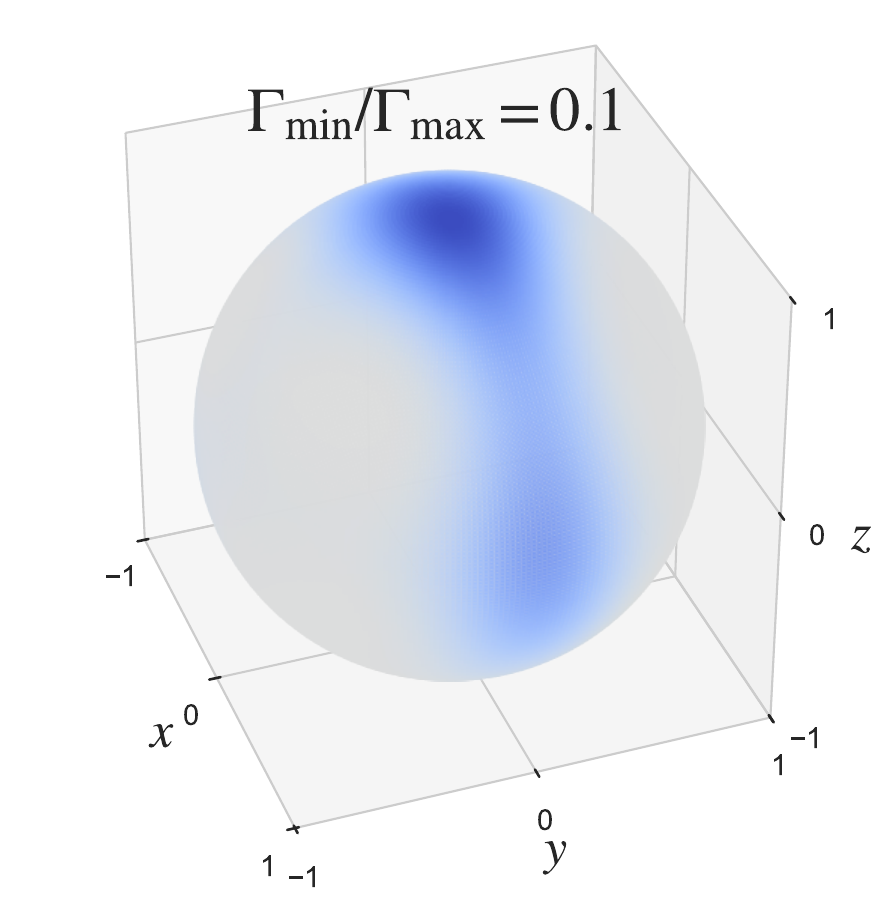}
        \caption{}
    \end{subfigure}\hfill
    \begin{subfigure}[t]{0.25\textwidth}
        \includegraphics[width=\linewidth]{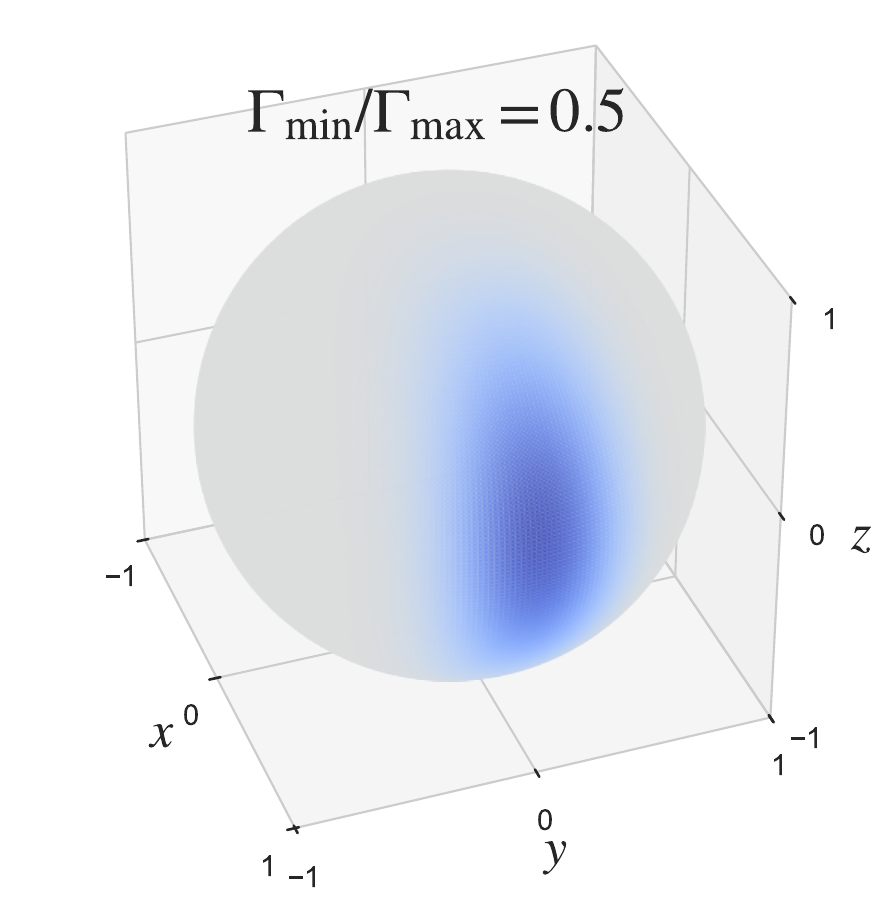}
        \caption{}
    \end{subfigure}\hfill
    \begin{subfigure}[t]{0.25\textwidth}
        \includegraphics[width=\linewidth]{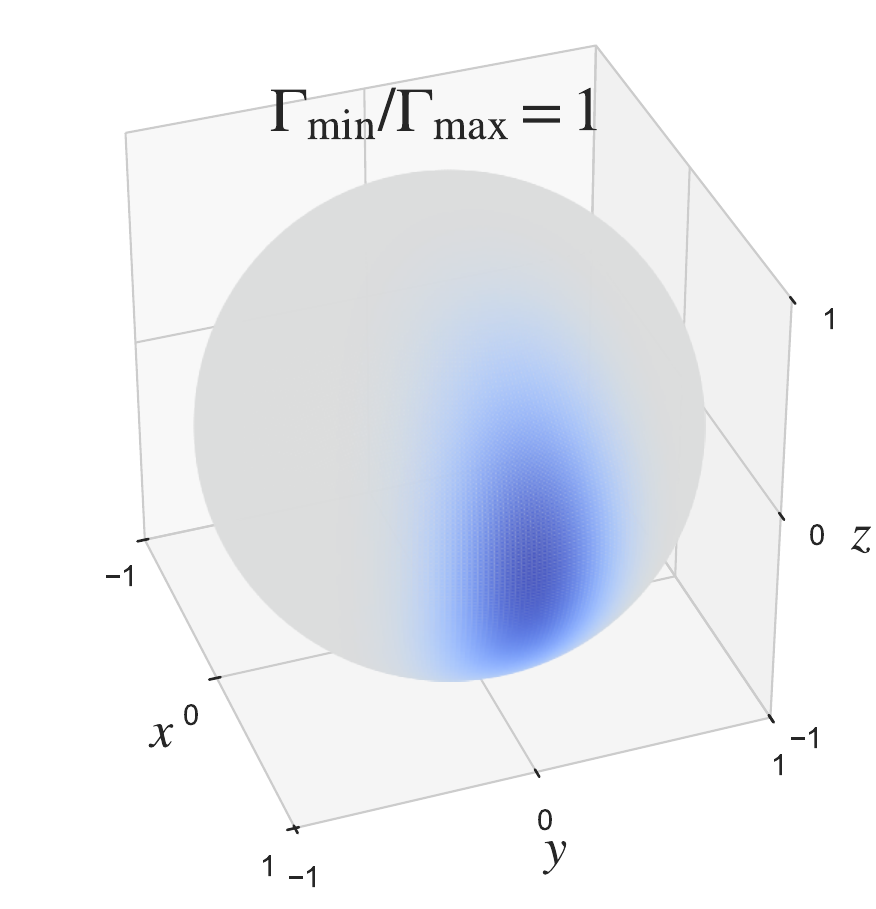}
        \caption{}
    \end{subfigure}

    \vspace{0.5em}

    \begin{subfigure}[t]{0.25\textwidth}
        \includegraphics[width=\linewidth]{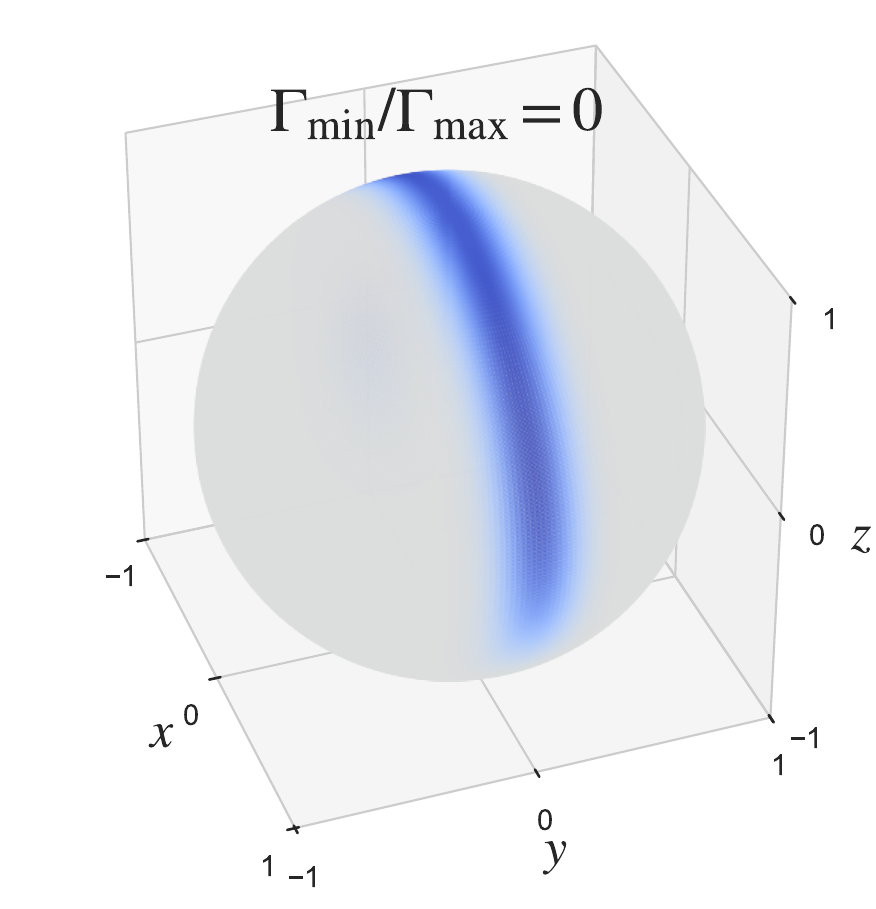}
        \caption{}
    \end{subfigure}\hfill
    \begin{subfigure}[t]{0.25\textwidth}
        \includegraphics[width=\linewidth]{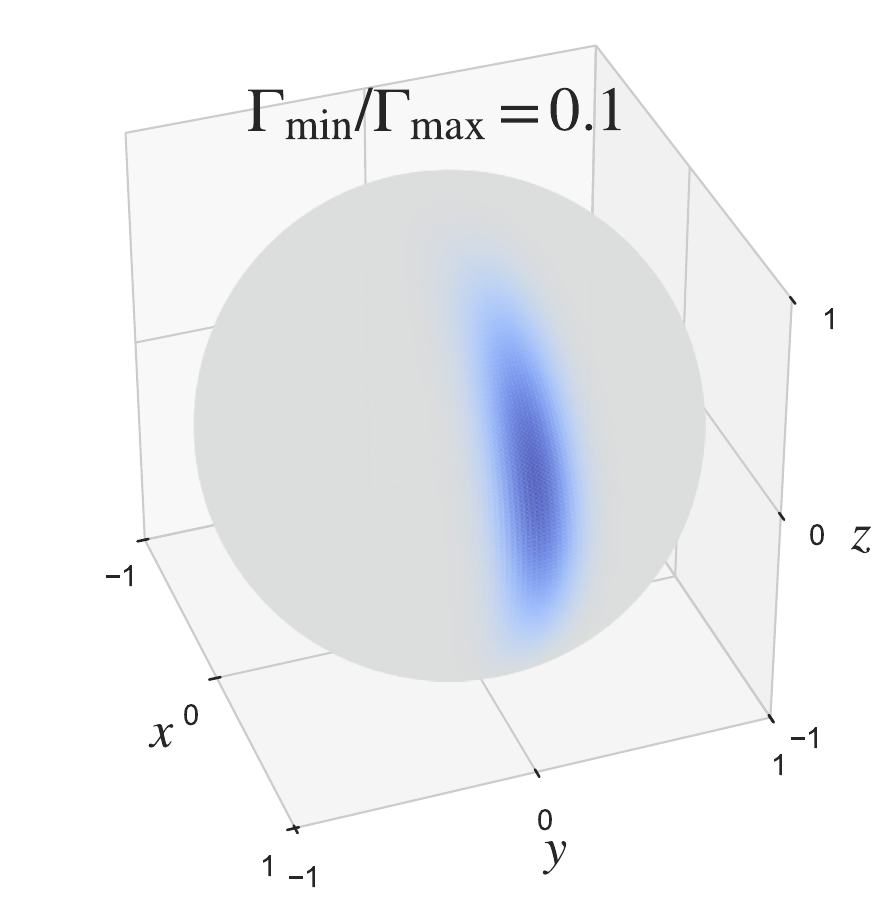}
        \caption{}
    \end{subfigure}\hfill
    \begin{subfigure}[t]{0.25\textwidth}
        \includegraphics[width=\linewidth]{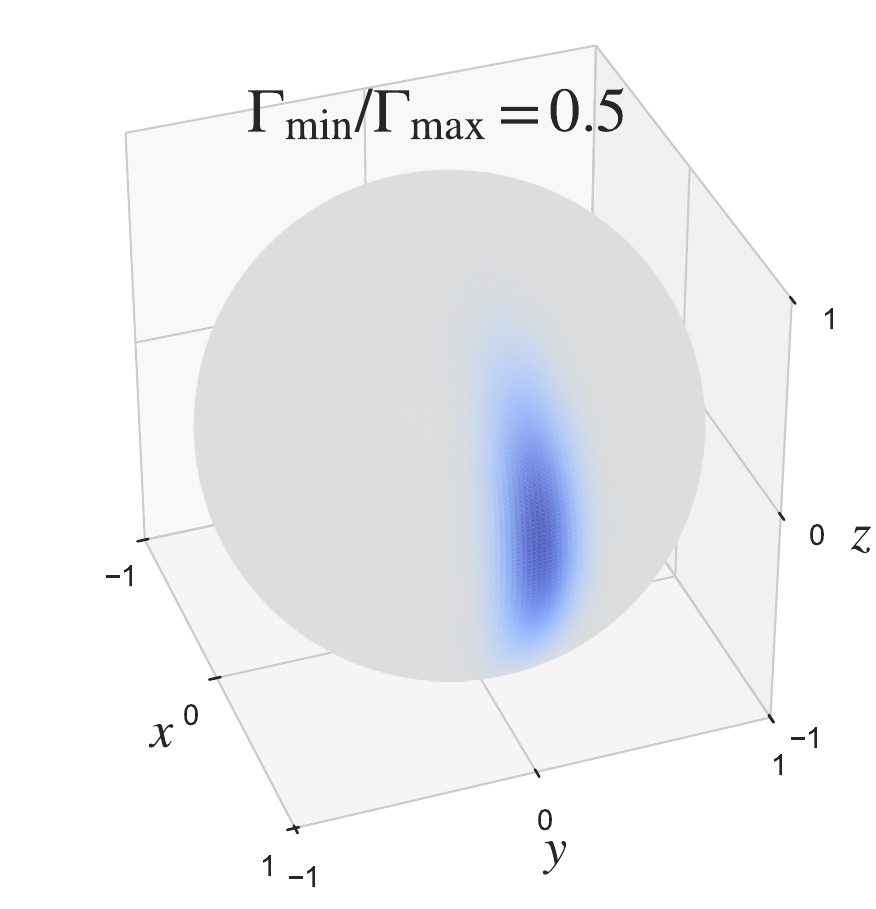}
        \caption{}
    \end{subfigure}\hfill
    \begin{subfigure}[t]{0.25\textwidth}
        \includegraphics[width=\linewidth]{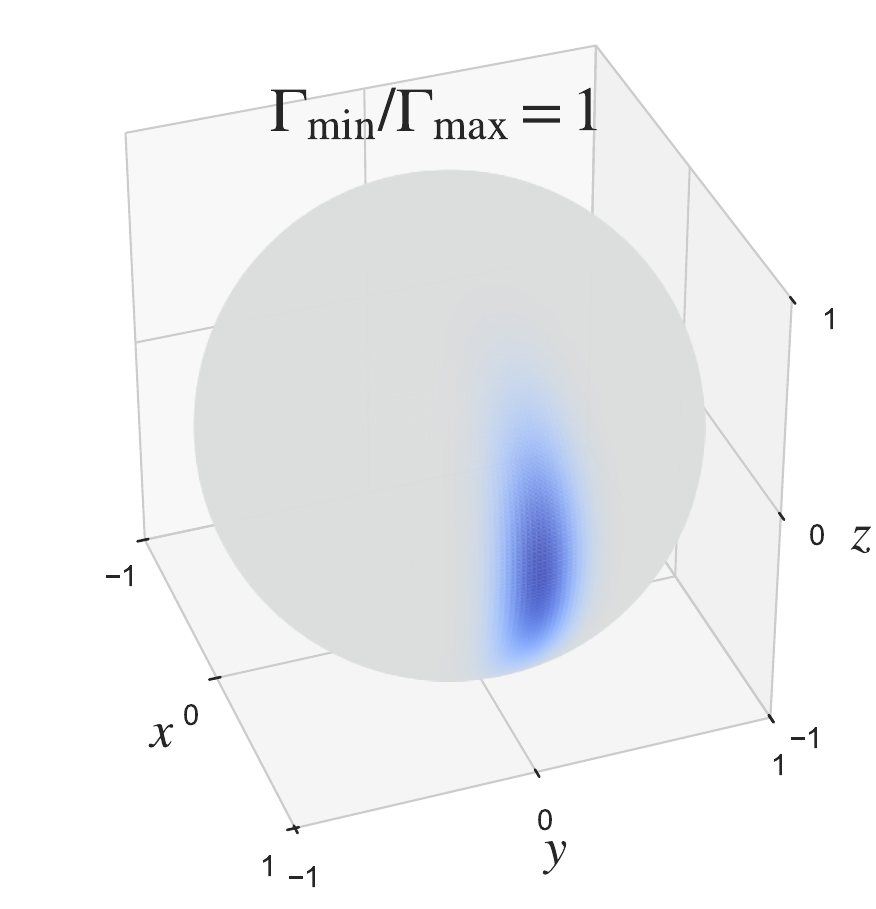}
        \caption{}
    \end{subfigure}

    \caption{(a-d) Optimal entangled states for $N=10$ obtained via ISS calculation, and (e-h) optimized spin squeezed states for $N=40$, shown for different $\Gamma_{\min}/\Gamma_{\max}$ ratios.}
    \label{fig:bloch_8}
\end{figure}

The analysis above corresponds to the asymptotic behavior of the entangled strategy. We further evaluated $\Ient$ for finite $N$. The performance of optimal entangled states and optimal spin squeezed states is shown in Fig. \ref{fig:opt_ent}. We calculated the optimal entangled strategy for up to $N=10$ numerically using ECQFI and iterative see-saw (ISS) method \cite{macieszczak2013quantum, kolodynski2013efficient, dulian2025qmetro++}.
For larger $N,$ ISS becomes intractable and we use a tensor-network based approach to evaluate $\Ient$, the ISS-tnet of QMETRO++ package \cite{dulian2025qmetro++}. 
As shown in Fig. \ref{fig:opt_ent}, the ISS-tnet results are not exact, yet they provide a good approximation of $\Ient.$
The performance of optimal spin squeezed states was obtained by optimizing the QFI rate over the squeezing rate $\chi$, the  rotation angle $\phi,$ (see Eq. \ref{eq:squeezed_state_params_def}) and the measurement time $t$. It can be observed that the optimal spin squeezed state strategy yields QFI that matches the values obtained with ISS-tnet for large $N$.  This provides a numerical evidence for the optimality of spin squeezed states already for $N \geq 30.$ 
The Husimi function of the optimal entangled states and optimal spin squeezed states for different $N,\Gamma_1,\Gamma_2$ is illustrated in Fig. \ref{fig:bloch_8}.
We can observe that for $N=10$ for $\Gamma_{\min}$ close to $\Gamma_{\max}$ the optimal entangled state is a spin squeezed state, while in the asymmetric case of $\Gamma_{\min} \ll \Gamma_{\max}$ the optimal state is not a spin squeezed state. It will become a spin squeezed only for larger $N$.  

The channel QFI and ISS calculations were performed using an analogy between our $N$ erasure qubits and a lossy two bosonic modes interferometer.
This analogy was previously studied  in Refs. \cite{demkowicz2014using, demkowicz2015quantum}.
Instead of working in the $2^{N}$ dimensional Hilbert space of $N$ qubits, the problem can be reduced to the Hilbert space of two bosonic modes with a total number of photons $\leq N$, which is $(N+1)(N+2)/2$-dimensional. 
Note that the input state and the unitary dynamics are restricted to the symmetric subspace of $N$-qubits, which is $N+1$ dimensional. The erasure noise takes us out of this subspace, since the state after an erasure decay is not symmetric anymore.
However, since there is no information in the erasure state we can trace out the qubits that decayed. The state after tracing out an erased qubit corresponds to a symmetric $\left(N-1\right)$-qubits state. This implies that the relevant Hilbert space corresponds to $\underset{k=0}{\overset{N}{\bigoplus}}\mathcal{H}_{\text{symm},k}$, where $\mathcal{H}_{\text{symm},k}$ is the symmetric subspace of $k$ qubits.
Furthermore the dynamics is isomorphic to the dynamics of two bosonic modes in a lossy interferometer. The space $\mathcal{H}_{\text{symm},k}$ is isomorphic to the space of $k$ photons in two modes. The rest of the isomorphism is given as follows:
\begin{align}
\omega J_{z}\leftrightarrow\omega\left(a_{1}^{\dagger}a_{1}-a_{2}^{\dagger}a_{2}\right),\;\Gamma_{k}\underset{j=1}{\overset{N}{\sum}}D\left[\sigma_{k}^{\left(j\right)}\right]\leftrightarrow\Gamma_{k}D\left[a_{k}\right],    
\end{align}
and the number of photons in bosonic mode $1$ ($2$) corresponds to the number of qubits in state $|1\rangle$ ($|2\rangle$). 

Hence the $N$ qubits dynamics can be reduced to the following two-bosonic modes evolution: 
\begin{align}
\frac{d\rho}{dt}=-i\omega\left[a_{1}^{\dagger}a_{1}-a_{2}^{\dagger}a_{2},\rho\right]+\underset{i=1,2}{\sum}\Gamma_{i}\left(a_{i}\rho a_{i}^{\dagger}-\frac{1}{2}\left\{ \rho,a_{i}^{\dagger}a\right\} \right).    
\end{align}
This simplified dynamics allows us to perform ECQFI calculations up to $N=7$ and ISS calculations up to $N=10.$ The results are shown in Figs. \ref{fig:opt_ent} and \ref{fig:bloch_8}.

\section{Optimal quantum error correction protocols for \texorpdfstring{$G \perp \sigma_{z}$}{G perpendicular to sigma\_z}}
\label{app:optimal_qec_for_sigma_x}

The protocols in this section involve continuous approximate quantum error correction and follow the general construction in Ref. \cite{zhou2020optimal}.
In all protocols we define an error correction code $\left\{ |0_{L}\rangle,|1_{L}\rangle\right\},$
which constitute a logical qubit. 
The probe is subject to continuous syndrome measurement and recovery operations.
Let us denote the projection onto code space as $\Pi:=|0_{L}\rangle\langle0_{L}|+|1_{L}\rangle\langle1_{L}|,$ and the projection superoperator as $\mathcal{P}\left(\cdot\right):=\Pi\left(\cdot\right)\Pi.$
The superoperator of the projection to the orthogonal complement of the code space is denoted by \rev{$\mathcal{P}_{\perp}(\cdot):=(\mathbbm{1}-\Pi) (\cdot) (\mathbbm{1}-\Pi),$}
and the recovery superoperator is denoted by $\mathcal{R}.$
Hence a measurement and recovery channel of $\mathcal{P}+\mathcal{R}\mathcal{P}_{\perp}$ is sequentially applied to the probe.
This channel 
leads
to the following effective dynamics inside the code space \cite{ZhouNC18AchievingHeisenberg, zhou2020optimal}
\begin{align}
\begin{split}
&\frac{d\rho}{dt}=-i\left[\mathcal{P}\left(H\right),\rho\right]+\gamma_{j}\left(\mathcal{R}\mathcal{P}_{\perp}\left(L_{j}\rho L_{j}^{\dagger}\right)+\mathcal{P}\left(L_{j}\rho L_{j}^{\dagger}\right)\rev{-}1/2\left\{ \mathcal{P}\left(L_{j}^{\dagger}L_{j}\right),\rho\right\} \right).
\end{split}
\label{eq:effective_master_eq}
\end{align}
We therefore get an effective Hamiltonian of 
\begin{align}
H_{\text{eff}}=\mathcal{P}\left(H\right)=\left(\begin{array}{cc}
\langle0_{L}|H|0_{L}\rangle & \langle0_{L}|H|1_{L}\rangle\\
\langle1_{L}|H|0_{L}\rangle & \langle1_{L}|H|1_{L}\rangle
\end{array}\right).
\label{eq:effective_hamiltonian}
\end{align}
The dissipative part is also modified as shown in Eq. \ref{eq:effective_master_eq}, where the non-Hermitian Hamiltonian becomes $\mathcal{P}\left(L_{j}^{\dagger}L_{j}\right),$
and the jump operators are dictated by $\mathcal{R}\mathcal{P}_{\perp}\left(L_{j}\rho L_{j}^{\dagger}\right)+\mathcal{P}\left(L_{j}\rho L_{j}^{\dagger}\right)$.

\subsection{Optimal Heisenberg limit protocols for \texorpdfstring{$G \perp \sigma_z$}{G perpendicular to sigma\_z}} 
As mentioned in the main text, for $G\perp \sigma_z$ HL can be achieved iff for every decay level $e_j$:
$\min\left(\gamma_{1,j},\gamma_{2,j}\right)=0.$
For all of these cases the noiseless QFI limit of $4N^2T^2$ is saturated using the following ancilla-free QEC code
\begin{align*}
 \left\{ |0_{L}\rangle=|+\rangle|+\rangle,|1_{L}\rangle=|-\rangle|-\rangle\right\}.    
\end{align*}
Let us show this using the effective dynamics of Eq. \ref{eq:effective_hamiltonian}.
The effective Hamiltonian 
is  $H_\text{eff}=\mathcal{P} \left(H \right)=2\omega \sigma_z^{L},$ 
where $\sigma_z^{L}=|0_L\rangle\langle0_L|-|1_L\rangle\langle1_L|.$
The syndrome measurement is detecting whether any of the sensors decayed to the erasure states $\left\{ |e_{j}\rangle\right\} _{j=1}^{k}$.
The recovery operation is 
\begin{align}
&\left\{ |e_{j}\rangle|+\rangle,|+\rangle|e_{j}\rangle\right\} \rightarrow|+\rangle|+\rangle,\;\left\{ |e_{j}\rangle|-\rangle,|-\rangle|e_{j}\rangle\right\} \rightarrow\left(-1\right)^{f_{j}}|-\rangle|-\rangle,\\
& \text{where } f_{j}=\begin{cases}
2 & \gamma_{1,j}\geq\gamma_{2,j}\\
1 & \text{otherwise.}
\end{cases}
\label{supp_eq:recovery_HS}
\end{align}
Formally, the recovery channel is given by
\rev{\begin{align}
\mathcal{R}\left(\cdot\right)=\underset{j}{\sum}\underset{l=1}{\overset{2}{\sum}}R_{j,l}\left(\cdot\right)R^{\dagger}_{j,l},
\label{supp_eq:recovery_HS_2}
\end{align}
with Kraus operators $R_{j,1}=|0_{L}\rangle\langle e_{j},+|+\left(-1\right)^{f_{j}}|1_{L}\rangle\langle e_{j},-|,$
$R_{j,2}=|0_{L}\rangle\langle+,e_{j}|+\left(-1\right)^{f_{j}}|1_{L}\rangle\langle-,e_{j}|.$}
Observe that $\mathcal{P}\left(L_{j}^{\dagger}L_{j}\right)=\mathbbm{1},$ 
$\mathcal{P}\left(L_{j}\rho L_{j}^{\dagger}\right)=0,$
$\mathcal{R}\mathcal{P}_{\perp}\left(L_{j}\rho L_{j}^{\dagger}\right)=\rho.$
Hence the effective dynamics of Eq. \ref{eq:effective_master_eq}
is the noiseless $\frac{d\rho}{dt}=-i2\omega\left[\sigma_{z}^{L},\rho\right],$
such that the noiseless QFI $I=4N^2T^2=16T^2$ ($N=2$ in this case) is retrieved.

More generally, for any $N$-qubit code space 
\begin{align*}
\left\{ |0_{L}\rangle=|+\rangle^{N},|1_{L}\rangle=|-\rangle^{N}\right\},     
\end{align*}
the effective master equation is 
\begin{align*}
\frac{d\rho}{dt}=-iN\omega\left[\sigma_{z}^{L},\rho\right],    
\end{align*}
and thus the optimal noiseless QFI of $4N^2T^2$ is achieved.

\subsection{Optimal standard quantum limit protocols for \texorpdfstring{$G \perp \sigma_z$}{G perpendicular to sigma\_z}}
Let us consider the case of a single erasure state $|e\rangle$. 
As shown in the main text and in Appendix~ \ref{app:sec_sequential_ecqfi_bounds} the ultimate sequential bound is $\Iult=\frac{4}{\min\{ \gamma_{1},\gamma_{2}\}}NT$.
Assuming without loss of generality that $\gamma_1 \leq \gamma_2,$
we show that the optimal approximate QEC code is given by
\begin{align*}
\left\{ |0_{L}\rangle=\left(\epsilon|1\rangle+\sqrt{1-\epsilon^{2}}|2\rangle\right)|0\rangle_a,|1_{L}\rangle=\left(\epsilon|1\rangle-\sqrt{1-\epsilon^{2}}|2\rangle\right)|1\rangle_a\right\} \text{ with } \epsilon \rightarrow 0,     
\end{align*}
such that $\Iult$ is asymptotically attainable using an optimal logical spin squeezed state of this code.
The recovery channel is given by
\begin{align*}
\mathcal{R}\left(\cdot\right)=\left(|0_{L}\rangle\langle e,0_a|-|1_{L}\rangle\langle e,1_a|\right)\left(\cdot\right)\left(|e,0_a\rangle\langle0_{L}|-|e,1_a\rangle\langle1_{L}|\right).    
\end{align*}
Let us now derive the effective dynamics. The effective Hamiltonian is (\ref{eq:effective_hamiltonian})
\begin{align*}
H_{\text{eff}}=\omega\left(\begin{array}{cc}
\langle0_{L}|\sigma_{x}|0_{L}\rangle & \langle0_{L}|\sigma_{x}|1_{L}\rangle\\
\langle1_{L}|\sigma_{x}|0_{L}\rangle & \langle1_{L}|\sigma_{x}|1_{L}\rangle
\end{array}\right)=\omega\left(\begin{array}{cc}
2\epsilon\sqrt{1-\epsilon^{2}} & 0\\
0 & -2\epsilon\sqrt{1-\epsilon^{2}}
\end{array}\right).    
\end{align*} 
To derive the effective dissipative dynamics, note that
\begin{align}
\begin{split}
&(i)\;\mathcal{P}\left(\gamma_{2}\sigma_{2}^{\dagger}\sigma_{2}\right)=\gamma_{2}(1-\epsilon^{2})\mathbbm{1}, \; \mathcal{P}\left(\gamma_{1}\sigma_{1}^{\dagger}\sigma_{1}\right)=\gamma_{1}\epsilon^{2}\mathbbm{1} \\
& (ii)\; \mathcal{R}\circ\mathcal{P}_{\perp}\left(\gamma_{2}\sigma_{2}\rho\sigma_{2}^{\dagger}\right)=\gamma_2(1-\epsilon^2)\rho\\
& (iii) \; \mathcal{R}\circ\mathcal{P}_{\perp}\left(\gamma_{1}\sigma_{1}\rho\sigma_{1}^{\dagger}\right)=\gamma_{1} \epsilon^2\sigma_{z}^{L}\rho\sigma_{z}^{L},
\label{supp_eq: qec_sigma_x_1}
\end{split}
\end{align}
and $\mathcal{P}\left(\gamma_{j}\sigma_{j}\rho\sigma_{j}^{\dagger}\right)=0.$
$(ii)$ follows from the fact that $\mathcal{R}$ corrects the dominant $\sigma_2$ error, and $(iii)$ is due to the fact that $\mathcal{R}$ does not correct $\sigma_1$ errors and induces a $\sigma_z^{L}$ error after the recovery.

Inserting Eq. \ref{supp_eq: qec_sigma_x_1} in Eq. \ref{eq:effective_master_eq} we obtain the following effective master equation
\begin{align}
\frac{d\rho}{dt}=-i\omega\left[2\epsilon\sqrt{1-\epsilon^{2}}\sigma_{z}^{L},\rho\right]+\gamma_{1}\epsilon^2\left(\sigma_{z}^{L}\rho\sigma_{z}^{L}-\rho\right).    
\end{align}
We thus obtain an effective Hamiltonian of $\omega2\epsilon\sqrt{1-\epsilon^{2}}\sigma_{z}^{L}$ and an effective dephasing noise of $D\left[\sqrt{\gamma_{1}\epsilon}\sigma_{z}^{L}\right].$
For this dephasing model the sequential ECQFI is
\begin{align*}
\Iult=\frac{4\epsilon^{2}\left(1-\epsilon^{2}\right)}{\epsilon^{2}\gamma_{1}}T=4\frac{1-\epsilon^{2}}{\gamma_{1}}T \underset{\epsilon\rightarrow 0}{\longrightarrow} \frac{4 T}{\gamma_1}.
\end{align*}
\rev{The resulting QFI approaches the ultimate bound as $\epsilon \rightarrow 0.$ Because the effective logical noise is dephasing, this bound can be attained asymptotically using an appropriate logical spin-squeezed state \cite{kitagawa1993squeezed,zhou2021asymptotic}.}
Note that this optimal code corresponds to the limit of vanishing signal and vanishing noise.

\section{Applications to qubits with thermal noise: optimal protocols}

We derive in this Appendix an erasure conversion scheme for achieving $\Iult$ in the $N\rightarrow \infty$ limit for qubits with thermal noise and $G=\sigma_z$. This scheme uses erasure conversion and then applies the same optimal spin squeezing of Appendix \ref{sec:appendix_squeezed_spin_erasure}
on the effective erasure qubits.
We consider qubits evolving under the following standard thermal Lindblad master equation
\begin{align}
\frac{d\rho}{dt}=-i\left[\omega\sigma_{z},\rho\right]+\Gamma_{2}D\left[\sigma_{-}\right]\rho+\Gamma_{1}D\left[\sigma_{+}\right]\rho,    
\end{align}
where $\omega$ is the parameter to be estimated, and $\sigma_{-}=|1\rangle\langle2|$, $\sigma_{+}=\sigma_{-}^{\dagger}$.
The case of $\Gamma_{2}=0$ corresponds to amplitude damping noise.
$\Iult$ of this problem is the same as with erasure decay rates of $\Gamma_{1},\Gamma_{2}$, i.e., Eq. 2 in the main text. 
Let us define the two-qubit code space (we change the notation to $|1_L\rangle,|2_L\rangle$ just for this case): $\{|1_L\rangle=|11\rangle,|2_L\rangle=|22\rangle\}.$
This logical code-space is the effective erasure qubit. 
The jump operators $\sigma_{-,+}^{(j)}$ map  this code space into the orthogonal erasure subspace spanned by $\left\{ |12\rangle,|21\rangle\right\} $.
Consider a protocol in which a continuous parity measurement of the physical qubits is performed, indicating whether they are in the code-space or decayed to the erasure subspace. Whenever an erasure is detected we simply discard the relevant two physical qubits.
We can use Eq. \ref{eq:effective_master_eq} to derive an effective logical master equation for this protocol.
Note that for any jump operators $L_j$, $\mathcal{P}\left(L_{j}\rho L_{j}^{\dagger}\right)=0,$ and that $\mathcal{P}\left(\sigma_{-}^{\left(j\right)}\sigma_{+}^{\left(j\right)}\right)=|1_{L}\rangle\langle1_{L}|,$ $\mathcal{P}\left(\sigma_{+}^{\left(j\right)}\sigma_{-}^{\left(j\right)}\right)=|2_{L}\rangle\langle2_{L}|,$
$\mathcal{R}\mathcal{P}_{\perp}\left(\sigma_{-}^{\left(j\right)}\rho\sigma_{+}^{\left(j\right)}\right)=\sigma_{2}\rho\sigma_{2}^{\dagger}$ , $\mathcal{R}\mathcal{P}_{\perp}\left(\sigma_{+}^{\left(j\right)}\rho\sigma_{-}^{\left(j\right)}\right)=\sigma_{1}\rho\sigma_{1}^{\dagger}.$
Applying this to Eq. \ref{eq:effective_master_eq} yields the following effective erasure dynamics of each logical qubit
\begin{align}
\frac{d\rho}{dt}=-i\left[2\omega\sigma_{z}^{L},\rho\right]+2\Gamma_{1}D\left[\sigma_{1}\right]\rho+2\Gamma_{2}D\left[\sigma_{2}\right]\rho,   
\end{align}
where  $\sigma_{j}=|e\rangle\langle j|$
are the erasure jump operators, and $\sigma_{z}^{L}=|2_{L}\rangle\langle2_{L}|-|1_{L}\rangle\langle1_{L}|$.
We thus obtained a logical dynamics of $L_1=\sqrt{2 \Gamma_{1}} \sigma_{1},L_2=\sqrt{2 \Gamma_{2}} \sigma_{2},$ $H=2\omega\sigma_{z}^{L},$
with $N/2$ logical qubits.
The corresponding $\Iult$ is the same as the original thermal noise problem: 
\rev{$\Iult=4\frac{16(N/2)T}{(\sqrt{2\Gamma_{1}}+\sqrt{2\Gamma_{2}})^{2}}=\frac{16NT}{(\sqrt{\Gamma_{1}}+\sqrt{\Gamma_{2}})^{2}}.$}
Since optimal spin squeezed states saturate $\Iult$ given erasure noise, then an identical logical spin squeezed state saturates $\Iult$ in this case.

We remark that this protocol is a simpler alternative to other protocols in the literature \cite{zhou2021asymptotic} that convert the noise into an effective dephasing noise. In the effective dephasing protocol of Ref. \cite{zhou2021asymptotic} each logical qubit consists of one physical qubit with two noiseless ancillas. In the erasure conversion protocol, each logical qubit consists of two physical qubits without any noiseless ancillas. 
\rev{The implementation considered here uses continuous parity monitoring} \cite{kubica2023erasure,levine2024demonstrating}.

\section{Single qubit ECQFI bounds and protocols}
We calculate the optimal product state QFI, $\Iprod$, using extended channel QFI (ECQFI) bounds.
Given a quantum channel, $\mathcal{E},$
the ECQFI optimizes the QFI over all possible input states (including input states that have entanglement of the probe with a noiseless ancilla)
$$\mathcal{I}(\mathcal{E})=\underset{\rho_0}{\max}I(\mathcal{E}(\rho_0)).$$
Let $\textbf{K}=\{K\}_{i=1}^m$ be the $d_{in} \times d_{out}$ Kraus operators of the channel,
the ECQFI is then given by
the following minimization problem \cite{demkowicz2012elusive,kolodynski2013efficient}
\begin{align}
4\;\underset{h}{\min}||\left(\mathbf{\overset{.}{K}}-ih\mathbf{K}\right)^{\dagger}\left(\mathbf{\overset{.}{K}}-ih\mathbf{K}\right)||, 
\label{eq:ecqfi_minimization}
\end{align}
where the minimization is over all hermitian $m\times m$ $h$ matrices.
This problem can be recast as an SDP,
denoting $\dot{\tilde{\textbf{K}}} = \dot{\textbf{K}} - ih \textbf{K} $, the SDP reads \cite{kolodynski2013efficient}
\begin{equation*}
    \mathcal{I}(\mathcal{E}) = 4 \underset{h}{\min}{\lambda},  \;\;\;\; \text{subject to} \;\;
\begin{pmatrix}
\sqrt{\lambda} I_{d_{in}} & \dot{\tilde{K}}^{\dagger}  \\
\dot{\tilde{K}} & \sqrt{\lambda} I_{m\cdot d_{out}}
\end{pmatrix} \geq 0.
\end{equation*}
While in general the optimal input state may require entanglement with a noiseless ancilla, we will show that this is not the case in our erasure noise problems. Therefore $\Iprod$ is given by $$\Iprod=\underset{t}{\max}\frac{NT}{t}\mathcal{I}\left(\mathcal{E}\left(t\right)\right)$$. 

\subsection{\texorpdfstring{$\Iprod$}{Iprod} for erasure qubits with \texorpdfstring{$G \parallel \sigma_z$}{G parallel to sigma\_z}}

\begin{figure}[h]
    \centering
    \begin{subfigure}[t]{0.4\textwidth}
        \centering
        \includegraphics[width=\linewidth]{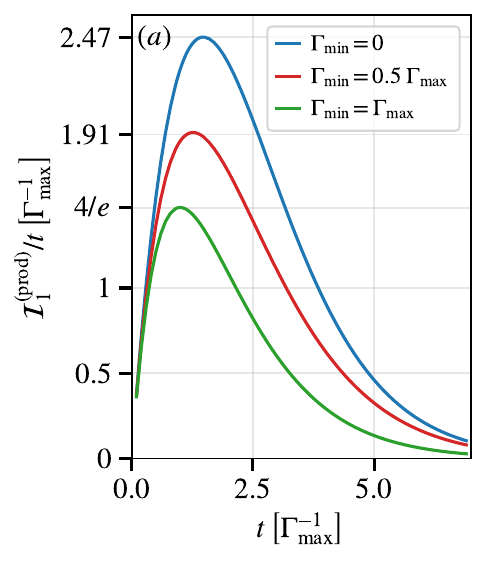}
        \caption*{}
        \label{fig:sz_cqfi_a}
    \end{subfigure}%
    \begin{subfigure}[t]{0.4\textwidth}
        \centering
        \includegraphics[width=\linewidth]{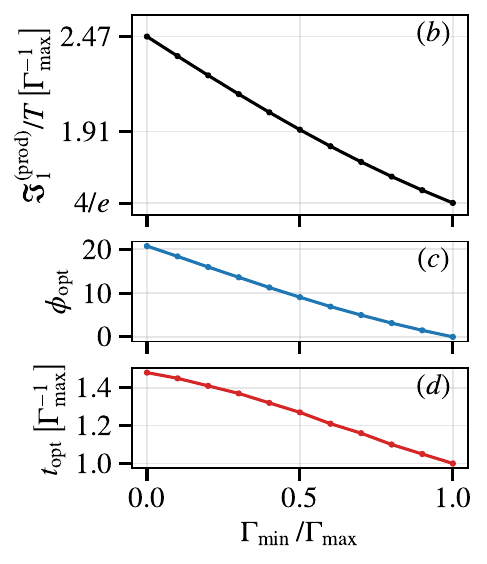}
        \caption*{}
        \label{fig:sz_cqfi_bcd}
    \end{subfigure}
    \caption{ Single qubit ECQFI bounds and protocols for $G=\sigma_{z}$ for various erasure decay configurations.
    (a) ECQFI as a function of time for different $\Gamma_{\min}$. (Eq. \ref{eq:ecqfi_sz_multi_erasure}) (b) The $t$-optimized ECQFI as a function of $\Gamma_{\min}/\Gamma_{\max}$. (c) and (d) present the optimal protocol. (c) Optimal $\phi$ as a function of the ratio $\Gamma_{\min}/\Gamma_{\max}.$ (Eq. \ref{supp_eq:optimal_phi_sz}) (d) Optimal measurement time, $t_{\text{opt}},$ as a function of $\Gamma_{\min}/\Gamma_{\max}.$} \label{fig:sz_cqfi}
\end{figure}

Let us first consider the case of a single erasure state $|e\rangle.$
The Kraus operators for an initial state with $d_{in}=2$ are given by
\begin{align*}
\textbf{K}=\left[\begin{pmatrix}\sqrt{\eta_{1}}e^{-i2\omega t} & 0\\
0 & \sqrt{\eta_{2}}\\
0 & 0
\end{pmatrix},\begin{pmatrix}0 & 0\\
0 & 0\\
\sqrt{1-\eta_{1}} & 0
\end{pmatrix},\begin{pmatrix}0 & 0\\
0 & 0\\
0 & \sqrt{1-\eta_{2}}
\end{pmatrix}\right]^{T},
\end{align*}
where $\eta_1=e^{-\gamma_1t},$
$\eta_2=e^{-\gamma_2t}.$
The derivatives
\begin{align*}
\overset{\cdot}{K_{1}}=\begin{pmatrix}-2it\sqrt{\eta_{1}}e^{-2i\omega t} & 0\\
0 & 0\\
0 & 0
\end{pmatrix},\end{align*}
and the others are zero.
The minimization problem of Eq. \ref{eq:ecqfi_minimization} can then be solved analytically. 
It is simple to see that the minimum is obtained for $h$ that takes the form of $h=\left(\begin{array}{ccc}
\lambda & 0 & 0\\
0 & 0 & 0\\
0 & 0 & 0
\end{array}\right)$, with some $\lambda\in\mathbb{R}$.
With this choice of $h$ we have 
\begin{align*}
\left(\mathbf{\overset{.}{K}}-ih\mathbf{K}\right)^{\dagger}\left(\mathbf{\overset{.}{K}}-ih\mathbf{K}\right)=\left(\begin{array}{cc}
\left(2t+\lambda\right)^{2} \eta_1\\
 & \lambda^{2}\eta_2
\end{array}\right).    
\end{align*}
The minimum is obtained when $\left(2t+\lambda\right)^{2}\eta_1=\lambda^{2}\eta_2.$
This equation has two solutions: $\lambda=\frac{2t\sqrt{\eta_{1}}}{\sqrt{\eta_{2}}-\sqrt{\eta_{1}}},-\frac{2t\sqrt{\eta_{1}}}{\sqrt{\eta_{2}}+\sqrt{\eta_{1}}}$, where the minimal operator norm is obtained for the latter: $\lambda=-\frac{2t\sqrt{\eta_{1}}}{\sqrt{\eta_{2}}+\sqrt{\eta_{1}}}$. The ECQFI is thus
\begin{align}
\mathcal{I}=\frac{16t^{2}\eta_{1}\eta_{2}}{\left(\sqrt{\eta_{1}}+\sqrt{\eta_{2}}\right)^{2}}=\frac{16t^{2}}{\left(e^{\gamma_{1}t/2}+e^{\gamma_{2}t/2}\right)^{2}}.
\label{supp_eq:ecqfi_sz_single_erasure}
\end{align}
This ECQFI is \rev{verified} with numerical SDP calculation for $\gamma_1=1$ and $\gamma_2\in\{0, 0.5, 1\}$ and is plotted in Fig. \ref{fig:sz_cqfi}.

We now generalize this analysis to an arbitrary number of erasure levels. 
In the multi-erasure case the Kraus operators are

\begin{align*}
\left\{ K_{1}=e^{-\frac{t}{2}\left(4i\omega+\Gamma_{1}\right)}\ket{1}\bra{1}+e^{-\frac{t}{2}\left(4i\omega+\Gamma_{2}\right)}\ket{2}\bra{2}\right\} \cup\left\{ \sqrt{\frac{\gamma_{1,j}}{\Gamma_1}(1-e^{-\gamma_{1,j}}})\ket{e_j}\bra{1},\sqrt{\frac{\gamma_{2,j}}{\Gamma_2}(1-e^{-\gamma_{2,j}})}\ket{e_j}\bra{2}\right\} _{j}.  
\end{align*}
Similar to the single erasure case only $\overset{\cdot}{K_{1}}$ is non-zero. Addressing the optimization problem of Eq. \ref{eq:ecqfi_minimization} the optimal $h$ is $h_{i,j}=\begin{cases}
\lambda & i=j=1\\
0 & \text{otherwise}
\end{cases}$ for some $\lambda \in \mathbb{R}.$ We then obtain
\begin{align*}
\left(\mathbf{\overset{.}{K}}-ih\mathbf{K}\right)^{\dagger}\left(\mathbf{\overset{.}{K}}-ih\mathbf{K}\right)=\left(\begin{array}{cc}
\left(2t+\lambda\right)^{2}e^{-\Gamma_{1}t}\\
 & \lambda^{2}e^{-\Gamma_{2}t}
\end{array}\right).    
\end{align*}
Minimizing the operator norm  over $\lambda$ leads to the ECQFI
\begin{align}
\mathcal{I}=\frac{16t^{2}}{\left(e^{\Gamma_{1}t/2}+e^{\Gamma_{2}t/2}\right)^{2}},
\label{eq:ecqfi_sz_multi_erasure}
\end{align}
which is the same as the single erasure state where $\gamma_{1} \mapsto \Gamma_{1}, \gamma_{2} \mapsto \Gamma_{2}.$

{\bf{The optimal initial state}}---
    For an initial state $\rho_0 = \frac{1}{2}\left( \mathbbm{1} + \cos{\phi}\,\sigma_x + \sin{\phi}\, \sigma_z\right),$
the evolved state, obtained by solving the master equation, is

\begin{align*}
\rho(t)=
\begin{pmatrix}
\frac{1}{2} e^{-\gamma_1 t} \big(1 + \sin\phi\big) &
\frac{1}{2} e^{-\frac{1}{2} t (\gamma_1 + \gamma_2 + 4i\omega)} \cos\phi &
0 \\
\frac{1}{2} e^{-\frac{1}{2} t (\gamma_1 + \gamma_2 - 4i\omega)} \cos\phi &
\frac{1}{2} e^{-t \gamma_2} \big(1 -\sin\phi\big) &
0 \\
0 & 0 &
\frac{1}{2} e^{-t(\gamma_1 + \gamma_2)} \big[e^{\gamma_1 t} \left( 2 e^{t \gamma_2} + \sin\phi-1\right)-e^{t \gamma_2} (1 + \sin\phi) 
\big]
\end{pmatrix}.
\end{align*}

The resulting QFI is:
$$I=\frac{8 t^2 \cos^2\phi}{
 e^{\gamma_1 t} (1 - \sin \phi) + 
  e^{t \gamma_2} (1 + \sin\phi)}.$$

Solving for the optimal $\phi$ yields
\begin{align}
\sin \phi_{\text{opt}} = \tanh{\frac{t \gamma_1}{4} \left(1 - \frac{\gamma_2}{\gamma_1}\right)}.
\label{supp_eq:optimal_phi_sz}
\end{align}
Inserting this $\phi_{\text{opt}}$ we obtain the ECQFI of Eq. \ref{supp_eq:ecqfi_sz_single_erasure}.
This optimal state can be also written as $\ket{\psi_{\text{opt}}} = \sqrt{p_{\text{opt}}}\ket{1}  + \sqrt{1-p_{\text{opt}}}\ket{2}$
with 
$p_{\text{opt}}=1/2\left(1+\tanh\dfrac{t_{\text{opt}}}{4}(\Gamma_{1}-\Gamma_{2})\right).$
This confirms that in this case the ECQFI is saturated without any noiseless ancilla.
For multi-erasure case the same results are obtained where each $\gamma_{i}$ is replaced by $\Gamma_{i}=\sum_{j} \gamma_{i,j}.$

{\bf{Time optimized ECQFI}}---
The ECQFI (Eq. \ref{supp_eq:ecqfi_sz_single_erasure}) represents an optimization over all initial states. To obtain $\Iprod$ we need to further optimize this ECQFI over measurement time.  
Let us assume without loss of generality $\gamma_1 \leq \gamma_2$. If $\gamma_1 = 0$, the optimal time is

\begin{equation}
    t_{\text{opt}} = \dfrac{1}{\gamma_2}\left[ 1 + 2 W_0\left(\dfrac{1}{2\sqrt{e}}\right) \right] \approx \dfrac{1.48}{\gamma_2},
\end{equation}
where $W_0$ is the Lambert function. Hence, $\Iprod$ in this case is

$$\Iprod=\frac{64 W_0^2\left(\frac{1}{2\sqrt{e}}\right)}{1 + 2 W_0\left( \frac{1}{2\sqrt{e}} \right)}  \dfrac{NT}{\gamma_2}
\approx \dfrac{2.47 N T}{\gamma_2}$$

For comparison, if the initial state $\rho_0$ is located at the equator of the Bloch sphere, which is the typical scenario in experiments, then $t_{\text{opt}} \approx 1.27\, \gamma_2^{-1}$ and $ I \approx 2.22\,T/\gamma_2$. Thus, $\phi_{\text{opt}}$ improves the QFI by around $11\%$. 

In the symmetric case of $\gamma_2=\gamma_1$, we have $t_{\text{opt}} = \gamma_2^{-1}$ and $\Iprod$ is thus $\Iprod=4NT/(\gamma_2 e).$

For arbitrary $\gamma_2,\gamma_1,$ we solve numerically and the results are shown in Fig. \ref{fig:sz_cqfi}.

\subsection{\texorpdfstring{$\Iprod$}{Iprod} for erasure qubits with \texorpdfstring{$G \perp \sigma_z$}{G perpendicular to sigma\_z}}\label{app:C2}

We show in this section that the single qubit QFI in this case is upper bounded by
\begin{align}
I\leq\frac{16}{\max\left(\gamma_{1},\gamma_{2}\right)}T=\frac{16}{\gamma_{2}}T,    
\end{align}
where without loss of generality  $\gamma_1\leq \gamma_2.$
To this end, we study the single qubit ECQFI of this problem for different values of $\omega,\gamma_1, \gamma_2,$
and the relevant optimal protocols.
This upper bound establishes the significant gain that can be obtained with QEC in this case (the QFI with QEC is given in \ref{supp_eq:asym_bound_sigma_x}).
Let us first study the case of a single erasure noise $D\left( \sqrt{\gamma_2} \sigma_{2} \right)$ and show that the ECQFI in this case satisfies $I \leq \frac{16}{\gamma_2}T,$ with equality when $\omega \rightarrow 0.$



\begin{figure}[h]
    \centering
    \begin{subfigure}[t]{0.67\textwidth}
        \centering
        \includegraphics[width=\linewidth]{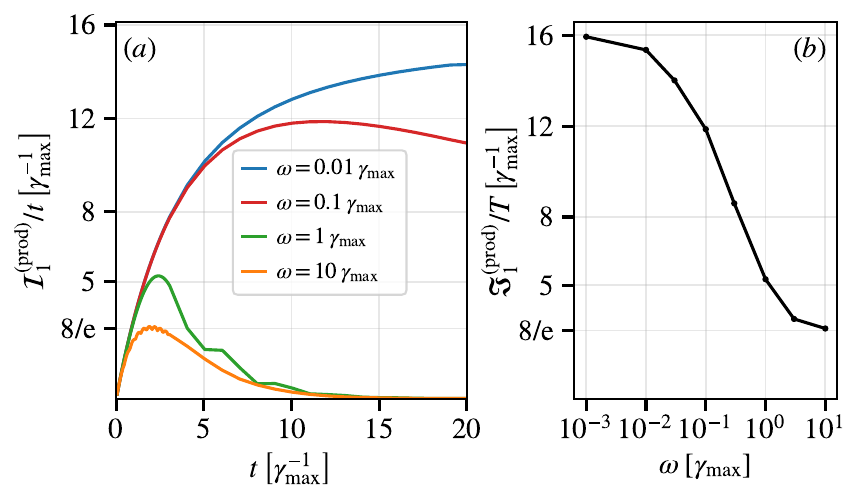}
    \end{subfigure}%
    \begin{subfigure}[t]{0.33\textwidth}
        \centering
        \includegraphics[width=\linewidth]{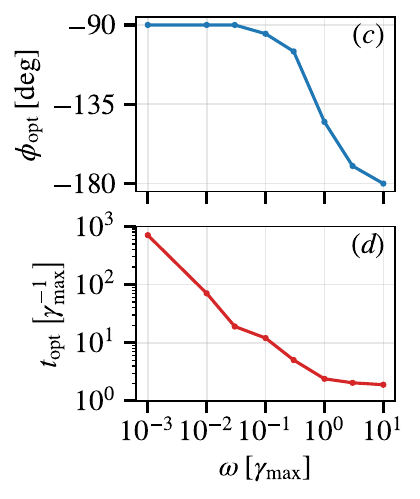}
    \end{subfigure}
    \caption{
    Single qubit ECQFI bounds and protocols for $G=\sigma_{x}$ and a single erasure decay configuration: $\gamma_{\min}=0.$
    (a) ECQFI as a function of time for different $\omega$. (b) The $t$-optimized ECQFI as a function of $\omega$. The optimal ECQFI rate is obtained as $\omega \rightarrow 0$, such that the $t$-optimized ECQFI converges to $16 T/\gamma_{\max}$ in this limit. (c) and (d) present the optimal protocol. (c) optimal $\phi$ as a function of $\omega.$ (d) optimal measurement time, $t_{\text{opt}}$ as a function of $\omega.$} \label{supp_fig:ECQFI_sigma_x}
\end{figure}

{\textbf{Single erasure decay:}} Let us consider the case of a single Lindblad term of $\gamma_{2}\left(\sigma_{2}\rho\sigma_{2}^{\dagger}-1/2\left\{ \sigma_{2}^{\dagger}\sigma_{2},\rho\right\} \right).$
We compute the Kraus operators numerically, and use them to calculate the ECQFI with SDP. The ECQFI  for different values of $\omega$ is shown in the Fig. \ref{supp_fig:ECQFI_sigma_x} (solid lines). We observe from the numerical results
that for every $t$ the largest ECQFI is obtained for $\omega \rightarrow 0$:
$\mathcal{I}\left(\omega,t\right)\leq \mathcal{I}\left(\omega\rightarrow0,t\right).$ Specifically, the $t$-optimized ECQFI lies within the range of $\frac{8}{\gamma_{2}e}T\leq\mathfrak{I}_{1}^{(\text{prod})}\leq\frac{16}{\gamma_{2}}T,$
where the upper bound is attainable in the limit of $\omega \ll \gamma_{2}$
and the lower bound corresponds to the limit of $\omega \gg \gamma_{2}.$ The transition between the two regimes is shown in Fig. \ref{supp_fig:ECQFI_sigma_x}.

\begin{figure}[t!]
\includegraphics[width=8cm]{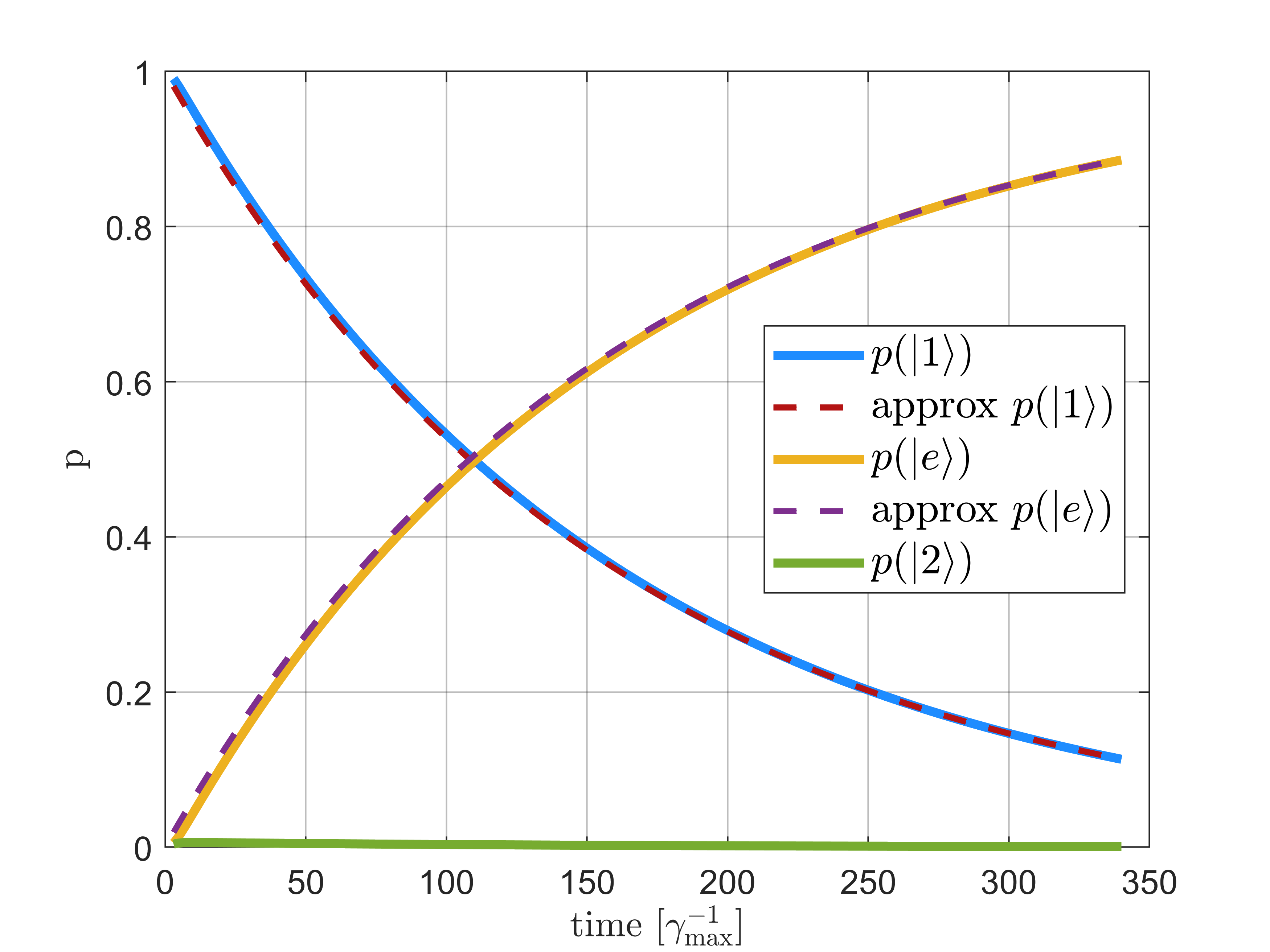}
\caption{Measurement probabilities of $|1\rangle,|2\rangle,$ and $|e\rangle$ as a function of time for a single erasure noise setting ($\gamma_{1}=0$), given an input state of $|1\rangle.$ In this illustration $\omega=0.04 \gamma_2 \ll\gamma_{2}.$
Dashed lines show the adiabatic elimination approximation,
which agrees with the exact dynamics.}
\label{fig:dynamics_sx_erasure_adiabatic}
\end{figure}

Given $\omega \ll \gamma_{2},$  the optimal initial state is $|\psi_0\rangle=|1\rangle.$
The optimality of this state is intuitive since it does not suffer from erasure noise, only $|2\rangle$ decays to $|e \rangle$. With this initial state, the decay of the dynamics is induced from the coupling of $|1\rangle$ to $|2\rangle$ through $H$ and the erasure noise of $|2\rangle.$
This is thus a coupling mediated decay, and since $\omega \ll \gamma_2$ (because $\omega \rightarrow 0$), we can obtain an effective dynamics using the theory of adiabatic elimination \cite{reiter2012effective,azouit2016adiabatic,guillaud2023quantum}.
The idea is that due to the fast decay of $|1\rangle,$ we can adiabatically eliminate it and derive an effective dynamics that involves only the subspace of $\left\{ |1\rangle,|e\rangle\right\} .$
Using Eqs. (3), (4) of Ref. \cite{reiter2012effective} we get that the effective jump operator is $L_{\text{eff}}=LH_{\text{NH}}^{-1}V_{+},$
where in our case $V_{+}=\omega|2\rangle\langle1|$, $H_{\text{NH}}=\frac{\gamma_{2}}{2}|2\rangle\langle2|$, $L=\sqrt{\gamma_{2}}\sigma_{2}$, hence we obtain
$L_{\text{eff}}=2\frac{\omega}{ \sqrt{\gamma_{2}} }\sigma_{1}.$
The effective master equation is thus
\begin{align*}
\overset{\cdot}{\rho}=4\frac{\omega^2}{\gamma_2}\left(\sigma_{1}\rho\sigma_{1}^{\dagger}-1/2\left\{ \sigma_{1}^{\dagger}\sigma_{1},\rho\right\} \right).
\end{align*}
We show in Fig. \ref{fig:dynamics_sx_erasure_adiabatic} that this effective dynamics indeed fits the exact time evolution for $t\gg 1/\gamma_2.$ 
The $t$-optimized ECQFI given this effective dynamics is $16T/\gamma_2$ \cite{sekatski2022optimal,gardner2025lindblad}.
The numerical results of Fig. \ref{supp_fig:ECQFI_sigma_x} confirm that  this expression matches the $t$-optimized ECQFI obtained from the full dynamics.

Let us show this more explicitly and evaluate the optimal $t$ needed to saturate $t$-optimized ECQFI.
we solve the  master equation given an initial state of $|\psi_0\rangle=|1\rangle,$ 
and expand $\rho(t)$ to second order in $\omega/\gamma_\rev{2}$.
we find
\begin{align}
\rho(t) &
=
\begin{bmatrix}
\frac{4 \omega^2}{\gamma_2^2} + \frac{4\omega^2}{\gamma_2^2}e^{-t \gamma_2} - \frac{8\omega^2}{\gamma_2^2}e^{-\frac{t \gamma_2}{2}} &
-\frac{2 i \omega}{\gamma_2} + \frac{2 i  \omega}{\gamma_2}e^{-\frac{t \gamma_2}{2}} &
0 \\[8pt]
\frac{2 i \omega}{\gamma_2} - \frac{2 i \omega}{\gamma_2}e^{-\frac{t \gamma_2}{2}} &
1- \frac{4 t \omega^2}{\gamma_2}  + \frac{8 \omega^2}{\gamma_2^2} - \frac{8  \omega^2}{\gamma_2^2}e^{-\frac{t \gamma_2}{2}} &
0 \\[8pt]
0 &
0 &
\frac{4 t \omega^2}{\gamma_2}
- \frac{12 \omega^2}{\gamma_2^2} - \frac{4 e^{-t \gamma_2} \omega^2}{\gamma_2^2} + \frac{16  \omega^2}{\gamma_2^2} e^{-\frac{t \gamma_2}{2}}
\end{bmatrix}
+ \mathcal{O}((\omega/\gamma_2)^3) \\
& \overset{(*)}{\approx}
\begin{bmatrix}
\frac{4 \omega^2}{\gamma_2^2} & -\frac{2 i \omega}{\gamma_2} & 0 \\
\frac{2 i \omega}{\gamma_2} & e^{-\frac{4 t \omega^2}{\gamma_2}} + \frac{8 \omega^2}{\gamma_2^2} & 0 \\
0 & 0 & 1 - e^{-\frac{4 t \omega^2}{\gamma_2}} - \frac{12 \omega^2}{\gamma_2^2}
\end{bmatrix},
\label{suppeq: rho_adiabatic_elimination}
\end{align}
where in $(*)$ we eliminated the terms that decay as $e^{-\gamma_{2}t}$ and approximated  $1- \frac{4 t \omega^2}{\gamma_2} $ by $e^{-\frac{4 t \omega^2}{\gamma_2}}$.
We can neglect in Eq. \ref{suppeq: rho_adiabatic_elimination}
all terms that go as $\omega^2/\gamma_{2}^2$ and do not grow with $t.$
we obtain that all $\left\{|2\rangle\langle j|, |j\rangle \langle2|  \right\}$ can be neglected such that we are left with the following effective state: $e^{-\frac{4t\omega^{2}}{\gamma_{2}}}|1\rangle\langle1|+\left(1-e^{-\frac{4t\omega^{2}}{\gamma_{2}}}\right)|e\rangle\langle e|.$
This is exactly the adiabatic elimination approximation and the approximated probabilities in Fig. \ref{fig:dynamics_sx_erasure_adiabatic}.
Given this effective state the QFI is  
\begin{align}
I=\frac{(\partial_{\omega}p(|e\rangle)^{2}}{p(|e\rangle) \rev{(1-p(|e\rangle))}}\approx\frac{64\omega^{2}t^{2}e^{-t\frac{8t\omega^{2}}{\gamma_{2}}}}{\gamma_{2}^{2}\left(1-e^{-\frac{4t\omega^{2}}{\gamma_{2}}}\right)}\underset{t\ll\gamma_{2}/\omega^{2}}{\approx}16\frac{t}{\gamma_{2}}, 
\label{supp_eq:QFI_sx_adiabatic_elimination_app}
\end{align}
which is the bound that we observe numerically. Eq. \ref{supp_eq:QFI_sx_adiabatic_elimination_app} also implies that this QFI is saturable by measuring the erasure state $|e\rangle.$

While the adiabatic elimination approximation suggests that the bound is achieved for any measurement time $t\ll \gamma_2/\omega^2,$ the optimal time analysis is more subtle. 
To determine the optimal measurement time and the leading order correction to the QFI due to finite $\omega,$ we must also include the stationary term in $p(|e\rangle)$: $p\left(|e\rangle\right)=1-e^{-4t\frac{\omega^{2}}{\gamma_{2}}}-12\omega^{2}/\gamma_{2}^{2} +\mathcal{O} \left( (\omega/\gamma_2)^3 \right).$
Inserting this expression into the QFI calculation of Eq. \ref{supp_eq:QFI_sx_adiabatic_elimination_app} leads to 
$I\approx\frac{16t}{\gamma_{2}}-\frac{48}{\gamma_{2}^{2}}\left(1+2\omega^{2}t^{2}\right)+\mathcal{O}\left(\left(\omega/\gamma_{2}\right)^{3}\right).$
Optimizing $I/t$ shows that the optimal measurement time is: $t = \frac{1}{\sqrt{2} \, \omega}$.
The resulting $t$-optimized ECQFI, including the leading order correction in $\omega$ is: $\mathfrak{I}^{(\text{prod})}_{1}\approx\frac{16}{\gamma_{2}}T\left(1-6\sqrt{2}\frac{\omega}{\gamma_{2}}\right).$

   In the limit of $\omega \gg \gamma_{2},$ 
   we observe that the $t$-optimized ECQFI equals to $8T/\left(\gamma_2 e\right)$ (see Fig. \ref{supp_fig:ECQFI_sigma_x}).
   In this limit the dynamics is equivalent to that of a symmetric erasure decay rate of $\gamma_2/2,$
   since
   the state rotates fast between $|2\rangle$ and $|1\rangle$ and thus experiences an effective symmetric erasure rate of $\gamma_{2}/2.$
   The $t$-optimized ECQFI in this symmetric case is $4T/\left( e \gamma_{2}/2 \right)=8T/ \left( e \gamma_{2} \right),$
   which explains the $\omega \gg \gamma_2$ limit. 
   
  

\begin{figure}[h]
    \centering
    \begin{subfigure}[t]{0.66\textwidth}
        \centering
        \includegraphics[width=\linewidth]{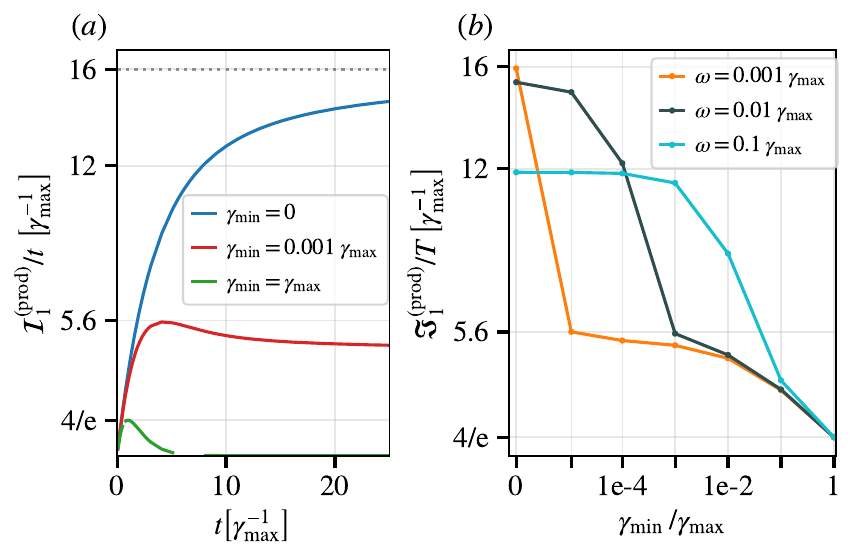}
    \end{subfigure}%
    \begin{subfigure}[t]{0.33\textwidth}
        \centering
        \includegraphics[width=\linewidth]{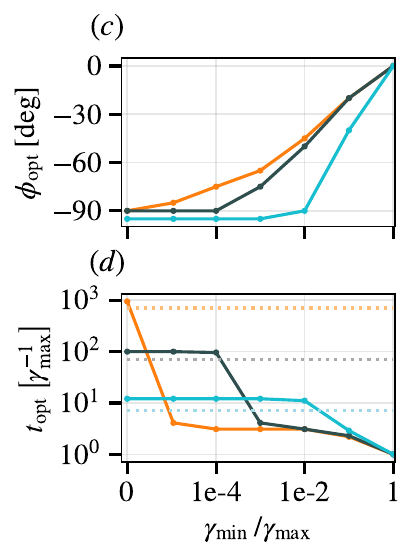}
    \end{subfigure}
    \caption{Single qubit ECQFI bounds and protocols for $G=\sigma_{x}$ for different $\gamma_{\min}/\gamma_{\max}$ in the small $\omega$ limit ($\omega \ll \gamma_{\max}$). 
    (a) ECQFI as a function of time for $\omega=0.01 \gamma_{\max}$ and for different $\gamma_{\min}$.
    The $t$-optimized ECQFI ranges from $4T/(e\gamma_{\max})$ in the symmetric case of $\gamma_{\max}=\gamma_{\min}$ to $16T/\gamma_{\max}$ for $\gamma_{\min}=0$.
    (b) The $t$-optimized ECQFI as a function of $\gamma_{\min}$ for different values of $\omega.$ For $\gamma_{\min}< \omega^2/\gamma_{\max}$ the limit of $16T/\gamma_{\max} $ is achieved. For $\gamma_{\min}> \omega^2/\gamma_{\max}$ the $t$-optimized ECQFI converges to $5.6 T/\gamma_{\max}.$ (c) and (d) present the optimal protocol. (c) optimal $\phi$ as a function of the ratio $\gamma_{\min}/\gamma_{\max}.$ (d) optimal measurement time, $t_{\text{opt}}$ as a function of $\gamma_{\min}/\gamma_{\max}.$ Dashed lines indicate $t_{\text{opt}} = \frac{1}{\sqrt{2}\,\omega}$, the small-$\omega$ limit at $\gamma_{\min} = 0$.
} \label{fig:channel_qfi_x_g2}
\end{figure}

{\textbf{Two erasure decays:}} Let us extend the analysis to the case of $\gamma_{\text{min}}=\gamma_1 > 0.$
We numerically calculate the ECQFI for $\omega=0.001 \gamma_{\text{max}}, 0.01 \gamma_{\text{max}}, 0.1 \gamma_{\text{max}}$ and for different values $\gamma_{\text{min}}$. 
The results are shown in Fig. \ref{fig:channel_qfi_x_g2}.
We want to consider the $t$-optimized ECQFI in the limit of $\omega ,\gamma_{\min} \ll \gamma_{\max}.$
Note that this expression depends on the order of limits.
The behavior of the $t$-optimized ECQFI depends on ratio between $\omega^{2}/\gamma_{\max}$ and $\gamma_{\min}.$
If $\gamma_{\min}$ is small enough such that $\omega^{2}/\gamma_{\max} > \gamma_{\min}$ then the optimal QFI of $16 T/\gamma_{\max}$ is retrieved.
In the other limit of  $\omega^{2}/\gamma_{\max} < \gamma_{\min},$
the QFI does not saturate this bound and converges to $\approx 5.6 T/\gamma_{\max}$. 



\section{Continuous erasure detection protocols}
We derive the precision limits obtained with continuous erasure detection, i.e., by continuously detecting whether the qubit decayed to one of the erasure states and resetting the state accordingly.
We remark that in these protocols no entanglement  or ancillary qubits are used, and the only assumptions are: 1. ability to perform continuous erasure detection. 2. ability to perform a fast reset of the state. 

The optimal continuous erasure protocols depend on $G,$ hence we split into the two cases of $G \parallel \sigma_{z}$ or $G \perp \sigma_z,$ and analyze several continuous erasure detection strategies.

\subsection{\texorpdfstring{$\Icd$}{Iprod} for erasure qubits with \texorpdfstring{$G \parallel \sigma_z$}{G parallel to sigma\_z}}

\textit{Strategy 1: fixed $\tau$ after decay: }
In this strategy we initialize the qubit to a state $|\psi_{0}\rangle=\alpha|1\rangle+\beta|2\rangle$ and let it evolve, where the erasure states are being continuously detected.
If an erasure is detected we reset the state to $|\psi_{0}\rangle.$ 
In this case we lose the information about $\omega$ that was encoded in the qubit. 
If no erasure has been detected by time $\tau$ since the last reset, we perform an optimal measurement on the qubit.
This measurement provides the information about $\omega.$
The analysis in this section assumes for simplicity a single erasure state with decay rates $\gamma_{1},\gamma_{2}$. The generalization to multi-erasure states is immediate: $\gamma_{i}$ just needs to be replaced by $\Gamma_{i}=\sum_{j} \gamma_{i,j}.$ 

Let us first calculate the FI obtained with this strategy for the identical erasure rates case $\gamma:=\gamma_{1}=\gamma_{2}$.
In this case the optimal initial state is $|\psi_{0}\rangle=\frac{1}{\sqrt{2}}\left(|1\rangle+|2\rangle\right)$, and the FI is given by
\begin{align}
I=N_{\text{cycles}} \, p\left(\text{success}\right) \, 4\tau^{2},
\label{supp_eq:FI_in_michal_method}
\end{align}
where $N_{\text{cycles }}$ is the number of independent experiments (each one terminated by either decay or qubit measurement) and $p\left(\text{success}\right)$ is the probability that no decay occurred in the cycle.
Note that $p\left(\text{success}\right)=e^{-\gamma\tau}$, since the decay corresponds to a Poisson process with decay rate $\gamma$.
In the limit of $T\rightarrow\infty$ , $N_{\text{cycles}}\rightarrow\frac{T}{\langle\tau_{\text{cycle}}\rangle}$, where $\langle\tau_{\text{cycle}}\rangle$ is the average cycle time.
$\langle\tau_{\text{cycle}}\rangle$ is given by
\begin{align}
\langle\tau_{\text{cycle}}\rangle=e^{-\gamma\tau}\tau+\int_{0}^{\tau_{\text{\text{meas}}}}te^{-\gamma t}\gamma\,dt=\frac{1-e^{-\gamma\tau}}{\gamma}.
\label{supp_eq:average_cycle_time}
\end{align}
Inserting Eq. \ref{supp_eq:average_cycle_time} in 
Eq. \ref{supp_eq:FI_in_michal_method}, using $N_{\text{cycles}}\rightarrow\frac{T}{\langle\tau_{\text{cycle}}\rangle},$
we obtain
\begin{align}
I=\frac{\gamma T}{1-e^{-\gamma\tau}}e^{-\gamma\tau}4\tau^{2}. 
\label{eq:michal_method_FI}
\end{align}
Optimizing over $\tau$ we obtain 
\begin{align}
\underset{\tau}{\text{max }}\frac{\gamma T}{1-e^{-\gamma\tau}}e^{-\gamma\tau}4\tau^{2} \approx \boxed{\frac{2.6}{\gamma}T}. 
\label{eq:product_state_continuous_erasure_bound}
\end{align}

This FI expression can be proven rigorously using the renewal-reward theorem \cite{ross1995stochastic}. 
Let us define the random variables $\left(X_{n},R_{n}\right)_n,$
where $X_n$ is the duration of the $n$-th cycle and $R_n$ is the FI obtained from this cycle
\begin{align*}
R_{n}=\begin{cases}
4\tau^{2} & X_{n}=\tau\\
0 & X_{n}<\tau.
\end{cases}    
\end{align*}
Both $\left(X_{n}\right)_{n}$ and $\left(R_{n}\right)_{n}$ are i.i.d random variables with averages 
$\langle R_{j}\rangle=e^{-\gamma\tau}4\tau^{2},$
$\langle X_{j}\rangle=\langle\tau_{\text{cycle}}\rangle=\frac{1-e^{-\gamma\tau}}{\gamma}.$
Hence according to the renewal-reward theorem \cite{ross1995stochastic}
\begin{align*}
\underset{T\rightarrow\infty}{\text{lim}}\frac{\sum_{n}R_{n}}{T}\rightarrow\frac{\langle R_{j}\rangle}{\langle X_{j}\rangle}=\frac{\gamma}{1-e^{-\gamma\tau}}e^{-\gamma\tau}4\tau^{2},    
\end{align*}
which is the QFI rate of Eq. \ref{eq:michal_method_FI}.

Let us generalize this to the case of unequal erasure decay rates.
This analysis requires also optimization over the input state. Without loss of generality we can assume that the optimal input is on the $\sigma_{x}-\sigma_{z}$ plane: $|\psi_{0}\rangle=\cos\left(\pi/4-\phi/2\right)|1\rangle+\sin\left(\pi/4-\phi/2\right)|2\rangle,$
where the optimal $\phi$ will be found later.
Let us calculate the average cycle time and average FI per cycle.
The probability of having no decay until time $t$ is $\cos\left(\pi/4-\phi/2\right)^{2}e^{-\gamma_{1}t}+\sin\left(\pi/4-\phi/2\right)^{2}e^{-\gamma_{2}t},$
and the probability of decaying at time $t$ is
$\cos\left(\pi/4-\phi/2\right)^{2}e^{-\gamma_{1}t}\gamma_{1}dt+\sin\left(\pi/4-\phi/2\right)^{2}e^{-\gamma_{2}t}\gamma_{2}dt.$
Hence the average cycle time is
\begin{align*}
\begin{split}
&\langle\tau_{\text{cycle}}\rangle=\left(\cos\left(\pi/4-\phi/2\right)^{2}e^{-\gamma_{1}\tau}+\sin\left(\pi/4-\phi/2\right)^{2}e^{-\gamma_{2}\tau}\right)\tau\\
&+\int_{0}^{\tau}t\left(\cos\left(\pi/4-\phi/2\right)^{2}e^{-\gamma_{1}t}\gamma_{1}+\sin\left(\pi/4-\phi/2\right)^{2}e^{-\gamma_{2}t}\gamma_{2}\right)dt=\\
&=\frac{\left(1-e^{-\gamma_{2}\tau}\right)}{\gamma_{2}}\sin\left(\pi/4-\phi/2\right)^{2}+\frac{\left(1-e^{-\gamma_{1}\tau}\right)}{\gamma_{1}}\cos\left(\pi/4-\phi/2\right)^{2}.
\end{split}
\end{align*}
Let us now calculate the average FI per cycle. The QFI given a successful cycle is
\begin{align*}
I_{\text{cycle}}=4\tau^{2}\frac{e^{-(\gamma_{1}+\gamma_{2})\tau}\cos^{2}(\phi)}{\Big(e^{-\gamma_{2}\tau}\sin\left(\pi/4-\phi/2\right)^{2}+\;e^{-\gamma_{1}\tau}\cos\left(\pi/4-\phi/2\right)^{2}\Big)^{2}}.    
\end{align*}
The average FI per cycle is thus
\begin{align*}
p\left( \text{success} \right) I_{\text{cycle}}=4\tau^{2}\frac{e^{-(\gamma_{1}+\gamma_{2})\tau}\cos^{2}(\phi)}{e^{-\gamma_{2}\tau}\sin\left(\pi/4-\phi/2\right)^{2}+\;e^{-\gamma_{1}\tau}\cos\left(\pi/4-\phi/2\right)^{2}}.    
\end{align*}
The total FI is thus
\begin{align}\label{supp_eq:cd_sz}
I=4\tau^{2}T\frac{e^{-(\gamma_{1}+\gamma_{2})\tau}4p\left(1-p\right)}{\left(e^{-\gamma_{2}\tau}\left(1-p\right)+\;e^{-\gamma_{1}\tau}p\right)\left(\frac{\left(1-e^{-\gamma_{2}\tau}\right)}{\gamma_{2}}\left(1-p\right)+\frac{\left(1-e^{-\gamma_{1}\tau}\right)}{\gamma_{1}}p\right)},    
\end{align}
where we denoted $p=\cos\left(\pi/4-\phi/2\right)^{2},$
$1-p=\sin\left(\pi/4-\phi/2\right)^{2}$ for brevity.
With this notation $|\psi_{0}\rangle=\sqrt{p}|1\rangle+\sqrt{1-p}|2\rangle.$
The optimal FI is then obtained by optimizing over $p$ and $\tau.$

\begin{figure}[h!]
    \centering
    \begin{subfigure}[t]{0.45\textwidth}
        \centering
        \includegraphics[width=\linewidth]{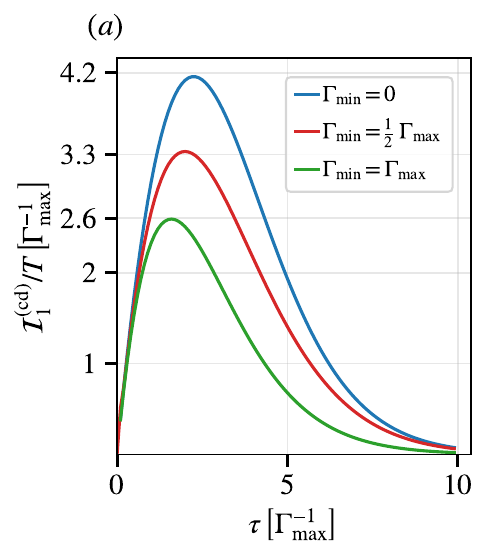}
        \caption*{}
        \label{fig:e_detection_qfi_z_a}
    \end{subfigure}%
    \begin{subfigure}[t]{0.41\textwidth}
        \centering
        \includegraphics[width=\linewidth]{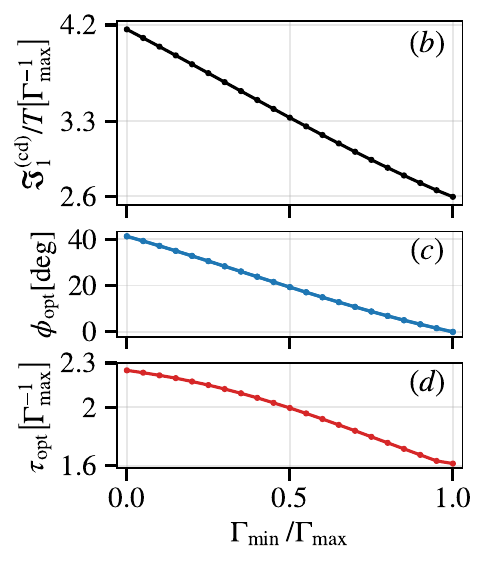}
        \caption*{}
        \label{fig:e_detection_qfi_z_bcd}
    \end{subfigure}
    \caption{QFI achieved with continuous erasure detection strategy (strategy 1: fixed $\tau$ after measurement) for $G=\sigma_{z}$ and different values of $\Gamma_{\min}$ (Eq. \ref{supp_eq:cd_szf}). (a) Continuous erasure detection QFI rate as a function of $\tau$ for different $\Gamma_{\min}, \Gamma_{\max}.$
    (b) QFI rate achieved with this continuous erasure strategy for different $\Gamma_{\min}/\Gamma_{\max}.$ (c) and (d) present the optimal protocol: optimal initial state ($\phi_{\text{opt}}$, extracted from \ref{supp_eq:cd_sz_opt_state}) and optimal evolution time $\tau.$
    } \label{fig:e_detection_qfi_z}
\end{figure}

Optimizing over $p$ (for any $\tau$) yields \begin{equation}\label{supp_eq:cd_sz_opt_state}
p_{\text{opt}}=\frac{1}{1+\sqrt{\frac{l_1 m_1}{l_2 m_2}}},\end{equation}
where $l_i=e^{-\gamma_i \tau},$
$m_i=(1-e^{-\gamma_i \tau})/\gamma_{i}.$
Inserting this $p_{\text{opt}}$ into Eq. \ref{supp_eq:cd_sz} we obtain
\begin{align}\label{supp_eq:cd_szf}
\begin{split}
&I=16\tau^{2}T\frac{\sqrt{\frac{\int_{0}^{\tau}e^{\gamma_{1}t}\,\text{d}t}{\int_{0}^{\tau}e^{\gamma_{2}t}\,\text{d}t}}}{\left(1+\sqrt{\frac{\int_{0}^{\tau}e^{\gamma_{1}t}\,\text{d}t}{\int_{0}^{\tau}e^{\gamma_{2}t}\,\text{d}t}}\right)\left(\int_{0}^{\tau}e^{\gamma_{1}t}\,\text{d}t+\sqrt{\int_{0}^{\tau}e^{\gamma_{1}t}\,\text{d}t}\sqrt{\int_{0}^{\tau}e^{\gamma_{2}t}\,\text{d}t}\right)}\\
&=\boxed{\frac{16\tau^{2}T}{\left(\sqrt{\int_{0}^{\tau}e^{\gamma_{1}t}\,\text{d}t}+\sqrt{\int_{0}^{\tau}e^{\gamma_{2}t}\,\text{d}t}\right)^{2}}. }
\end{split}    
\end{align}
$\Icd$ is then obtained by optimizing this expression over $\tau$, which matches the expression in the main text.
This bound has the same functional form as the single‑qubit ECQFI (Eq.\ref{eq:ecqfi_sz_multi_erasure}), but with effectively reduced decay rates. To see this, note that the single qubit ECQFI  is retrieved by replacing $\int_{0}^{\tau}e^{\gamma_{j}t}\,\text{d}t/\tau$  with $e^{\gamma_{j}\tau}$. We therefore obtain effective decay rates of \rev{$\frac{1}{\tau}\text{log}\left(\int_{0}^{\tau}e^{\gamma_{j}t}\,\text{d}t/\tau\right)$}. Since $\int_{0}^{\tau}e^{\gamma_{j}t}\,\text{d}t/\tau\leq e^{\gamma_{j}\tau}$ the effective decay rate is indeed smaller than the original decay rate $\gamma_{j}$.
The optimal values of $p,\tau$ and the resulting FI are shown in Fig. \ref{fig:e_detection_qfi_z}.

\textit{Strategy 2: fixed $\tau$: }
This strategy underperforms strategy 1, but we provide it here for completeness. 
In this strategy we divide the experiment into periods of $\tau,$
where at the end of every such period
the qubit is being measured in the optimal basis.
Unlike the previous strategy, here the measurement is done in constant, predetermined time lapses.
This strategy also leads to a considerable improvement compared to the ECQFI, yet it underperforms strategy 1. Let us calculate the optimal QFI with this strategy for the case of $\gamma\coloneqq\gamma_1=\gamma_2.$ 
In this case the optimal initial state is $|\psi\rangle=\frac{1}{\sqrt{2}}\left(|1\rangle+|2\rangle\right)$
due to the identical decay rates.
The FI given a total time $T$ is
\begin{align}
I=\frac{T}{\tau}I_{\tau},
\label{eq:continuous_erasure_QFI_rate}
\end{align}
where $I_{\tau}$ is the FI obtained with this protocol given that the measurement time is $\tau.$
$I_{\tau}$ includes all possible trajectories, i.e., all possible outcomes of the erasure measurements and the final qubit measurement at $\tau.$
Let us denote the outcomes of the erasure measurements as $\vec{x}_{e}$ and the outcome of the final qubit measurement as $x_{q}.$
$I_{\tau}$ then equals to
\begin{align*}
  I_{\tau}=
  \underset{\vec{x}_{e},x_{q}}{\sum}\frac{\left(\partial_{\omega}p\left(\vec{x}_{e},x_{q}\right)\right)^{2}}{p\left(\vec{x}_{e},x_{q}\right)}=\underset{\vec{x}_{e}}{\sum}p\left(\vec{x}_{e}\right)\underset{x_{q}}{\sum}\frac{\left(\partial_{\omega}p\left(x_{q}|\vec{x}_{e}\right)\right)^{2}}{p\left(x_{q}|\vec{x}_{e}\right)},
\end{align*}
where the last equality is due to the fact that $p\left(\vec{x}_{e}\right)$ is independent of $\omega$
and thus $\partial_{\omega}p\left(\vec{x}_{e},x_{q}\right)=p\left(\vec{x}_{e}\right)\partial_{\omega}p\left(x_{q}|\vec{x}_{e}\right).$
It can be seen that the term $\sum_{x_{q}}\frac{\left(\partial_{\omega}p\left(x_{q}|\vec{x}_{e}\right)\right)^{2}}{p\left(x_{q}|\vec{x}_{e}\right)}$ depends only on the \rev{time of the most recent decay}.
Denoting this time as $t_{\text{last}}$ we have that
\begin{align*}
\begin{split}
&\sum_{x_{q}}\frac{\left(\partial_{\omega}p\left(x_{q}|\vec{x}_{e}\right)\right)^{2}}{p\left(x_{q}|\vec{x}_{e}\right)}=4\left(\tau-t_{\text{last}}\right)^{2}\\
&\Rightarrow I_{\tau}=\underset{\vec{x}_{e}}{\sum}p\left(\vec{x}_{e}\right)4\left(\tau-t_{\text{last}}\left(\vec{x}_{e}\right)\right)^{2}\end{split}   
\end{align*}
The probability for a trajectory with $k$ decays , where the last decay is at a fixed time $t_{\text{last}}$ ($t_{1},...,t_{k-1}$ represent the other decay times) is
\begin{align*}
\begin{split}
&p\left(k\text{ decays},t_{k}=t_{\text{last}}\right)=\gamma \,dt_{\text{last}} \, e^{-\gamma\tau}\int_{0}^{t_{\text{last}}}\gamma\;dt_{k-1}...\int_{0}^{t_{3}}\gamma\;dt_{2}\int_{0}^{t_{2}}\gamma\;dt_{1}=\\
&\gamma \,dt_{\text{last}} \, e^{-\gamma\tau}\frac{\left(\gamma t_{\text{last}}\right)^{k-1}}{\left(k-1\right)!}.   \end{split} 
\end{align*}
Hence the FI from all trajectories with $k$ decays is
\begin{align*}
\gamma e^{-\gamma\tau}\int_{0}^{\tau}\frac{\left(\gamma t_{\text{last}}\right)^{k-1}}{\left(k-1\right)!}4\left(\tau-t_{\text{last}}\right)^{2}dt_{\text{last}}=\frac{8}{\left(\gamma\right)^{2}}e^{-\gamma\tau}\frac{\left(\gamma\tau\right)^{k+2}}{\left(k+2\right)!}    
\end{align*}
$I_{\tau}$ obtained by summing over all trajectories is thus
\begin{align*}
I_{\tau}=\frac{8}{\left(\gamma\right)^{2}}e^{-\gamma\tau}\underset{k=0}{\overset{\infty}{\sum}}\frac{\left(\gamma\tau\right)^{k+2}}{\left(k+2\right)!}=\frac{8}{\left(\gamma\right)^{2}}e^{-\gamma\tau}\left[e^{\gamma\tau}-\left(\gamma\tau+1\right)\right].    
\end{align*}
We now want to optimize $I$  (Eq. \ref{eq:continuous_erasure_QFI_rate}) over $\tau$ which leads to
\begin{align*}
\underset{\tau}{\text{max }}\frac{8}{\gamma}\frac{1}{\gamma\tau}\left[1-e^{-\gamma\tau}\left(\gamma\tau+1\right)\right]T\approx\boxed{\frac{2.4}{\gamma}T.}    
\end{align*}

\textbf{Continuous erasure detection with entangled multi-probe strategies:}
\rev{We briefly explore whether} combining continuous erasure detection with entangled states leads to improved FI, i.e., FI larger than the product state bound of Eq. \ref{eq:product_state_continuous_erasure_bound}.
We focus in this part on the equal decay rates case $\gamma:=\gamma_1=\gamma_2.$
We first show that with GHZ states the same product state bound of Eq. \ref{eq:product_state_continuous_erasure_bound} is obtained, \rev{hence it does not lead to an improvement}. 

Given a GHZ state $\frac{1}{\sqrt{2}}\left(|1\rangle^{N}+|2\rangle^{N}\right),$
Using strategy 1, the average FI per cycle is
$e^{-N\gamma\tau}4N^{2}\tau^{2}$
and the average cycle time is $\frac{1-e^{-N\gamma\tau}}{N\gamma}.$
Defining $\gamma'=N\gamma$ we obtain that the FI is given by 
\begin{align*}
I=N^{2}\frac{T}{1-e^{-\gamma'\tau}}e^{-\gamma'\tau}4\gamma'\tau^{2}.    
\end{align*}
The optimization over $\tau$ is thus identical to \rev{that in} Eq. \ref{eq:product_state_continuous_erasure_bound}, \rev{upon} replacing $\gamma$ \rev{by} $\gamma'=N\gamma$ and \rev{including an overall} factor of $N^2$.
\rev{This yields}
\begin{align}
\underset{\tau}{\text{max }}N^{2}\frac{T}{1-e^{-\gamma'\tau}}e^{-\gamma'\tau}4\gamma'\tau^{2} \approx N\frac{2.6}{\gamma}T,
\label{eq:cd_GHZ_time_optimized}
\end{align}
which \rev{coincides with the} bound \rev{of} Eq. \ref{eq:product_state_continuous_erasure_bound}, and thus no improvement is obtained. 

\rev{For $N=2$ qubits we can optimize this strategy over the input state.} From symmetry, we can restrict ourselves to states that take the form of
\begin{align*}
|\psi_{0}\rangle=\frac{\cos\left(\theta\right)}{\sqrt{2}}\left(|11\rangle+|22\rangle\right)+\frac{\sin\left(\theta\right)}{\sqrt{2}}\left(|12\rangle+|21\rangle\right).
\end{align*}
Let us calculate the average FI per cycle. 
If no decay occurred until $\tau$ we obtain the FI $4\text{var}_{|\psi_0\rangle}\left(  \sigma_{z,1}+\sigma_{z,2}\right)=16\cos\left(\theta\right)^{2}\tau^{2}.$
If a decay occurred in one of the qubits we immediately measure the remaining qubit in its optimal basis and reset the two qubits afterwards.
\rev{Hence the state of the surviving qubit is $\rho_{\text{surv}}=\frac{1}{2}\left(\begin{array}{cc}
1 & \sin\left(2\theta\right)e^{-2i\omega t}\\
\sin\left(2\theta\right)e^{2i\omega t} & 1
\end{array}\right),$}
and the FI obtained after a decay is thus
\rev{$4\sin\left(2\theta\right)^{2}t^{2}.$}
The average FI per cycle is thus
\rev{
\begin{align*}
\begin{split}
&e^{-2\gamma\tau}\tau^{2}16\cos\left(\theta\right)^{2}+4\sin\left(2\theta\right)^{2}\int_{0}^{\tau}t^{2}e^{-2\gamma t}2\gamma dt=\\
&e^{-2\gamma\tau}\tau^{2}16\cos\left(\theta\right)^{2}+\frac{2\sin\left(2\theta\right)^{2}}{\gamma^{2}}\left[1-e^{-2\gamma\tau}\left(1+2\gamma\tau+2\gamma^{2}\tau^{2}\right)\right].
\end{split}
\end{align*}
}
The average cycle time is $\langle \tau_{\text{cycle}}\rangle=\frac{1-e^{-2\gamma\tau}}{2\gamma}. $
We thus want to optimize \rev{the FI rate} 
\rev{
\begin{align*}
\frac{e^{-2\gamma\tau}\tau^{2}16\cos\left(\theta\right)^{2}+\frac{2\sin\left(2\theta\right)^{2}}{\gamma^{2}}\left[1-e^{-2\gamma\tau}\left(1+2\gamma\tau+2\gamma^{2}\tau^{2}\right)\right]}{\frac{1-e^{-2\gamma\tau}}{2\gamma}}    
\end{align*}}
over \rev{ both $\theta$ and $\tau.$}
\rev{The optimum is obtained at $\theta=0,$ which corresponds to the GHZ state, and yields the same FI as in Eq. \ref{eq:cd_GHZ_time_optimized}.
Hence the optimal state with this strategy is the GHZ state and the optimal FI coincides with the product state CD FI, $\Icd$. We thus obtain no improvement for $N=2$ even after optimizing over the input state.
We did not identify a two qubit entangled CD protocol that outperforms $\Icd$. Determining whether more general CD strategies, particularly with entangled inputs for larger $N$, can provide an advantage is left for future work.}

\subsection{\texorpdfstring{$\Icd$}{Icd} for erasure qubits with \texorpdfstring{$G \perp \sigma_z$}{G perp to sigma\_z}}


We focus on the single erasure noise decay. Without loss of generality, let $\gamma_1=0.$ and $\gamma \coloneqq\gamma_2.$
We consider again strategy 1, fixed $\tau$ after reset, that was described earlier.
Unlike the $G=\sigma_{z}$ case, now the decay vs. no-decay events provide information about $\omega$.
The optimal protocol employs the input state $|\psi_{0}\rangle=|1\rangle$, i.e., the noiseless state in the $\omega =0$ limit, 
and $\tau \rightarrow \infty.$
Hence in this case
the protocol consists of the following repeated cycles:
initialize the state in $|\psi_{0}\rangle=|1\rangle,$
let it evolve with continuous erasure detection, wait until an erasure is detected and reset the state.  
We show that this protocol yields $I=16T/\gamma$
for every $\omega$.

Let us denote the probability density function of the decay time as $p_{\text{dec}}(t).$
Given the input state $|\psi_{0}\rangle=|1\rangle$ we have
\begin{align*}
p_{\text{dec}}\left(t\right)=\gamma\rho_{2,2}(t)=\frac{16e^{-\frac{t\gamma}{2}}\gamma\omega^{2}\sinh^{2}\left(\tfrac{t}{4}\sqrt{\gamma^{2}-16\omega^{2}}\right)}{\gamma^{2}-16\omega^{2}}.
\end{align*}
Using the renewal-reward theory we get that the total FI equals to: $I=\frac{T}{\langle\tau_{\text{cycle}}\rangle}I_{\text{cycle}},$
where $I_{\text{cycle}}$ is the FI per cycle and $\langle\tau_{\text{cycle}}\rangle$ is the average time of a cycle. In this case
\begin{align*}
\begin{split}
&I_{\text{cycle}}=\int_{0}^{\infty}\frac{\left(\partial_{\omega}p_{\text{dec}}\left(t\right)\right)^{2}}{p_{\text{dec}}\left(t\right)}\text{d}t=\frac{4\gamma^{2}+32\omega^{2}}{\omega^{2}\gamma^{2}},\\
&\langle\tau_{\text{cycle}}\rangle=\int_{0}^{\infty}tp_{\text{dec}}\left(t\right)\,\text{d}t=\frac{\gamma^{2}+8\omega^{2}}{4\omega^{2}\gamma}.
\end{split}
\end{align*}
Therefore
\begin{align*}
I=\frac{T}{\langle\tau_{\text{cycle}}\rangle}I_{\text{cycle}}=\frac{16}{\gamma}T.    
\end{align*}
Continuous erasure detection provides a modest improvement compared to the product protocols in this case:
while product-state protocols require asymptotically small $\omega$ to reach this limit, this protocol achieves it for any $\omega.$

\begin{figure}[h!]
   \centering
   \includegraphics[width=.45\linewidth]{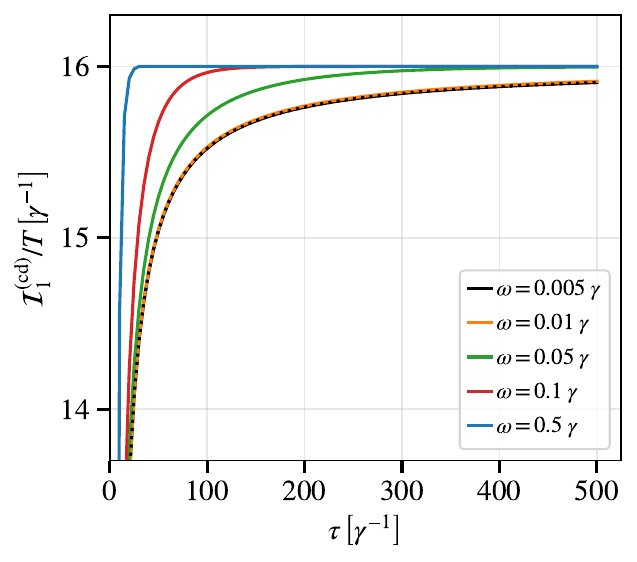}
   \caption{FI achieved with continuous erasure detection strategy for $G=\sigma_{x}$ with single erasure noise, i.e., $\gamma_{\min}=0$ and $\gamma_{\max}\equiv\gamma$, for different values of $\omega$. For every $\omega,$ $\tau_{\text{opt}}\rightarrow \infty$ and $\mathfrak{I}_1^{\text{(cd)}}=\frac{16T}{\gamma}.$} \label{supp_fig:erasure_detection_x}
\end{figure}

{\textbf{Finite $\tau$ analysis:}} In the analysis above we chose $\tau \rightarrow \infty.$
The optimality of this protocol is illustrated in Fig. \ref{supp_fig:erasure_detection_x}, where the FI is plotted as a function of $\tau$. Let us prove the optimality of $\tau \rightarrow \infty.$
To this end, we calculate the FI as a function of $\tau.$ For any finite $\tau$ the protocol is: reset after any erasure, if no erasure occurred for a period of $\tau$ since the last reset, perform optimal measurement on the qubit and reset the state.
The outcomes of each cycle are thus: $\left\{ \text{decay at }t<\tau,\text{ no decay until } \tau \text{ followed by a qubit measurement}\right\} .$
The probability of no decay until $\tau$  is $p_{\text{nodec}}\left(\tau\right)=\rho_{1,1}\left(\tau\right)+\rho_{2,2}\left(\tau\right)=\frac{e^{-\frac{\gamma\tau}{2}}\left(-16\omega^{2}+\gamma^{2}\cosh\!\left(\tfrac{\tau}{2}\sqrt{\gamma^{2}-16\omega^{2}}\right)+\gamma\sqrt{\gamma^{2}-16\omega^{2}}\sinh\left(\tfrac{\tau}{2}\sqrt{\gamma^{2}-16\omega^{2}}\right)\right)}{\gamma^{2}-16\omega^{2}}$. Hence the probabilities of the different outcomes are $\left\{ p_{\text{dec}}\left(t\right)\right\} _{t<\tau}\cup\left\{ p_{\text{nodec}}\left(\tau\right)p_{1},p_{\text{nodec}}\left(\tau\right)p_{2}\right\}$ , where $\left\{ p_{1},p_{2}\right\}$  are the outcomes of the optimal qubit measurement.
The FI per cycle is thus: 
\begin{align}
I_{\text{cycle}}=\int_{0}^{\tau}\frac{\left(\partial_{\omega}p_{\text{dec}}\left(t\right)\right)^{2}}{p_{\text{dec}}\left(t\right)}\;\text{d}t+\frac{\left(\partial_{\omega}p_{\text{nodec}}\left(\tau\right)\right)^{2}}{p_{\text{nodec}}\left(\tau\right)}+p_{\text{nodec}}\left(\tau\right)I_{\text{no-decay}},    
\end{align}
where $I_{\text{no-decay}}$ is the QFI of the qubit state conditioned on no decay until $\tau$ . The average time of a cycle is now
\begin{align}
\tau_{\text{cycle}}=\int_{0}^{\tau}tp_{\text{dec}}\left(t\right)\;\text{d}t+p_{\text{nodec}}\left(\tau\right)\tau.    
\end{align}
The total FI is given by $I=\frac{I_{\text{cycle}}}{\tau_{\text{cycle}}}T$. The numerical results of this calculation are shown in Fig. \ref{supp_fig:erasure_detection_x}





\clearpage
\twocolumngrid
\renewcommand{\addcontentsline}[3]{}
\bibliography{refs}
\end{document}